\documentclass[]{aastex63}

\usepackage{hyperref}

\def\pfrac#1#2{\left( \frac{#1}{#2} \right)}
\def\avg#1{\langle #1 \rangle}
\def\msol{M_\odot}\def\msol{M_\odot}

\def\fluence{{\cal F}}

\def\iso#1#2{\mbox{${}^{#2}{\rm #1}$}}
\def\be1#1{\iso{Be}{1#1}}
\def\al2#1{\iso{Al}{2#1}}
\def\ar3#1{\iso{Ar}{3#1}}
\def\ca4#1{\iso{Ca}{4#1}}
\def\k4#1{\iso{K}{4#1}}
\def\mn5#1{\iso{Mn}{5#1}}
\def\fe6#1{\iso{Fe}{6#1}}
\def\rb8#1{\iso{Rb}{8#1}}
\def\nb9#1{\iso{Nb}{9#1}}
\def\zr9#1{\iso{Zr}{9#1}}
\def\tc9#1{\iso{Tc}{9#1}}
\def\pd10#1{\iso{Pd}{10#1}}
\def\i12#1{\iso{I}{12#1}}
\def\cs13#1{\iso{Cs}{13#1}}
\def\sm14#1{\iso{Sm}{14#1}}
\def\gd15#1{\iso{Gd}{15#1}}
\def\dy15#1{\iso{Dy}{15#1}}
\def\hf18#1{\iso{Hf}{18#1}}
\def\gd15#1{\iso{Gd}{15#1}}
\def\pb20#1{\iso{Pb}{20#1}}
\def\bi21#1{\iso{Bi}{21#1}}
\def\u23#1{\iso{U}{23#1}}
\def\np23#1{\iso{Np}{23#1}}
\def\pu24#1{\iso{Pu}{24#1}}
\def\cm24#1{\iso{Cm}{24#1}}
\def\th23#1{\iso{Th}{23#1}}
\def\re18#1{\iso{Re}{18#1}}

\shorttitle{$r$-process, Near-Earth Supernovae \& Kilonovae}
\shortauthors{Wang et al.}
\graphicspath{{./}{figures/}}

\begin{document}

\title{{\em r}-Process Radioisotopes from Near-Earth Supernovae and Kilonovae}

\correspondingauthor{Xilu Wang}
\email{wangxl@ihep.ac.cn, xlwang811@gmail.com} 

\author[0000-0002-5901-9879]{Xilu Wang}
\affil{Department of Physics, University of California, Berkeley, CA 94720, USA}
\affil{Department of Physics, University of Notre Dame, Notre Dame, IN 46556, USA}
\affil{Key Laboratory of Particle Astrophysics, Institute of High Energy Physics, Chinese Academy of Sciences,  Beijing, 100049, China}
\affil{N3AS Collaboration}

\author[0000-0002-2881-7982]{Adam M. Clark}
\affiliation{Department of Physics, University of Notre Dame, Notre Dame, IN 46556, USA}

\author[0000-0002-7399-0813]{John Ellis}
\affiliation{Theoretical Physics and Cosmology Group, Department of Physics, King's College London, London WC2R 2LS, UK}
\affiliation{NICPB, R\"avala pst.~10, 10143 Tallinn, Estonia}
\affiliation{Theoretical Physics Department, CERN, CH-1211 Geneva 23, Switzerland}

\author[0000-0002-3876-2057]{Adrienne F. Ertel}
\affiliation{Department of Astronomy, University of Illinois, Urbana, IL 61801, USA}
\affiliation{Illinois Center for the Advanced Study of the Universe, University of Illinois, Urbana, IL 61820}

\author[0000-0002-4188-7141]{Brian D. Fields}
\affiliation{Department of Astronomy, University of Illinois, Urbana, IL 61801, USA}
\affiliation{Department of Physics, University of Illinois, Urbana, IL 61801, USA}
\affiliation{Illinois Center for the Advanced Study of the Universe, University of Illinois, Urbana, IL 61820}

\author[0000-0002-2786-5667]{Brian J. Fry}
\affiliation{Department of Physics, United States Air Force Academy, Colorado Springs, CO 80840, USA}

\author[0000-0002-8056-2526]{Zhenghai Liu}
\affiliation{Department of Astronomy, University of Illinois, Urbana, IL 61801, USA}
\affiliation{Illinois Center for the Advanced Study of the Universe, University of Illinois, Urbana, IL 61820}

\author[0000-0001-5071-0412]{Jesse A. Miller}
\affiliation{Department of Astronomy, University of Illinois, Urbana, IL 61801, USA}
\affiliation{Illinois Center for the Advanced Study of the Universe, University of Illinois, Urbana, IL 61820}

\author[0000-0002-4729-8823]{Rebecca Surman}
\affiliation{Department of Physics, University of Notre Dame, Notre Dame, IN 46556, USA}
\affiliation{N3AS Collaboration}

\begin{abstract}
The astrophysical sites where $r$-process elements are synthesized remain mysterious: 
it is clear that neutron star mergers (kilonovae (KNe)) contribute,  
and some classes of core-collapse 
supernovae (SNe) 
are also possible sources of at least the lighter $r$-process species. 
The discovery of \fe60 on the Earth and Moon implies that one or more 
astrophysical explosions have occurred near the Earth within the last few million years, probably SNe. 
Intriguingly, \pu244 has now been detected, mostly  
overlapping with \fe60 pulses. However, the \pu244 flux may extend to before 12Myr ago, pointing to a different origin.  
Motivated by these observations and difficulties for $r$-process nucleosynthesis
in SN models, we propose that ejecta from a KN enriched 
the giant molecular cloud that gave rise to the Local  
Bubble, where the Sun resides. 
Accelerator mass spectrometry (AMS) measurements of \pu244 and searches for other live isotopes could probe  
the origins of the $r$-process and the history of the solar neighborhood, 
including triggers for mass extinctions, e.g., that at the end of the Devonian epoch, 
motivating the calculations of the abundances of  
live $r$-process radioisotopes produced in SNe and KNe that we present here. 
Given the presence of \pu244, other $r$-process species such as  
\zr93, \pd107, \i129, \cs135, \hf182, \u236, \np237, and \cm247 should be 
present. Their abundances and well-resolved time histories could distinguish between the SN and KN scenarios, 
and we discuss prospects for their detection in deep-ocean deposits and the lunar regolith. 
We show that AMS \i129 measurements in Fe-Mn crusts already constrain a possible nearby KN scenario.\\

KCL-PH-TH/2021-03, CERN-TH-2021-014, N3AS-21-007
\end{abstract}


\section{Introduction} 
\label{sec:intro}

Astrophysical explosions such as supernovae (SNe)
within ${\cal O}(10)$~pc would be close enough to endanger life on Earth~\citep{Ruderman1974, Ellis1995}, and SN
explosions within ${\cal O}(100)$~pc would have been close enough to deposit detectable
amounts of live (undecayed) radioactive isotopes~\citep{Ellis1996}. 
Over the past two decades, many
experiments have detected live \fe60 in deep-ocean sediments and ferromanganese (Fe-Mn) crusts~\citep{Knie1999, Knie2004, Fitoussi2008, Ludwig2016, Wallner2016, Wallner2020, Wallner2021}, in the lunar
regolith~\citep{Fimiani2016}, in cosmic rays~\citep{Binns2016}, and in Antarctic snow~\citep{Koll2019}. The \fe60 is
best understood as evidence
for explosion of one or more nearby and recent SNe,
and the deep-ocean data point to an epoch
$\sim 3$~Myr ago and a distance
$\la 120 \ \rm pc$ away 
\citep{Ellis1999,Fields2005,Fry2015}.
Recent evidence for \mn53 in Fe-Mn crusts~\citep{Korschinek2020} 
adds support to the picture of a nearby SN,
and we note that multiple nearby SNe are
postulated when modeling the Local Bubble~\citep{Smith2001, Breit2016, Schulreich2017}.
The wealth and variety of \fe60 detections establish that near-Earth explosions indeed occurred in the geological past.
These results bring home the environmental hazards facing citizens of star-forming galaxies such as ours.

Intriguingly, there also have been several reports
of non-anthropogenic \pu244 in deep-ocean deposits
from the past 25~Myr~\citep{Paul2001, Wallner2004, Raisbeck2007, Wallner2015, Wallner2021}, 
which are of particular interest because \pu244 originates in
the astrophysical $r$-process. 
Whereas previous \pu244 detections were tentative, the recent measurements by~\citet{Wallner2021}
 are definitive and presumably represent injections from one or more extrasolar explosions,
and it is important to consider
their profound implications for potential $r$-process sites in the solar neighborhood.
Therefore, in this paper we study potential
near-Earth {\em r}-process events that could possibly explain the \pu244 detections:  their astrophysical sources, means of delivery to Earth, and radioisotope signatures.

The most important sites for the $r$-process are currently
a subject of debate~\citep{Cowan2021}. Certainly neutron star mergers (kilonovae (KNe)) provide inevitable sites, 
and the recent observation of a KN associated with the GW190521 gravitational-wave signal~\citep{GBM:2017lvd,Monitor:2017mdv} 
suggests encouraging prospects for more detailed studies of similar events in
the future~\citep{Zhu2018, Barnes2020, Korobkin2020, Wang2020, Zhu2021}. However,
neutron star mergers may be only partially responsible for the Galactic tally of $r$-process elements~\citep{Kyutoku2016, Cote2019, Kobayashi2020, Yamazaki2021}, and a variety of other sites have been proposed, 
including both standard and rare types of core-collapse SNe (see,  e.g.,~\citet{Hoffman1997, Wanajo2006, Fujimoto2008, Winteler2012, Mosta2018, Siegel2019, Miller2020, Choplin2020, Reichert2021, Fujimoto2021}).
In support of this possibility, we note that
\citet{Yong2021} recently discovered an {\rm r}-process-enriched (and actinide-enhanced or -boosted) halo star with 
a very low metallicity  $[\rm Fe/H]=-3.5$, which implies very early production suggestive of an SN origin,
perhaps in magnetohydrodynamic jets.
The radioisotopes produced
 by various $r$-process sites have also been studied: see, e.g., early work by \citet{Seeger1970} and \citet{Blake1973}; \citet{Meyer1993} on pre-solar abundances; \citet{Goriely2016} on SN neutrino-driven winds;
 the \pu244 yields and ratios in \citet{Tsujimoto2017}; and recent work by \citet{Beniamini2020} and \citet{Cote2021}.
 
As \pu244 is among the heaviest of the $r$-process actinides, its production requires the most robust of
$r$-process conditions.  Modern simulations do not find these conditions in regular core-collapse SNe \citep{Fischer2010, Hudepohl2010, arc11}.
Indeed, in most recent models, {\em no} actinides are produced at all, and \pu244 is absent.  
On its face, this would indicate a KN scenario as the origin for the observed \pu244. 
However, there are still significant uncertainties in the extreme physical conditions of SNe, particularly
(1) in the physics of neutrinos and their impact on the neutron abundance \citep{McLaughlin1999, Duan2011, Roberts2012, Johns2020, Abbar2021}, and (2) in the effects of relativistic magnetohydrodynamic jets
that can expel neutron-rich material from the proto-neutron star \citep{Winteler2012, Mosta2018, Reichert2021}.  In some scenarios these effects can 
lead to SN actinide production. We study possible SN sources as well as KNe, and study the ability of the \pu244 detection to discriminate among these scenarios.
 
We have also been motivated to calculate the possible yields of other
$r$-process nuclei in both SN and KN sites and compare their abundances relative to
\pu244. These may be measurable in both deep-ocean deposits and the lunar regolith,
where \fe60 has already been detected~\citep{Fimiani2016}. Of particular interest are layers of ages 
$\sim 2.5$~and $\sim 7$ Myr where there are 
peaks in the deep-ocean \fe60 signal, as discovery of one or
more $r$-process isotopes there would confirm that at least one SN was an $r$-process site.
However, other layers are also of interest, particularly because there is some
evidence for deep-ocean \pu244 atoms 
deposited earlier than the two \fe60 peaks, which points to one or more other sources of
unknown astrophysical origin. Another motivation is the possibility that one or more 
other astrophysical explosions may have occurred at closer distances in the more distant past.
Specifically, it has been suggested~\citep{Fields:2020nmv} that one or more extinction
events toward the end of the Devonian $\sim 360$~Myr ago might have been caused by SN 
explosions. 
These events probably occurred too long ago
to have left detectable deposits of \fe60, in view of its relatively short
half-life of $2.6$~Myr, but might have left detectable deposits of longer-lived $r$-process
isotopes such as \pu244 (half-life 80~Myr).

In order to lay a basis for a systematic study of the live isotopes from
possible nearby astrophysical explosions and $r$-process sites, we survey all
the nuclear isotopes with half-lives between 1~Myr and 1~Gyr, identifying the
nucleosynthesis processes that might produce them and commenting on the results of 
previous searches on the Earth and Moon and on the prospects for their future detection.

Three timescales are of particular interest.
First, the widespread detection of live \fe60 
shows that at least one nearby SN
injected \fe60 into the interstellar medium (ISM) at a time
\begin{equation}
\label{eq:t-Plio}
    t_{\rm Plio} \simeq 3.2 \ \rm Mya \, ,
\end{equation}
where $\rm Mya = Myr \ ago$.
This is derived from the sediment data of~\citet{Wallner2016} and~\citet{Ludwig2016}, where~\citet{Wallner2016} report the earliest \fe60 detection. The peak of the \fe60 deposition on Earth due to this event
was $\sim 2.5$~Mya, around the end of the Pliocene 
epoch, and a linkage to a coincident mass extinction has been proposed in~\citet{Melott2017}. We
note also that the ferromanganese crust data from~\cite{Wallner2016, Wallner2021} provides evidence of a potential second peak at about 7~Mya that has not yet been
detected in sediment data.

A second important timescale
is the lifespan of Local Bubble, a $\gtrsim 100 \ \rm pc$ region of hot, low-density gas
in which the Sun resides \citep{Frisch1981Natur, Crutcher1982, Paresce1984, Frisch2011}.
Multiple supernovae are required to account for this structure \citep{Smith2001, Berghofer2002},
and the \fe60 pulses are likely to be among the most recent and nearest such events.
As we will discuss, the timescale for the creation 
of the bubble and the subsequent deaths of the massive stars within it can be as long as
\begin{equation}
\label{eq:t-LB}
    t_{\rm LB} \lesssim {\cal O}(50) \ \rm Mya
\end{equation} 
Possibly related, 
geological indications of live \pu244
imply a flux on Earth
that stretches farther back, with the earliest potential detection
in a layer deposited 12--25 Mya.
It is thus of interest to consider an event 
at least this long ago.  We will see {in Section~\ref{sect:129I}} that 
existing \pu244 and \i129 data suggest
a timescale comparable to that in eq.~(\ref{eq:t-LB}).

Finally, \citet{Marshall2020} have recently found evidence of a dramatic
loss of stratospheric ozone 359 Myr ago in the so-called Hangenberg
crisis, the last of several poorly understood mass extinction events that punctuated the end
of the Devonian period.  This raises
the possibility that one or more nearby SNe
were responsible for the Hangenberg event
and possibly others as well~\citep{Fields:2020nmv}, roughly at
\begin{equation}
\label{eq:t-Devo}
    t_{\rm Devo} \simeq 360 \ \rm Mya \, .
\end{equation}
We highlight in the following the yields and isotope ratios at these epochs. 
We note also that other extinction events may be connected to astrophysical explosions. For example, 
\citet{Melott2004} have suggested 
a gamma-ray burst origin for the late Ordovician mass extinction
$\sim 440 \ \rm Myr$ ago.

The layout of our paper is as follows.
In Section~\ref{sect:rad-inventory} we survey the radioisotopes
with half-lives between 1~Myr and 1~Gyr that are candidates for providing interesting signatures of nearby astrophysical explosions
due to SNe and/or KNe. In Section~\ref{sect:observability} we review the available measurements of
\fe60 and \pu244 in deep-ocean deposits, and in Section~\ref{sect:modelling} we introduce the SN and KN models
we use to illustrate the range of possible astrophysical $r$-process sites
that fit the data on isotope abundances of representative metal-poor stars and also match solar abundances. Then,
in Section~\ref{sect:ratios+time} we
present detailed calculations of the 
ratios of the abundances of live isotopes and
their time evolution in these models. In light of these results,
we discuss in Section~\ref{sect:models} the observability
of $r$-process isotopes in deep-ocean sediments and
crusts and on the Moon. Finally, in Section~\ref{sect:disc}
we discuss the prospects for terrestrial and lunar searches for
live isotopes, and how they might cast light on $r$-process sites and the history of the solar neighborhood,
and we summarize our conclusions and suggest directions
for future work in Section~\ref{sect:Conx}.

\section{A Survey of Radioisotope Signatures of Astrophysical Explosions}
\label{sect:rad-inventory}

An SN or KN within ${\cal O}(10)$ pc would be near enough to pose a threat to
life on Earth. Fortunately, SN explosions within ${\cal O}(10)$ pc of Earth are 
expected to occur only on intervals of a billion years or so, and nearby KN explosions
are thought to be even rarer. However, SNe are estimated
to occur within ${\cal O}(100)$ pc every few million years, 
and it was suggested in \citet{Ellis1996} that live radioisotopes
would be the premier signatures of such events. 
The detection of such signatures can cast 
light on the history of the Local Bubble and SN nucleosynthesis mechanisms~\citep{Breit2016, Schulreich2017}, refine estimates of the SN threat to life on Earth, and possibly serve as markers of 
past SN effects on the biosphere.

The existence of the well-studied \fe60 peak at 3~Myr ago,
and the \citet{Wallner2021} reports of another peak at 7~Myr ago
and \pu244 possibly extending to 10 Myr ago, focus interest on the search for other
live isotopes deposited around those times.
But there is also interest in looking
for earlier radioisotope deposits, e.g., from the epochs of mass extinctions,
many examples of which are known in the fossil record. The famous Cretaceous–Paleogene extinction,
which included the death of non-avian dinosaurs, was triggered by a different type of astrophysical
event, namely an asteroid impact, whereas the end-Permian extinction is thought
to be due to large-scale vulcanism. However, there are other events in the
fossil record whose origins are unknown as yet, and detection of coincident
radioisotope signatures could provide evidence for any astrophysical origins. Candidate 
extinctions whose origins could be explored in this way include those at the end 
of the Devonian epoch $\sim 360$~Mya~\citep{Fields:2020nmv}.

\begin{figure}[!htb]
\centering
    \includegraphics[width=0.85\textwidth]{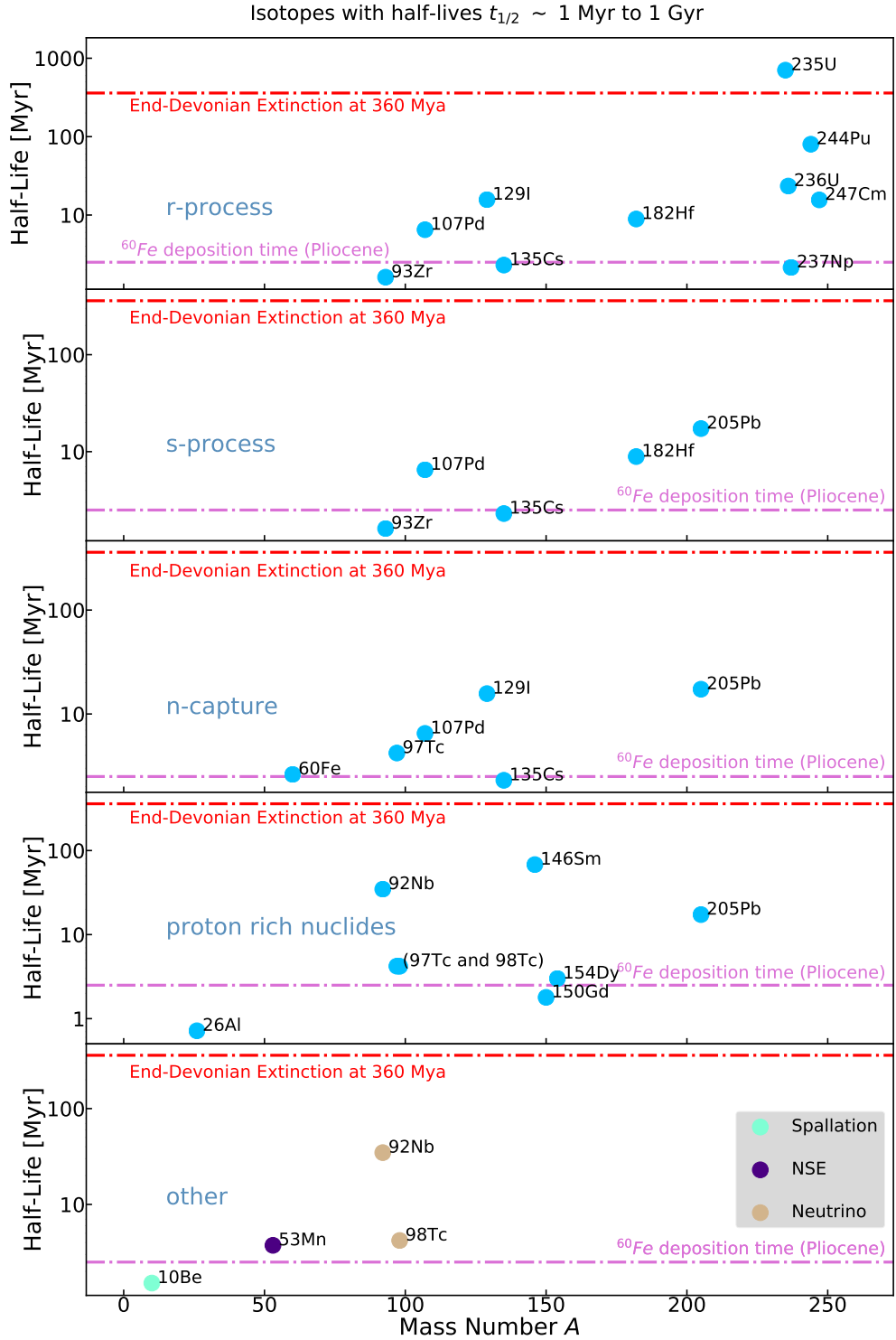}\\
\vspace{-2.5cm}
\caption{\it Radioisotopes with half-lives of interest to searches for nearby SN and KN explosions.  The horizontal orchid-colored
lines indicate the age of the SN explosion $\sim 3$~Mya
attested by discoveries of deposits of \fe60, and the horizontal red lines
mark the end-Devonian time scale.
These five panels represent different nucleosynthesis channels,
with the proton-rich nuclide panel including both $p$-process and $\gamma$-process isotopes, which are ordered by isotope mass number $A$.
	}
	\label{fig:radioisotope-cornucopia2}
\end{figure}

In view of the timings of these target events, radioisotopes of interest are those
with half-lives between 1~Myr and 1~Gyr.
We have therefore identified all nuclides with half-lives from $t_{1/2} = 1 \ \rm Myr$ to 1 Gyr.
We display in Fig.~\ref{fig:radioisotope-cornucopia2} scatterplots of all
radioisotopes with half-lives $> 10^6$~yr, ordered
by their atomic weights $A$, and
{separated} according to their respective dominant nucleosynthesis mechanisms.
More relevant information about these isotopes is given in 
Table~\ref{tab:radioisotope-inventory}, including their half-lives $t_{1/2}$,
their dominant decay modes, the nucleosynthesis mechanisms dominating their
production, and comments on the prospects for their detectability, which we develop in more detail later in this paper.
The half-lives are from the NUBASE2016 evaluation \citep{NUBASE2016} and, unless otherwise noted,
the nucleosynthesis processes are from \citet{Lugaro2018} and the accelerator mass spectrometry (AMS) detection information is from
\citet{Kutschera2013}.

\begin{table}[htb]
\caption{Radioisotopes with Half-lives $t_{1/2} \sim 1 \ \rm Myr$ to ${\rm 1 \ \rm Gyr}$: Astrophysical Production and Geological Detection }
\hspace{-2.7cm}
\scalebox{0.9}{\begin{tabular}{c|c|c|c|c|c|l}
    \hline \hline
     Isotope & Half-Life & Decay & Nucleosynthesis & AMS & Background &  Notes on Extrasolar Evidence \\
     & (Myr) & Mode & Process (Lug18) &  (Kut13) & Measured & and Terrestrial Backgrounds (Section~\ref{sect:observability}) \\
    \hline
    \be10 & 1.51${}^{*}$ & $\beta^-$ & CR (Gos01,Mas99) &  yes & yes & used as chronometer\\
    \hline
    \al26 & 0.717 & $\beta^+$, EC & proton capture & yes & yes & searches in Fe-Mn crusts \\
    \hline
    \mn53 & 3.74 & EC & NSE & yes & yes & evidence in Fe-Mn crusts (Kor20) \\
    \hline
    \fe60 & 2.62 & $\beta^-$ &  neutron capture & yes & no & detection in Fe-Mn crusts and nodules, deep-ocean   \\
    &&&&&&  sediments, Antarctic snow, lunar regolith, cosmic rays \\
    \hline
    \nb92 & 34.7 & $\beta^+$ & $\alpha$-rich freeze-out, $p$, $\nu$ & yes (Guo13)  & no &  \\
    \hline
    \zr93 & 1.61 & $\beta^-$ & $s$ (She20), $r$,  & yes & no & \\
    &&& $\alpha$-rich freeze-out & & &\\
    \hline
    \tc97 & 4.2 & EC & $p$ (Nis18), $n$ capture & no & no & possible SN $\nu$ production in Mo ore (Hax88,Ngu05)\\
    \hline
    \tc98 & 4.2 & $\beta^-$ & $p$ (Nis18), $\nu$ (Hay18) & no & no & possible SN $\nu$ production in Mo ore (Hax88,Ngu05) \\
    \hline
    \pd107 & 6.5 & $\beta^-$ & $s$, $r$, $n$ capture & yes & no &  \\
    \hline
    \i129 & 15.7${}^{*}$ & $\beta^-$ & $r$ (Dil08,Dav19) & yes & yes & pre-anthropogenic background seen in Fe-Mn crusts \\
    &&& $n$ capture &  &  &  \\
    \hline
    \cs135 & 1.33 & $\beta^-$ & $s$, $r$, $n$ capture & yes & no &  \\
    \hline
    \sm146 & 68 & $\alpha$ & $p$ (Nis18) & yes & no &  \\
    \hline
    \gd150 & 1.79 & $\alpha$ & $p$ (How93) & no & no & \\
    \hline
    \dy154 & 3.0 & $\alpha$ & $p$ & no & no &  \\
    \hline
    \hf182 & 8.9 & $\beta^-$ & $s$, $r$ (Voc04), $n$ capture & yes & no &  \\
    \hline
\pb205 & 17.3 & EC & $s$, $n$ capture & yes & no & \\
\hline
\u235 & 704 & $\alpha$ & $r$ & yes & yes & high natural background \\
\hline
    \u236 & 23.4 & $\alpha$ & $r$ & yes & yes & natural and anthropogenic background \\ 
    \hline
    \np237 & 2.14 & $\alpha$ & $r$ & yes & yes &  anthropogenic background seen \\
    \hline
    \pu244 & 80${}^{*}$ & $\alpha$ 99.88\% & $r$ & yes & yes & detection in Fe-Mn crusts \\
    && SF 0.12\% &&&& anthropogenic signature from global fallout (Ste13) \\
    \hline
    \cm247 & 15.6 & $\alpha$ & $r$ & yes & no & possible anthropogenic background \\
         \hline \hline
    \end{tabular} }\\
\noindent
\parbox{\textwidth}{
  {\em Notes}:\\
Our calculations use half-lives 
from NUBASE2016~\citep{NUBASE2016} as implemented in~\citet{Mumpower2018} and \citet{Sprouse2020}. As indicated by asterisks,
the recent NUBASE2020
update~\citep{Kondev2021}
has small changes to some values,
including those of the {\em r}-process species
\i129 and \pu244. \\
  Decay mode:  $\beta^-$ = $\beta$-decay, $\beta^+$ = positron emission, EC = electron capture, $\alpha$ = $\alpha$-decay, SF = spontaneous fission \\
      Nucleosynthesis process: CR = cosmic-ray spallation; NSE = nuclear statistical equilibrium; $s$ = weak/limited or main slow neutron capture ($s$) process; $p$ = $p$-process, synthesis of $p$-rich species by proton capture and/or $\gamma$-processes, $\nu$ = neutrino ($\nu$) process;
      $r$ = weak/limited or main rapid neutron capture ($r$) process; and $n$ capture = neutron captures on preexisting species. \\
      AMS:  Accelerator mass spectrometry demonstrated for this isotope. \\
      Background measured:  Natural or anthropogenic levels detected. \\
      {\em References}: [Dav19] \citet{Davila2019}, [Dil08] \citet{Dillmann},
      [Gos01] \citet{Gosse}, [Guo13] \citet{Guozhu2013}, [Hax88] \citet{Haxton1988}, [Hay18] \citet{Hayakawa2018}, [How93] \citet{Howard}, [Kor20] \citet{Korschinek2020}, [Kut13] \citet{Kutschera2013},  [Lug18] \citet{Lugaro2018}, [Mas99] \citet{Masarik1999}, [Ngu05] \citet{Nguyen2005}, [Nis18] \citet{Nishimura}, [She20] \citet{Shetye}, 
      [Ste13] \citet{Steier2013}, and
      [Voc04] \citet{Vockenhuber}.
      }
    \label{tab:radioisotope-inventory}
\end{table}

We now discuss relevant features of the various isotopes listed in Table~\ref{tab:radioisotope-inventory}.
As noted, there is a large background of
\be10 production by cosmic-ray interactions, so this is not a promising signature of a nearby
astrophysical explosion. There is expected to be copious ejection
of \mn53 and \fe60 by Type-Ia~\citep{Lugaro2018, Kobayashi2020} and core-collapse SNe, respectively, rather than
by the $r$-process, and these isotopes are not
expected to be prominent in KN debris. 
The main mechanism for producing the proton-rich
isotopes \nb92 and $^{97,}$\tc98 is expected to be the $p$-process,~\footnote{As mentioned in
Table~\ref{tab:radioisotope-inventory}, $^{97,}$\tc98 may be produced by SN neutrino interactions in molybdenum, 
via the reactions $\nu_e + ^{98}{\rm Mo} \to \tc97 + e^- + n$ and $\nu_e + ^{97,98}{\rm Mo} \to ^{97,98}{\rm Tc} + e^-$.
Searches for $^{97,}\tc98$ in molybdenum ore could be interesting complementary ways to search for evidence
of recent nearby SN explosions~\citep{Haxton1988, Nguyen2005, Lazauskas2009}.} while there may be $p$-, $r$- and $s$-process contributions to \zr93 production. Most of the heavier isotopes with
$A > 100$ are expected to be produced mainly via the $r$-process, exceptions being \sm146, 
\gd150, \dy154 and \pb205.~\footnote{We do not include in Table~\ref{tab:radioisotope-inventory} or
in our subsequent considerations the long-lived state \bi210*,
an excitation lying 271 keV above the ground state, which is expected to have a low production rate
in all the models studied.}
In the cases of the actinide isotopes,
one must be mindful of the possible presence of terrestrial anthropogenic contamination by
nuclear accidents or bomb debris, which was an issue for the analysis of \u236 and \np237 in {\it Apollo}
lunar regolith samples~\citep{Fields1972, Fields1976}.
The ambient terrestrial level of \pu244 has been measured in \citet{Winkler2004}, and
the detection of \pu244 in deep-ocean deposits by~\citet{Wallner2015} is thought to be
free of this background, which was considered in detail in~\citet{Wallner2021}.

We focus in the following on the long-lived radioisotopes that could be synthesized through the $r$ process, as listed in the top panel of Fig.~\ref{fig:radioisotope-cornucopia2},
and their production by SNe and KNe alongside \pu244. 
Since several of these isotopes have multiple avenues of astrophysical production while \pu244 is an $r$-only species, we emphasize that the $r$-process production ratios we present in the following are lower limits.

\section{Searches for Explosion Ejecta on the Earth and Moon}
\label{sect:observability}

The Earth and the Moon serve as natural archives that store any debris from nearby explosions that reach within 1 au from the Sun.  
This provides a great opportunity to bring samples of ejecta to the laboratory and analyze their content, which can be realized
after finding suitable deposition sites and favorable samples by then identifying the signals within them.  
Live radioisotopes have the advantage of minimizing the 
natural background, which may render the search possible,
even if the measurements remain difficult.
As noted in Eqs.~(\ref{eq:t-Plio}) - (\ref{eq:t-Devo}) and the surrounding discussion, 
the three timescales of particular interest are $\sim (3,50, and 360) \ \rm Mya$,
corresponding to the best-observed \fe60 pulse, the \pu244 half-life (approximately), and the end
of the Devonian epoch.

\subsection{Sensitivities to Radioisotopes of Interest}
\label{sect:sensitivities}

A challenge common to terrestrial and lunar
searches is the tiny abundance of any radioisotope that one may wish to seek.
The widespread \fe60 detections summarized in the introduction provide a model for successful
detection of an extraterrestrial species.
As discussed above, the $r$-process components of any reasonable signal
are expected to have fluences smaller than the established \fe60 signal, implying that
only AMS techniques
may have the needed sensitivity, i.e., the capability of separating and identifying the isotopes of interest 
given the expected number of atoms per gram in a sample.
At present, \fe60 measurements can find isotope fractions with a sensitivity down to
$\fe60/{\rm Fe} \sim (0.3-1) \times 10^{-16}$ \citep{Wallner2020}. However, this
sensitivity may be impaired in the cases of isotopes with a significant cosmic-ray-induced background.
For example, in the case of the recently reported evidence for the terrestrial deposition of \mn53, 
the apparent excess of the signal over the background is $\mn53/{\rm Mn} \sim 1.5 \times 10^{-14}$
\citep{Korschinek2020}.~\footnote{See also~\citet{Feige2018} for a recent example of a study in deep-ocean sediments of \al26, an isotope with a significant background.}

When a natural terrestrial background is absent or small, AMS sensitivities are often limited by the 
ability to remove or discriminate an interfering stable isobar 
with the same $A$ and thus nearly the same mass as the species of interest; this is important for 
many of the {\em r}-process signals that may reside within these samples.
Removal of this interference is typically performed through both chemical processing and ion identification techniques, but limitations still remain.
Recent advances suggest that it is possible to reach $\zr93/\zr92 \sim 6 \times 10^{-11}$ \citep{Hain2018, Pavetich2019}, though
this hinges on successful removal of the stable isobar $\iso{Nb}{93}$.
In the case of \pd107, \citet{Korschinek1994} reported an AMS sensitivity
$\pd107/\pd106 \sim 10^{-8}$; here the interfering stable isobar is \iso{Ag}{107}.
In the case of \i129 there is a natural background level
$\i129/\i127 \sim 1.5 \times 10^{-12}$~\citep{ji2015crusts}.
Sensitivities down to $\i129/\i127 \lesssim 10^{-14}$ are possible \citep{Vockenhuber2015}
since the interfering stable isobar \iso{Xe}{129} is an inert gas, fortuitously, so it does not form negative ions
and will not interfere.
In the case of \cs135, \citet{Yin2015} were able to reduce stable isobar contamination
to $\iso{Ba}{135}/\cs133 \sim 9 \times 10^{-12}$, and we 
adopt the same value for the prospective sensitivity to \cs135, namely
$\cs135/\cs133 \sim 9 \times 10^{-12}$.
In the case of \hf182, \citet{Vockenhuber2004} reported a sensitivity
$(\hf182+{\iso{W}{182}})/\hf180 \sim 10^{-11}$; the ability to measure $\hf182/\hf180$ to this precision
or better requires that techniques be developed to suppress further the stable isobar \iso{W}{182}. In the case of \u236, there are no isobaric contaminants, and the detection limits are set by the ability to discriminate the neighboring abundant uranium isotopes \u235 and \u238. Efforts by \citet{Wilcken2008} estimate a detection limit of \u236/\u238 $
\sim 10^{-13}$.

In the cases of \np237, \pu244, and \cm247, there are no stable isotopes nor isobars, so searches
can first focus simply on extracting the element, guarding against anthropogenic contamination, which could be orders of magnitude greater than a stellar signal \citep{Wallner2004}.
In addition, these AMS samples must be ``spiked'' with a known quantity of shorter-lived isotopes of each species
in order to calibrate the response.  Both $\iso{Pu}{236}$ and \pu242 have been used as calibration standards for \pu244
\citep{Wallner2004, Raisbeck2007, Wallner2015}.  The other species have been studied less.
In the case of \np237, sector-field inductively coupled plasma mass spectrometry 
studies have sensitivities down to a mass fraction
of $10^{-15}$
within soil and sediment samples \citep{Rollin2009}, while systematic AMS studies suggest subfemtogram-per-sample detection limits as long as the \u238 content remains sufficiently low 
~\citep{Fifield1997, Lopez-Lora2019}.
Initial AMS studies of \cm247 made by \citet{Christl2014} suggest a detection limit of $<$0.1 femtogram in a typical sample, where limits were set by the impurities of the \cm244 spike added for reference.\\

\subsection{Plutonium Measurements}

\begin{figure}[htb]
    \centering
    \includegraphics[width=0.7\textwidth]{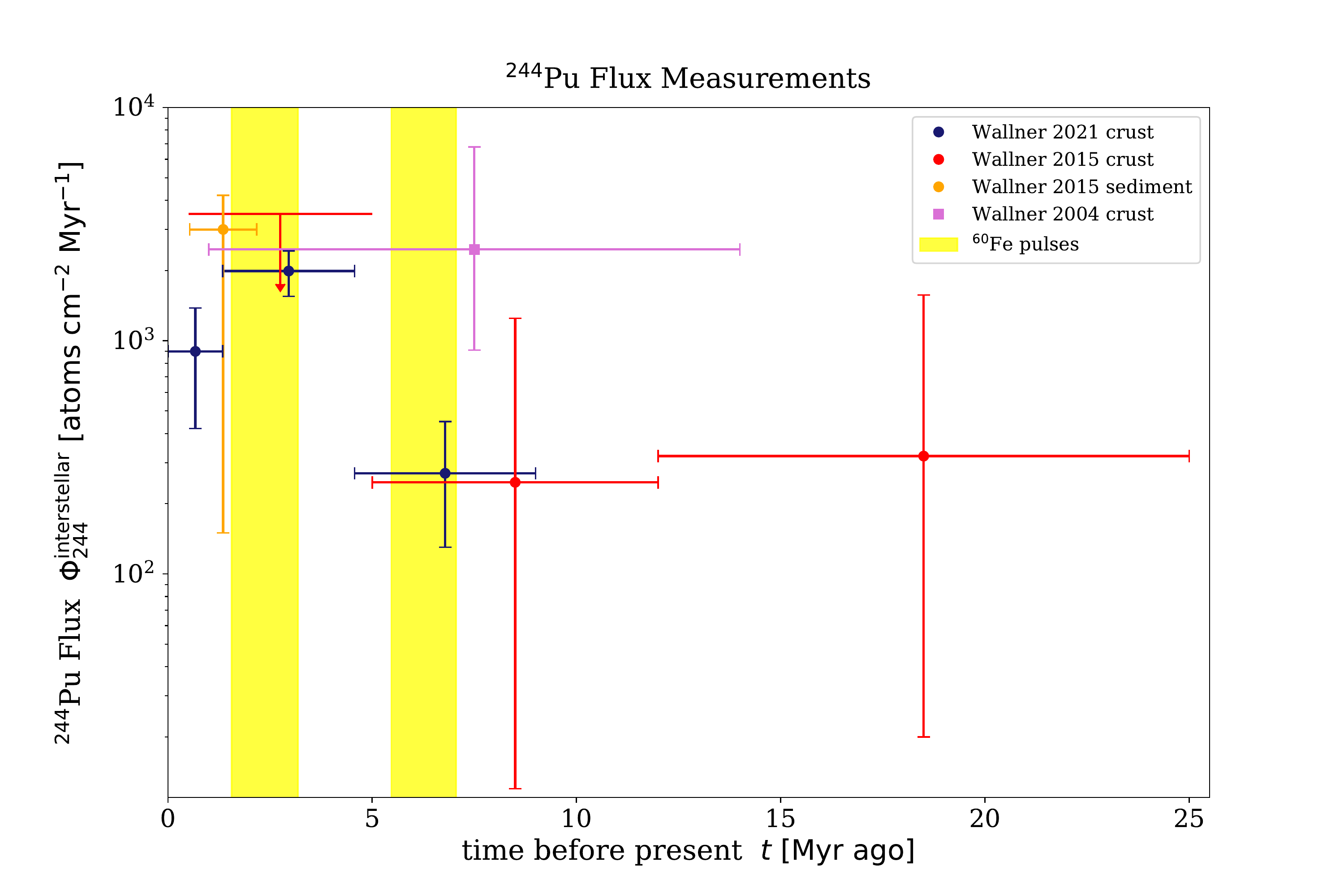}
    \caption{\it Geological searches for \pu244, in deep-ocean crusts
    \citep{Wallner2004,Wallner2015} and sediments \citep{Paul2001,Raisbeck2007,Wallner2015}.
    The \pu244 flux is} expressed as a measurement when 
    the count is nonzero, and as a limit for zero counts and for the first \citep{Raisbeck2007}  time bin as described in the text.
    \label{fig:Pu244-vs-time}
\end{figure}

We anchor our predictions for prospective $r$-process radioisotopes using
the results of geological searches for \pu244 in deep-ocean crusts and sediments 
that are displayed in Fig.~\ref{fig:Pu244-vs-time} and are summarized in Table~\ref{tab:244Pu-obs}.~\footnote{We also note
that also \citet{Hoffman1971} reported a signal in Precambrian bastn{\"a}site, but this claim
was not confirmed subsequently by~\citet{Lachner2012}.}
The most significant of these is the remarkable \citet{Wallner2021} study,
which not only presented solid detections of astrophysical \pu244 in an Fe-Mn crust from the deep Pacific,
but also identified the two \fe60 peaks.

Any claim of astrophysical \pu244 detection must contend with anthropogenic contamination,
and this is a major focus of \citet{Wallner2021}.
They searched not only for \pu244, which potentially contains an astrophysical signal,
but also for the short-lived \iso{Pu}{239}, \iso{Pu}{240}, and \iso{Pu}{241} isotopes,
which measure anthropogenic contamination.  
All of the \iso{Pu}{239}, \iso{Pu}{240} and \iso{Pu}{241} were found in the top layers of the crust,
and exhibited \iso{Pu}{239}/\pu240 and \iso{Pu}{239}/\pu241 ratios consistent with
anthropogenic fallout.  This shows that some uptake has occurred in modern times.
In the deeper layers corresponding to times ${\cal O}$(several) Mya, the \pu244/\iso{Pu}{239} ratio shows an excess 
over the value in the top layer.  In contrast, the \pu240/\iso{Pu}{239} and \pu241/\iso{Pu}{239}  ratios do not show significant
variation with depth.
The fact that, uniquely, \pu244 exhibits an excess points to a source for this isotope
distinct from anthropogenic production.  There being no significant natural \pu244 on Earth today, the signal
must be extraterrestrial.

\citet{Wallner2021} inferred the extraterrestrial \pu244 incorporation rate into the crust, after subtracting the anthropogenic \pu244 contribution.
An incorporation efficiency or uptake of $U_{\rm Pu} = 0.17$ is adopted, the same as that found for \fe60
in the same crust.
For the two layers below the top, fluxes are evaluated as follows (see also Table \ref{tab:244Pu-obs}):
\begin{eqnarray}
\label{eq:244pu-5Myr}
\Phi_{244}^{\rm interstellar}(0-4.76 \ \rm Myr) & = & (1.67 \pm 0.35) \times 10^3 \ \rm atoms \ cm^{-2} \ Myr^{-1} \,,\\
\label{eq:244pu-9Myr}
\Phi_{244}^{\rm interstellar}(0-9 \ \rm Myr) & = & (0.98 \pm 0.18) \times 10^3 \ \rm atoms \ cm^{-2} \ Myr^{-1}  \,.
\end{eqnarray}
These fluxes will be central inputs to our study.
We see that these timespans overlap with the \fe60 pulses at $\sim 3$ and $\sim 7$ Myr ago.
These data leave open the question of whether the flux is different in the earlier time bin versus the overall 
average; the reported difference of about $1.8 \sigma$ is not decisive.

Other important \pu244 measurements have been reported previously. An upper limit of  $\Phi_{244}^{\rm interstellar} < 2 \times 10^5 \ \rm atoms \ cm^{-2} \ Myr^{-1}$ on the rate of extraterrestrial deposition 
in young sediment was set by \citet{Paul2001}.
Subsequently, AMS measurements of crust VA13-2 by~\citet{Wallner2004} and of sediment 
MD90-0940 by~\citet{Raisbeck2007} each yielded
one event, dated to $ 1-14$~Mya and $2.4-2.7$~Mya, respectively.
\citet{Raisbeck2007} did not attribute their single event to a signal, but derived
upper limits on the fluence assuming one count in each of three
time bins. We follow this practice, noting that the flux inferred from the nonzero bin would vastly exceed the other limits and detections overlapping this time period.
More recently, a search for \pu244 by \citet{Wallner2015} yielded a possible signal in three
samples corresponding to three different epochs:  sediments spanning $0.53-2.17$~Mya,
and crust layers at $5-12$~Mya and $12-25$~Mya.  
In each of these samples, at most only a single \pu244
count was found in each time bin, so these results must be treated with great caution.

\begin{table}[htb]
    \caption{Geological Searches for Natural \pu244}
    \centering
    \begin{tabular}{l|l|cccc}
    \hline \hline
    Study& Sample & Time & \pu244 Counts & Flux $\Phi_{244}^{\rm interstellar}$ & Fluence ${\cal F}_{244}^{\rm interstellar}$ \\
    &  & [Mya] & [atoms] & $[\rm atoms \ cm^{-2} \, Myr^{-1}]$ & $[\rm atoms \ cm^{-2}]$ \\
    \hline \hline
    \citet{Paul2001} & Sediment 92SAD01 & $0-0.3$ & $1^{*}$ & $< 2 \times 10^{5}$ & $< 2 \times 10^{4}$ \\
    \hline
    \citet{Wallner2004} & Crust VA13-2 & 1--14 & 1 & 2500 & $1.6 \times 10^4$ \\
    \hline
    \citet{Raisbeck2007} & Sediment MD90-0940 & 2.4--2.7 & 1 & $< 3 \times 10^7$ & $< 3 \times 10^6$ \\
    \hline
    \citet{Wallner2015} & Crust 273KD & 0.5--5 & 0 & $<3500$ & $< 800$ \\
    &  & 5--12 & 1 & $247_{-235}^{+1000}$ &  \\
    & & 12--25 & 1 & $320_{-300}^{+1250}$ &  \\
    & Sediment TR149-217 & 0.53--2.17 & 1 & $3000_{-850}^{+12000}$ &  \\
    \hline
    \citet{Wallner2021} & Crust-3/A & 0 - 1.34 & $34 \pm 17$ & $930 \pm 480$ & $1200 \pm 600$ \\
 & Crust-3/B & 1.34 - 4.57 & $141 \pm 19$ & $1990 \pm 440$ & $6400 \pm 1400$ \\
 & Crust-3/C & 4.57 - 9.0 & $6.3^{+4.3}_{-3.2}$ & $270^{+180}_{-140}$ & $1200^{+800}_{-600}$ \\
\hline \hline
    \end{tabular}
    \parbox{0.9\textwidth}{${}^{*}$\citet{Paul2001} argue that their detection could be due to anthropogenic contamination.}
    \label{tab:244Pu-obs}
\end{table}

The detections reported in Fig.~\ref{fig:Pu244-vs-time} and Table~\ref{tab:244Pu-obs} 
show that extraterrestrial \pu244 deposition has occurred over at least over the last 9 Myr and possibly goes back
to as far as 25 Myr.  
Taking the measurements at face value,
it would seem that the \pu244 flux history differs from
the two \fe60 pulses, which are each limited in time (though a small \fe60 continues to the present).  
That is, the wide time ranges of the \citet{Wallner2021} detections
both overlap with the \fe60 pulses, but earlier indications of 
\pu244 flux extend from nearly the present back to at least 12 to 25~Mya.
Within the large uncertainties, it is unclear if the flux varies over this time.  
The data could accommodate--but within uncertainties do not demand--a larger flux around the time of the
\fe60 pulse(s) at $\sim 3$ Mya (and $\sim 7$ Mya).
The possible difference between the \fe60 and \pu244 deposition histories suggests a different origin for 
at least some of the \pu244, a point also made in previous studies, e.g., \citet{Wallner2015}.

\section{Modeling $r$-Process Production in SNe and KNe}

\label{sect:modelling}

We can link the observed \pu244 flux with that of other {\em r}-process radioisotopes through theoretical calculations of {\em r}-process nucleosynthesis.
These calculations depend on the nature of the candidate nucleosynthesis event, and on the nuclear inputs one adopts.  The results are constrained
by the observed {\em r}-process pattern in solar system material and in stars.  Here we describe our calculations and their uncertainties.

The production ratios of radioactive isotopes resulting from an $r$-process event can be estimated by the extraction and post-processing of ejected matter trajectories from astrophysical simulations of the event. 
The trajectories contain the time history of $(\rho,T, and Y_e)$, where $\rho$ is the density, $T$ is the temperature, and the electron fraction $Y_e = n_e/n_B = \avg{Z/A}$ 
measures the neutron richness $Y_n = 1-Y_e = \avg{N/A}$. The evolution of the nuclear material along this trajectory is then calculated using a network of the relevant nuclear reactions. 
This procedure brings with it significant challenges, starting from the identification and characterization of the appropriate nucleosynthesis sites within the candidate event.

SNe were the first type of event suggested for $r$-process production~\citep{B2FH}, and for many decades the core-collapse SN neutrino-driven wind was considered the leading candidate site. 
However, modern simulations show that the neutrino-driven wind is unlikely to be sufficiently neutron-rich to synthesize the actinides \citep{Fischer2010,Hudepohl2010,Arcones2013},
though it may produce $A\sim80$-100 species through a weak $r$-process \citep{Bliss2018} or a 
$\nu p$-process \citep{Frohlich2006}. 
The ultimate extent of nucleosynthesis in this environment depends on neutrino physics that is not fully understood \citep{Balantekin2005,Duan2011,Johns2020,Xiong2020}. 
Rare types of core-collapse events may also generate neutron-rich outflows, with promising candidates including magneto-rotational (MHD) SNe~\citep{Winteler2012,Mosta2018,Reichert2021} and collapsars~\citep{Pruet2003,Surman2006,Fujimoto2008,Siegel:2018zxq}. 

Studies of
galactic chemical evolution suggest that, whilst collapsars might have been important
early in the history of the universe during the epoch of Population III stars,
they were less relevant during the epochs of interest for this study.
For this reason, and given the uncertainty in how robust
collapsar $r$-process calculations might be~\citep{Miller2020}, we do not consider them further in this paper.~\footnote{We
note the suggestion that core-collapse SNe driven by the quark-hadron transition might
also be rare $r$-process sites~\citep{Fischer2020}, but also do not discuss this possibility here.}

While neutron star mergers have recently been confirmed to produce $r$-process elements \citep{Cowperthwaite2017, Kasen, AbbottGW170817, NSM}, exactly how, where, and how much have yet to be definitively worked out \citep[see reviews in ][and references therein]{Cowan1991, solar, Kajino2019, Cowan2021}. 
Possible nucleosynthetic environments within a merger include the prompt ejecta---cold, very neutron-rich tidal tails and/or shock-heated ejecta 
from the neutron star contact interface \citep{Bauswein2013, Hotokezaka2013, Rosswog2013, Endrizzi2016, Lehner2016, Sekiguchi2016, Rosswog2017}---and magnetic, viscous, and/or neutrino-driven outflows from the resulting accretion disk \citep{Chen2007, Surman2008, Dessart2009, Perego2014, Wanajo2014, Just2015, Martin2015, Siegel2018}. 
The composition and relative contributions of each type of mass ejection depend on quantities such as the physical parameters of the merging system 
and the still unknown microphysics of dense matter and its neutrino emission \citep[see, e.g.,][]{Caballero2012, Foucart2015, Malkus2016, Kyutoku2018}.

In view of the large astrophysical uncertainties in each candidate $r$-process site, our calculations of isotopic yields rely on illustrative models
that indicate the ranges of possibilities. We choose matter trajectories from modern simulations that capture the rough characteristics ($Y_e$, entropy $s/k$, and dynamical timescale $\tau$) expected for each site. Different combinations of ($Y_e$, $s/k$, and $\tau$) lead to distinct nucleosynthetic pathways through the neutron-rich side of the nuclear chart, leading to different amounts of individual isotopes even when the final elemental yields are similar. We choose at least two distinct trajectories for each type of event 
so as to ensure production of both main ($A>120$) and weak ($70<A<120$) $r$-process nuclei.

We adopt four illustrative $r$-process model combinations, using trajectories from a {\em forced} modification of a conventional
SN neutrino-driven wind scenario (SA), an MHD SN model (SB), and two neutron star merger disk and dynamical ejecta combinations (KA and KB).
We then combine and scale the resulting abundances to the elemental patterns of select $r$-process-enhanced stars. We scale to individual metal-poor stars rather than, e.g., the solar abundances, since these stars have experienced fewer generations of stellar nucleosynthesis and thus are cleaner representations of the yields from single $r$-process events. 
We choose
one of the few stars for which elements in all three $r$-process peaks have been detected ~\citep{Ian},
and J0954+5246, the star with the largest enhancement in actinide elements ever detected \citep{Erika}.
We use a variety of $r$-process species measured in the above-discussed metal-poor stars to normalize our estimates, including ytterbium, tellurium, cadmium and zirconium, and note the mixing fraction(s) $f$ of the total mass(es) of the weak $r$-process trajectory (trajectories) relative to the most neutron-rich $r$-process trajectory that appears in the combined model fit. The total mass of the $r$-process ejecta for each model is normalized to unity. The details of each model combination and constraint are described below and summarized in Table~\ref{tab:models}.

Our nucleosynthesis calculations are made with the nuclear reaction network code Portable Routines for Integrated nucleoSynthesis Modeling (PRISM; \citet{Mumpower2018,Sprouse2020}), implemented as in ~\citet{Wang2020} for the baseline calculation. 
We note that isotopic ratio estimates are shaped in addition by the nuclear physics properties of the thousands of exotic nuclei that participate in an $r$-process. Thus for each model we explore variations in the nuclear inputs for quantities for which experimental values are unavailable ({masses from \citet{HFB17PRL} (HFB), or $\beta$-decay rates from \citet{MKT} (MKT) for both SN and KN models, and fission yields from \citet{KT} for KN models}), 
as in \citet{Wang2019}. 
{Additionally, because of the general limitation of the network code for time step evolution at times $\gtrsim$ Myr, which results in large time steps comparable to the half-lives of the radioisotopes of interest in this work such as \zr93, we use PRISM to generate $r$-process abundance yields until 1 kyr for \hf182 and lighter radioisotopes, and until 1 Myr for actinides (except for \cm248 and 
\cm245, for which we run until 0.1 Myr), and then switch to a pure radioactive decay calculation for these radioisotopes.}
These calculations provide the relative abundance yields for the radioisotopes.

\begin{figure}[!htb]
	\centering
	\vspace{-5mm}
   \includegraphics[width=18cm]{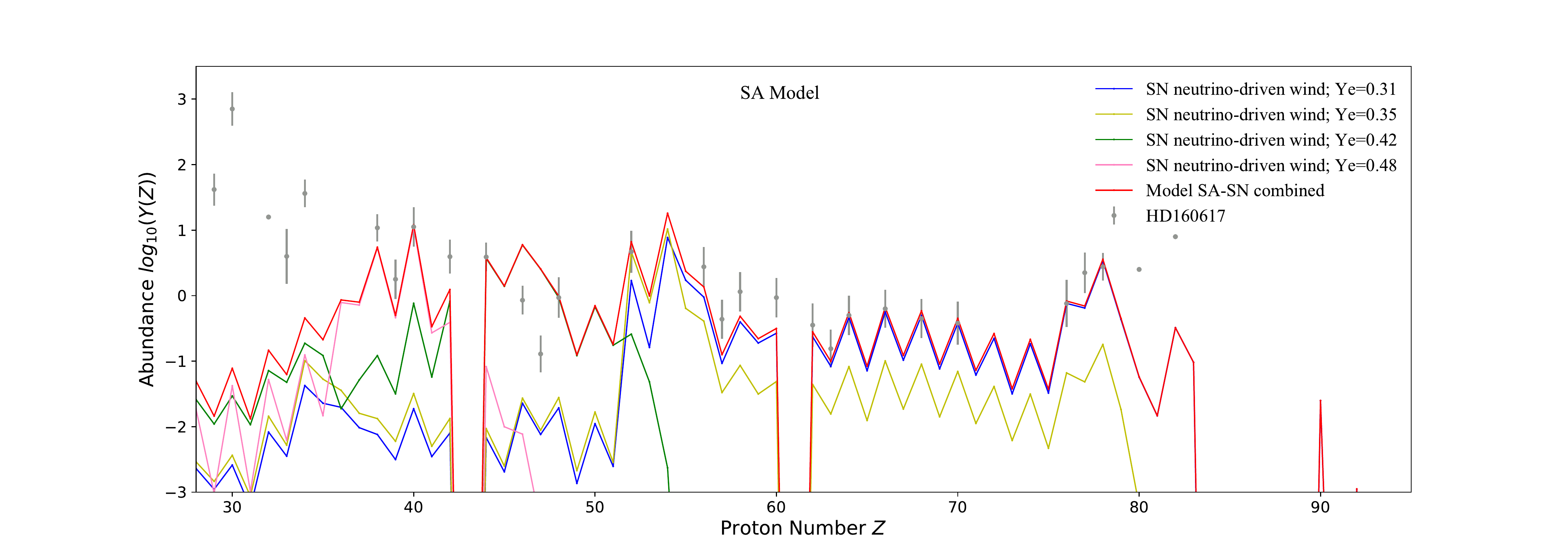}\\
      	\vspace{-4mm}
   \includegraphics[width=18cm]{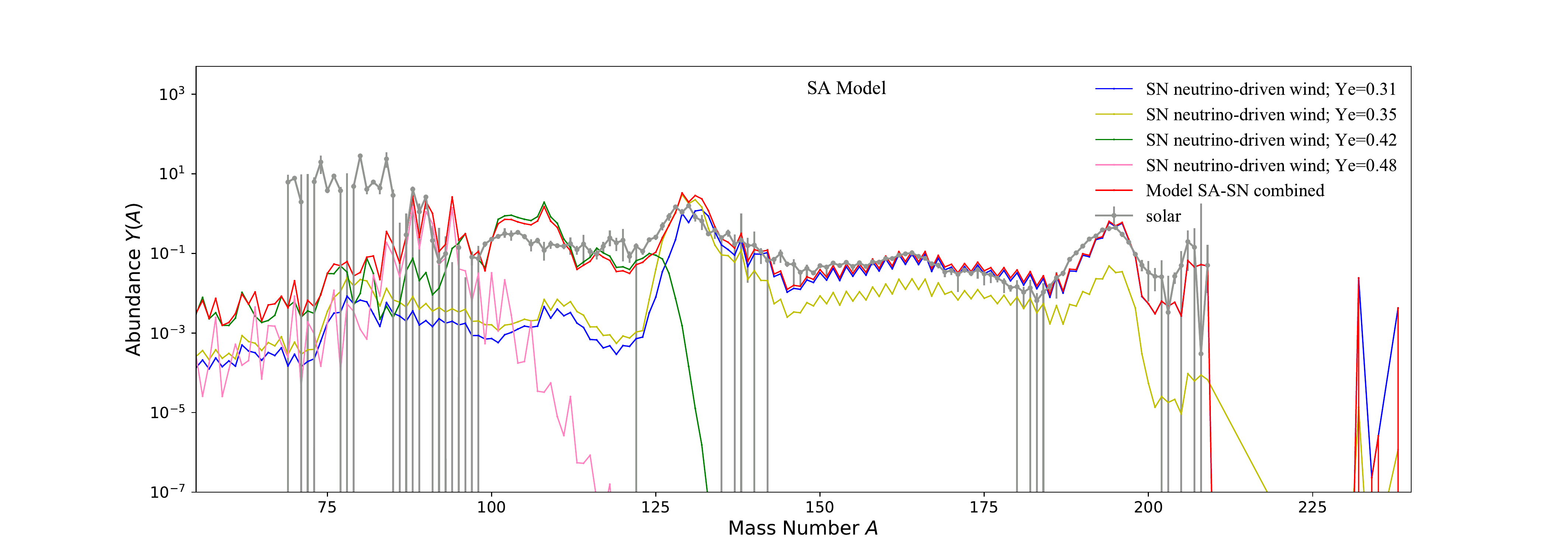}\\
	\caption{\it Upper panel: The abundances {at $t\sim10 \ \rm Gyr$} of $r$-process nuclei produced in SN {\em forced} modifications of neutrino-driven wind simulations from \citet{arc07} and \citet{arc11} with the electron fractions $Y_e = 0.31$ (blue), $0.35$ (yellow), $0.42$ (green), and  $0.48$ (pink), and a combination (SA, red) fitted to abundances measured in the metal-poor star HD 160617~\citep{Ian}, plotted as functions of the atomic number $Z$.
	Lower panel: A comparison with the corresponding solar abundance data~\citep{solar}, plotted as functions of the atomic weight $A$.
}
	\label{fig:SN-rpro-SA}
\end{figure}

As commented above, the SN neutrino-driven wind scenario has fallen out of favor as a primary $r$-process site because modern simulations do not show that
sufficiently neutron-rich conditions to reproduce the solar {\em r}-process pattern \citep{Fischer2010,Hudepohl2010,Arcones2013}.
Indeed, no actinides are produced at all.
We are however motivated by the intriguing possibility of
coincident identifications of \fe60 and \pu244 to consider here
a {\em forced} neutrino-driven wind scenario,
denoted in Table~\ref{tab:models} by "$\nu \star$". We start with the neutrino-driven wind simulations of \citet{arc07,arc11} and modify the initial $Y_e$ in order to produce different $r$-process yields. 
The upper panel of Fig.~\ref{fig:SN-rpro-SA} shows the {final abundance pattern results at $t\sim10\ \rm Gyr$}
for four values of $Y_e = 0.31$ (blue), 0.35 (yellow), 0.42 (green), and 0.48 (pink). We have found that results
for $Y_e \le 0.31$, following different
trajectories, and using different nuclear networks, make qualitatively similar
predictions for substantial production of isotopes with atomic numbers $A \gtrsim 130$, up to and
including the actinides. On the other hand, simulations with $Y_e \ge 0.38$
yield much less production of isotopes with $A \gtrsim 130$, as exemplified by
the results shown in green and pink. The results shown in red are for a mixture (model SA) of the simulations for SN {\em forced} neutrino-driven wind with four
different values of $Y_e$ (see Table~\ref{tab:models}), whose relative normalizations are scaled to fit data on the abundances of ytterbium, tellurium, cadmium, and zirconium in the metal-poor star HD 160617 \citep{Ian}. The range of $Y_{e}$ in the SA model is similar to that of the SN model with actinide production in \citet{Goriely2016}. 
These and other abundances are shown in gray in the upper panel of Fig.~\ref{fig:SN-rpro-SA}, and we see in the lower panel of Fig.~\ref{fig:SN-rpro-SA}
that model SA also matches the solar abundance data very well.

\begin{table}[htb]
    \scalebox{0.8}{ \hspace{-2.6cm}
    \begin{tabular}{|c|cc||cc|}
    \hline
     &
     \multicolumn{2}{c||}{SN Models} & \multicolumn{2}{c|}{KN Models} \\
      \hline
     Label & SA ($\nu \star$) & SB (MHD) & KA & KB  \\
     \hline
    Simulations & SN {\em forced} neutrino-driven wind: four trajectories & MHD SN:  & \multicolumn{2}{c|} {KN dynamical ejecta: two trajectories from \cite{Bovard}}\\    
     & from \citet{arc07,arc11} & two trajectories from  & \multicolumn{2}{c|}{diskwind: 2 trajectories from \cite{Just}} \\
     & with modified $Y_e=0.31,0.35,0.42, and 0.48$ &\citet{Mosta2018}  & \multicolumn{2}{c|}{} \\
    \hline
     Scaling  & HD 160617: Yb, Te, Cd and Zr  &  HD 160617: Yb and Zr  & HD 160617: Yb and Zr & J0954+5246: Yb and Zr \\  
    \hline 
     Mixing fractions $f$ & $f_{0.35}$=0.757, $f_{0.42}$=1.778, and $f_{0.48}$=0.770 & 3.137 & 3.980 & 0.819  \\ 
    \hline
    \end{tabular}}
    \vspace{0.3cm}
    \caption{\it Combinations of  {\em Forced} Modifications of Neutrino-driven SN Models~\citep{arc07,arc11} and MHD Models~\citep{Mosta2018} Constrained by Observations of the Metal-poor Star HD160617 \citep{Ian} (SA and SB, Respectively), and Combinations of KN Dynamical Ejecta Models~\citep{Bovard} with Disk Neutrino-driven Wind Models from~\citet{Just},
    Constrained by Observations of HD160617 (KA) or the Actinide-boost star J0954+5246 \citep{Erika} (KB).
    \label{tab:models}}
\end{table}

\begin{figure}[!htb]
	\centering
	\vspace{-7mm}
    \includegraphics[width=18cm]{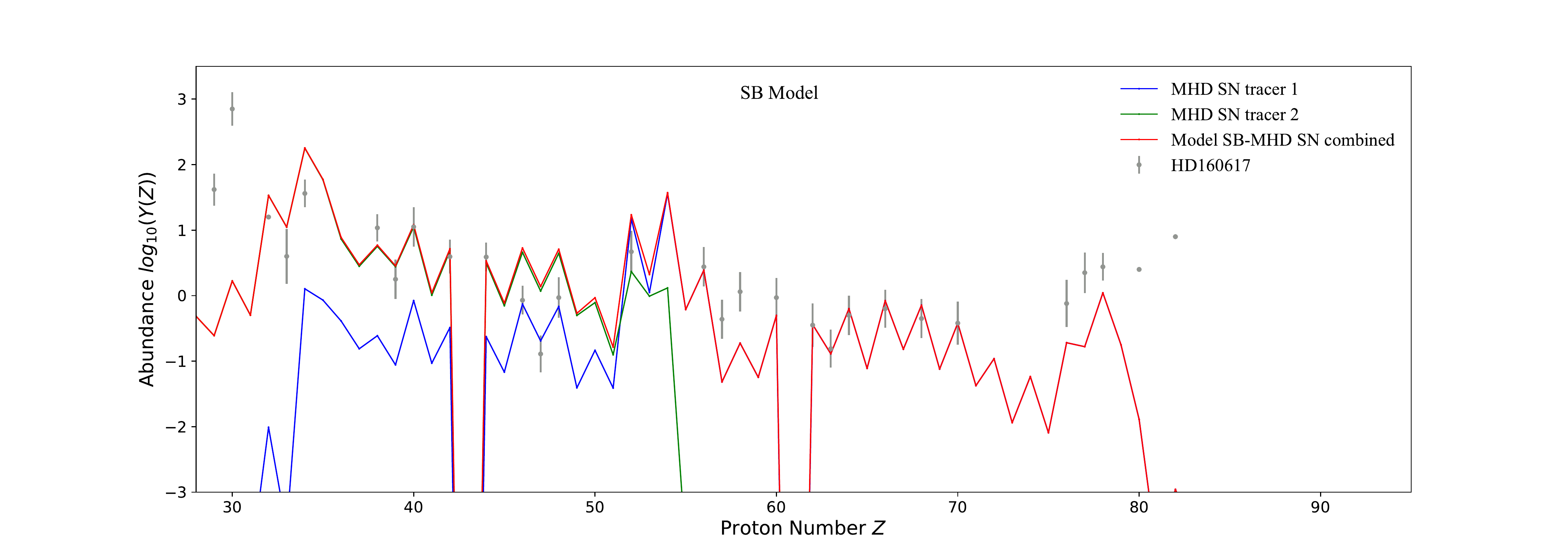}\\
      	\vspace{-4mm}
  \includegraphics[width=18cm]{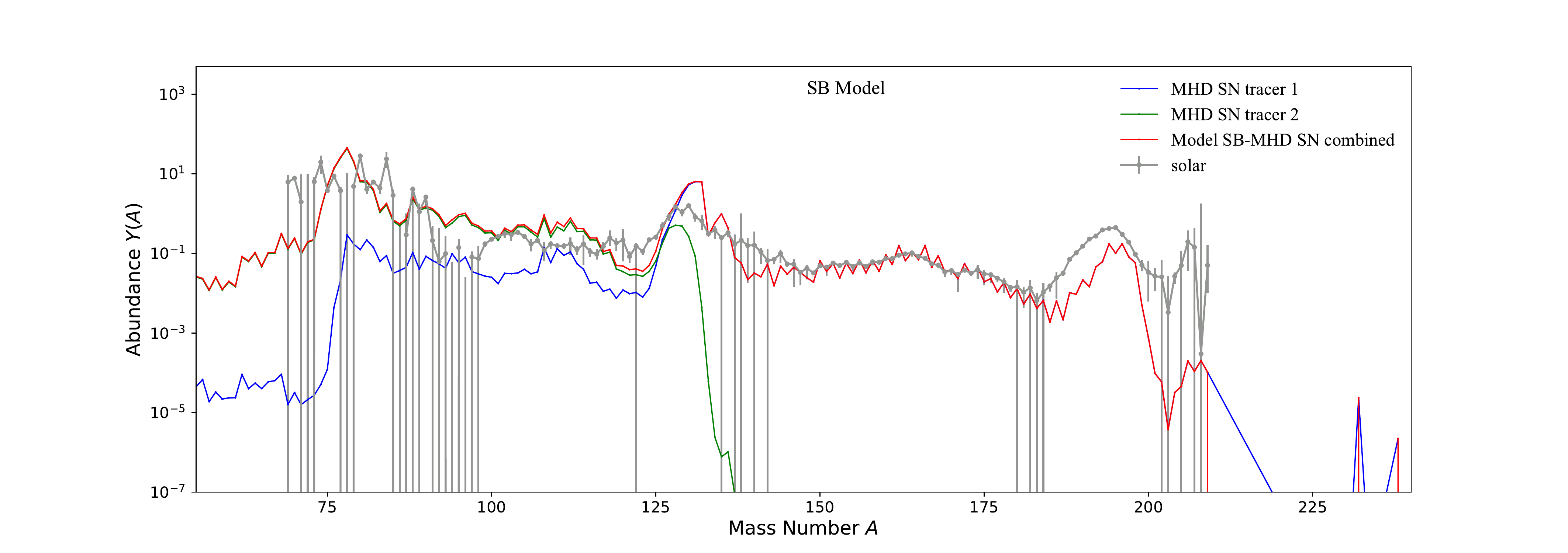}\\
	\caption{\it Upper panel: The abundances {at $t\sim10 \ \rm Gyr$} of $r$-process nuclei produced in MHD SN simulations from \citet{Mosta2018} with two different tracers 1 (blue) and 2 (green)
	and a combination (SB, red) fitted to the abundances of ytterbium and zirconium measured in the metal-poor star HD 160617~\citep{Ian},
	plotted as functions of the atomic number $Z$. Lower panel: A comparison with the corresponding solar abundance data~\citep{solar},
	plotted as functions of the atomic weight $A$.
	}
	\label{fig:SN-rpro-SB}
\end{figure}

Fig.~\ref{fig:SN-rpro-SB} shows analogous results 
using the \citet{Mosta2018} MHD SN model. In the upper panel, we show the abundance predictions of two trajectories (blue for the main r-process trajectory and green for the light r-process trajectory)
and a combination (SB, red) fitted to the abundances of ytterbium and zirconium measured in the metal-poor star
HD 160617 \citep{Ian} (see Table~\ref{tab:models}). These and other abundances are shown in gray in the upper panel of Fig.~\ref{fig:SN-rpro-SB}, and
we see in the lower panel of Fig.~\ref{fig:SN-rpro-SB}
that this mixture of simulations also matches the solar abundance data quite well in general, 
though it overestimates the
structure seen in the solar data for $A \sim 130$, and falls off more
rapidly for $A \gtrsim 190$. Indeed, almost all modern SN models struggle to produce actinides, displaying higher production of lighter $r$-process species relative to plutonium as described in the next section.

\begin{figure}[!htb]
	\centering
	\vspace{-5mm}
  \includegraphics[width=16cm]{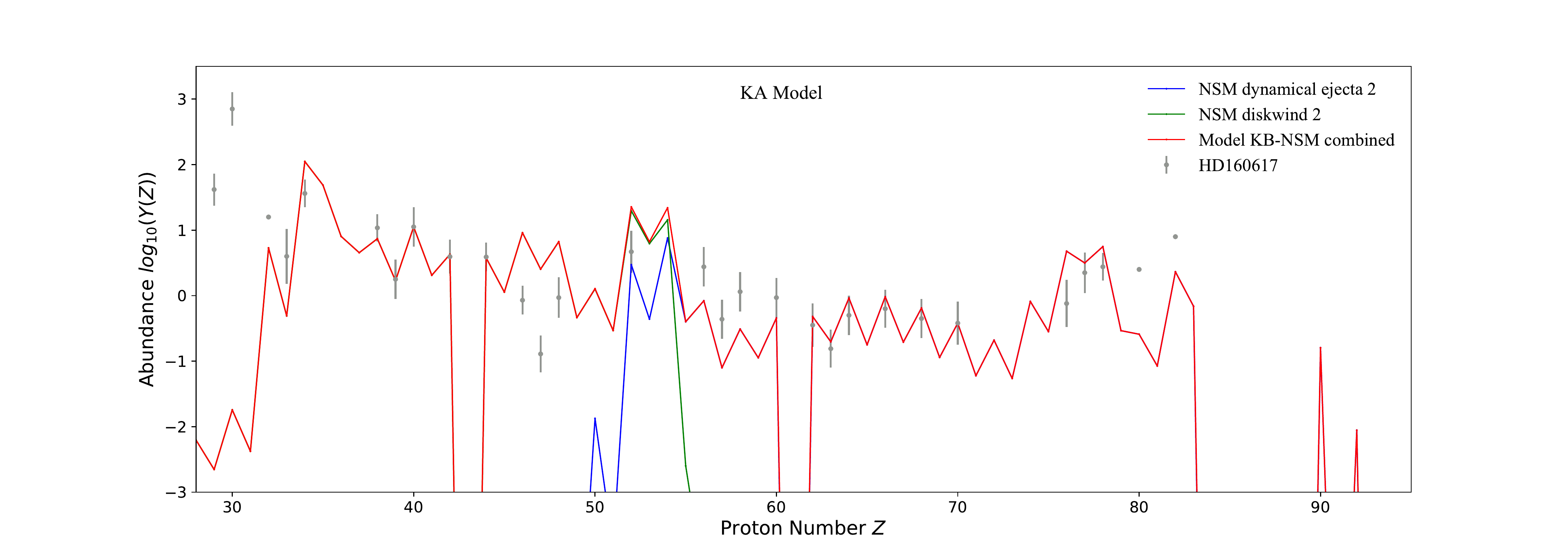}\\
  	\vspace{-4mm}
\includegraphics[width=16cm]{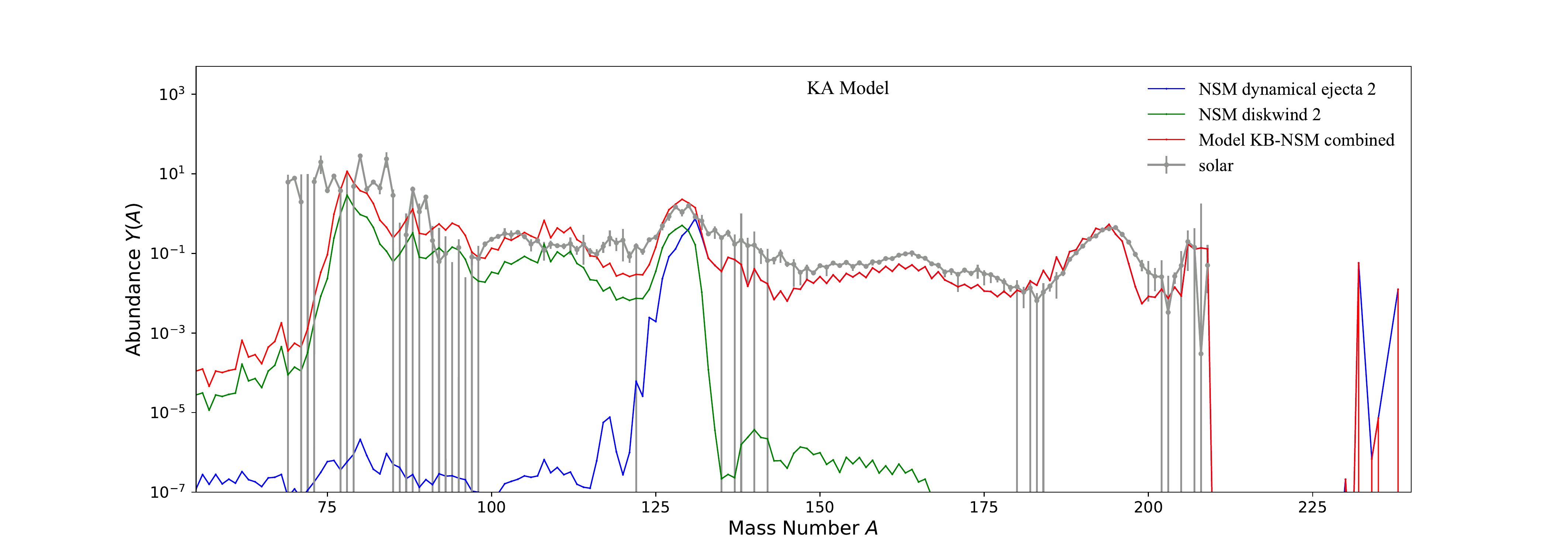}\\
	\caption{\it Upper panel: The abundances {at $t\sim10 \ \rm Gyr$} of $r$-process nuclei produced in simulations of KN dynamical ejecta from \citet{Bovard} (blue) and of disk neutrino-driven wind from \citet{Just} (green), and a mixture (KA, red) fitted to measurements of the metal-poor star HD 160617 \citep{Ian}, plotted as functions of the atomic number $Z$.
	Lower panel: The corresponding abundance pattern of model KA compared with solar abundance data~\citep{solar}, plotted as functions of the atomic weight $A$. 
	}
	\label{fig:KN-rpro-KA}
\end{figure}

The upper panel of Fig.~\ref{fig:KN-rpro-KA} shows results 
from representative simulations of the 
abundances of nuclei produced by the $r$-process {(at $t\sim10$~Gyr)} in
KN dynamical ejecta based on the work of \citet{Bovard} (blue), and in KN
disk neutrino-driven wind based on the work of~\cite{Just} (green). Also shown is a combination of these simulations
(KA, red) fitted to the abundances of ytterbium and zirconium in the metal-poor star HD 160617~\citep{Ian} (see Table~\ref{tab:models}).
These and other abundances are shown in gray in the upper panel of Fig.~\ref{fig:KN-rpro-KA}, and  
we see in the lower panel of Fig.~\ref{fig:KN-rpro-KA}
that the KA model also matches the solar abundance data quite well, though with some deviations for $A \sim 140$.

\begin{figure}[!htb]
	\centering
	\vspace{-5mm}
  \includegraphics[width=16cm]{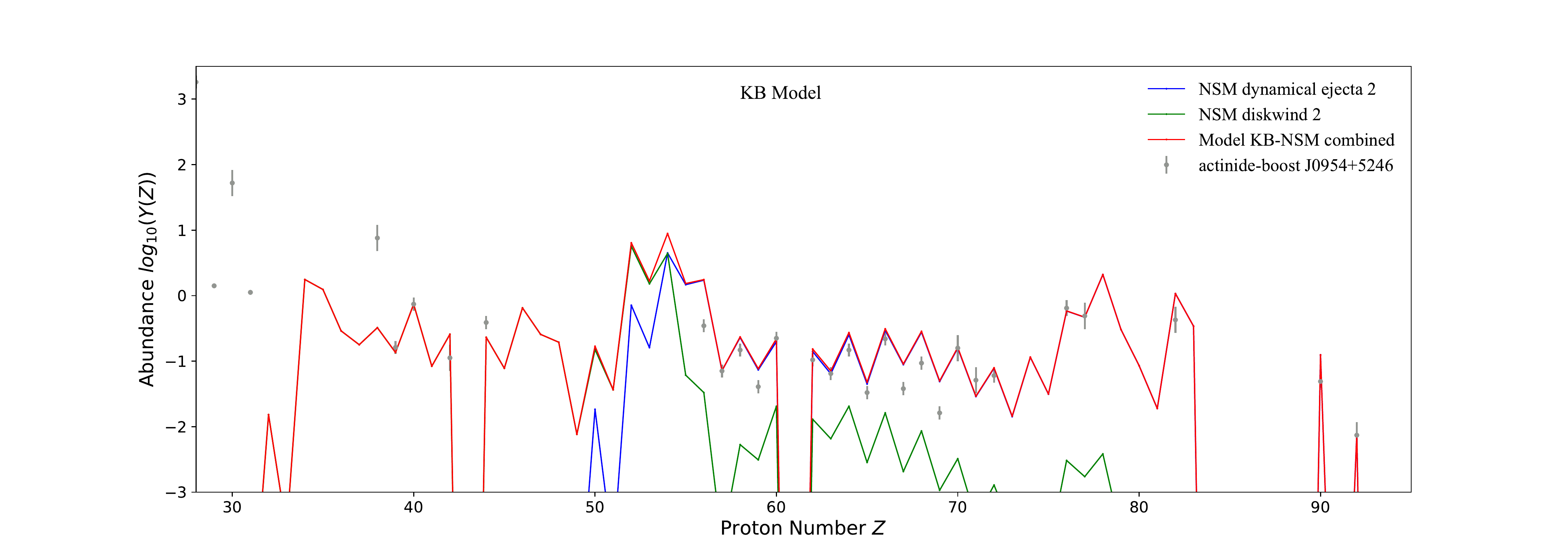}\\
  	\vspace{-4mm}
  \includegraphics[width=16cm]{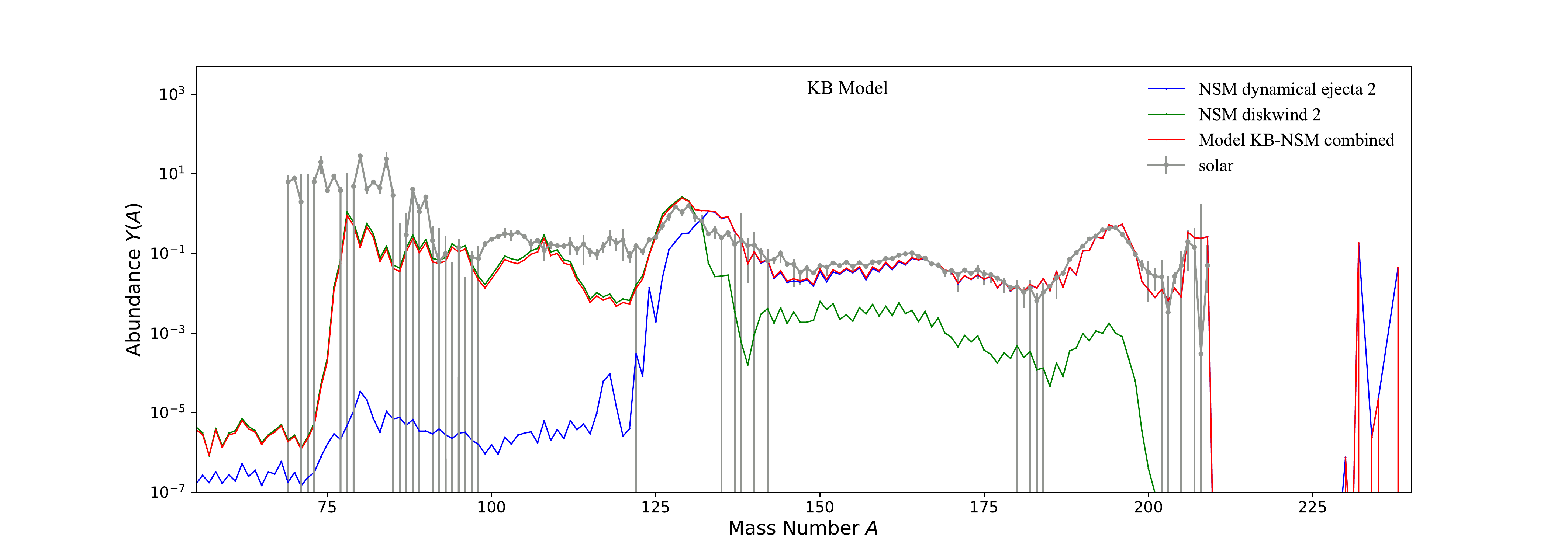}
	\caption{\it 
	Upper panel: The abundances {(at $t\sim10Gyr$)} of $r$-process nuclei produced in simulations of KN more neutron-rich dynamical ejecta from \citet{Bovard} (blue) and of disk neutrino-driven wind from \citet{Just} (green), and a mixture (KB, red) fitted to the measurements of the actinide-boost star J0954+5346 \citep{Erika}, plotted as functions of the atomic number $Z$.
	Lower panel: The corresponding abundance pattern of model KB compared with solar abundance data~\citep{solar},
	plotted as functions of the atomic weight $A$.
	}
	\label{fig:KN-rpro-KB}
\end{figure}

As compared with Figure.~\ref{fig:KN-rpro-KA}, Figure.~\ref{fig:KN-rpro-KB} shows analogous results obtained
using the KN dynamical model with a relatively more neutron-rich trajectory from \citet{Bovard} 
(blue), and the disk neutrino-driven wind model
of~\citet{Just}. Also shown are the predictions of a combination (KB, red) fitted to the abundances of 
ytterbium and zirconium measured in the actinide-boost star J0954+5346 \citep{Erika}.
These and other abundances are shown in gray in the upper panel of Fig.~\ref{fig:KN-rpro-KB}, 
and we see in the lower panel of Fig.~\ref{fig:KN-rpro-KB}
that this combination model also matches the solar abundance data quite well in general, though with some deviations
from the solar data for $A \sim 120$ {and $A \sim 135$}. The KA and KB models both exhibit robust production of actinides that subsequently fission, so fission yields play important roles in shaping the second peak in these models \citep{Eichler2015,Giuliani2020,Vassh2019,Vassh2020}. The fission yields of most neutron-rich actinides have not been experimentally determined, so in addition to the simple symmetric-split fission yields adopted in the baseline calculation, we also implement the wide-Gaussian fission yields from \citet{KT} to estimate the uncertainty range. We find that the fission yields from \citet{KT} would bring a small boost to the left and right sides of the second peak ($A\sim120$ and $A\sim133$) for the KA model, while leaving the yields of the interesting $r$-process radioisotopes listed in Table~\ref{tab:ratios-Plio} largely unchanged. For the KB model with boosted actinide production, fission deposition could potentially fill the gap around $A\sim120$ and lower the bump around $A\sim133$ to bring the abundance pattern closer to the solar pattern, thus resulting in a smaller \cs135 yield. In a more extreme example of a wider distribution of fission fragments, the neutron star merger nucleosynthesis calculations in \citet{Shibagaki2016} exhibit a fission-recycling $r$-process pattern without a second peak, which could bring even smaller abundance yields of $^{129}$I and \cs135. Additionally, spallation reactions that can occur when fast neutron star merger ejecta interact with the ISM may also affect the abundances of the radioisotopes located around $r$-process peaks \citep{Wang2019}. These details do not influence the overall conclusion, however, that the KN models are predicted to produce actinides robustly, leading to lower ratios of lighter $r$-process species relative to plutonium, as described in the next section.

\section{$r$-Process Radioisotope Ratios and Time Evolution}
\label{sect:ratios+time}

We have already seen in Figs.~\ref{fig:SN-rpro-SA}, \ref{fig:SN-rpro-SB}, ~\ref{fig:KN-rpro-KA}, and \ref{fig:KN-rpro-KB} that
the abundances of different $r$-process radioisotopes depend sensitively
on the model of the SN or KN that is adopted. The same holds for actinide
isotopes, even when attention is focused on hybrid models whose parameters are
adjusted to yield ratios of the radioisotopes with $A \lesssim 200$ that are similar
to 
those measured for the metal-poor star HD 160617 or the actinide-boost star J0954+5346. 
Tables~\ref{tab:ratios-prod}, \ref{tab:ratios-Plio}, \ref{tab:ratios-pu} and \ref{tab:ratios-Devo}
show the production ratios for the $r$-process isotopes of interest after $10^5$~yr
(i.e., after the decays of short-lived isotopes), 3 Myr (as in eq.~\ref{eq:t-Plio}, corresponding to the \fe60 detection from
near the end of the Pliocene era: the ratios after 7~Myr corresponding to the other \fe60
pulse reported in~\citet{Wallner2021} are similar), 50 Myr (as in eq.~\ref{eq:t-LB}, comparable with the half-life of \pu244), and 360 Myr (eq.~\ref{eq:t-Devo}; the time since the end-Devonian mass extinction(s)), respectively, as discussed in Section~\ref{sec:intro}. These have been
calculated from the SN and KN models studied in the previous Section~\ref{sect:modelling}, and are
expressed as the abundance ratios relative to \pu244. Included 
for information are the production ratios for
several $r$-process isotopes with half-lives $> 500$~Myr, namely \th232, \u235, and \u238.
However, in view of the backgrounds from astrophysical processes before the formation of the solar
system, we do not consider further these isotopes.

Also, we note that SNe in general produce \fe60 by other
nuclear mechanisms in addition to the $r$-process, such as by explosive burning, where the yield is expected to exceed greatly any possible $r$-process contribution, so the \fe60/\pu244 ratios for SN models SA and SB
in these and subsequent tables are in general underestimates and could be viewed as lower limits on the actual ratio. {If the ratio of the \fe60 synthesized through the $r$-process to the total \fe60 produced in a SN is $\sim f(\fe60, r)$, then the \fe60/\pu244 ratio for the SN model is boosted to $Y(\fe60)/(f(\fe60, r)Y(\pu244))$. Additionally, our SN models provide the estimates per $r$-process event $ {Y_i}({\rm SN, r})$, implicitly assuming that all SNe are similar $r$-process sites. {As only a small fraction of SNe may be $r$-process sites}, we may account for SN heterogeneity by assuming, in the crudest picture, that the the $r$-process occurs in only a fraction $f_{\rm SN,r}$ of SNe. In this case, the probability that a given SN will eject $r$-process material is $f_{\rm SN, r}$, implying that, for the $r$-process yield per SN event, $ {Y_i}({\rm SN})$ must be lower, i.e., $ {Y_i}({\rm SN})  ={f_{\rm SN,r}}  {Y_i}({\rm SN, r}) $.

 \begin{table}[htb]
    \centering
    \caption{{\em r}-Process Isotope Production Ratios after $10^5 \ \rm yr$ }
    \label{tab:ratios-prod}
 \hspace{-2cm}
 \begin{tabular}{c|cc|cc}
    \hline \hline
    Radioisotope & \multicolumn{2}{c|}{SN Model} & \multicolumn{2}{c}{KN Model} \\
    Ratio & SA & SB & KA & KB \\
    \hline
    \fe60/\pu244 &0.39 &$2.5\times10^{4}$ &$8.6\times10^{-3}$ &$4.5\times10^{-5}$\\
    \zr93/\pu244 &35 &$6.4\times10^{5}$ &28&1.1\\
    \pd107/\pu244 & $1.4\times10^{2}$ &$3.9\times10^{5}$ &18&1.8\\
    \i129/\pu244 &$7.1\times10^{2}$ &$4.5\times10^{6}$ &$1.8\times10^{2}$&41\\
    \cs135/\pu244 &48 &$1.2\times10^{6}$ &2.6 &13\\
    \hf182/\pu244 &7.5 &$1.2\times10^{4}$ &1.5 &0.28\\
    \th232/\pu244 & 2.7 & 24 & 1.7 & 0.65 \\
    \u235/\pu244 & 2.9 & 15 & 3.1 & 1.7 \\
    \u236/\pu244 &3.8 &23 &3.7 &2.4\\
    \u238/\pu244 & 2.2 & 9.6 & 2.4 & 1.5 \\
    \np237/\pu244 &3.3 &8.9 &3.4 &2.6\\
    \pu242/\pu244 &1.9 &2.6 &1.9 &1.9\\
    \cm247/\pu244 &1.1 &1.2 & 0.97 & 1.0\\
    \cm248/\pu244 &1.1 &1.5 &0.86 &1.1\\
    \hline \hline
 \end{tabular}
 \end{table}
 
Fig.~\ref{fig:plutonium} displays the subsequent time evolution of several
radionuclides of interest (\fe60, \zr93, \pd107, \i129, \cs135, \hf182, \pu244 and \cm247) from an $r$-process event only as calculated using
the SN and KN models discussed in the previous section.
The solid lines are obtained using our baseline $r$-process calculation,
and the shaded bands are the ranges that result from the nuclear data variations described in Section~\ref{sect:modelling}.
Also shown as vertical lines are the three timescales
of interest, namely the age of the well-attested SN explosion $\sim 3$~Mya, 
an age of $50$~Mya comparable with the half-life of \pu244, 
and the age of the end-Devonian extinction(s) $\sim 360$~Mya.
We see that there are substantial differences between the abundances
of the radioisotopes calculated in different models. 
We note that the relative production rates of light (second peak and lighter) and heavy (third peak and higher) $r$-process nuclei depend strongly on the astrophysical conditions of the $r$-process sites. On the other hand, the relative ratios of the actinides themselves are largely insensitive to the site and thus have less discriminating power. The uncertainties in the relative yields of the actinides are dominated by large nuclear physics uncertainties in this region, so their yields depend sensitively on the choice of nuclear data adopted (as seen in Figure~\ref{fig:plutonium}).
Therefore
measurements of radioisotope ratios, especially the ratios of light $r$-process nuclei to actinides, could provide useful diagnostic tools for the nature of
any astrophysical explosion that occurred near Earth within the last few hundred
million years. 

 \begin{figure}[!htb]
	\centering
   \includegraphics[width=7cm]{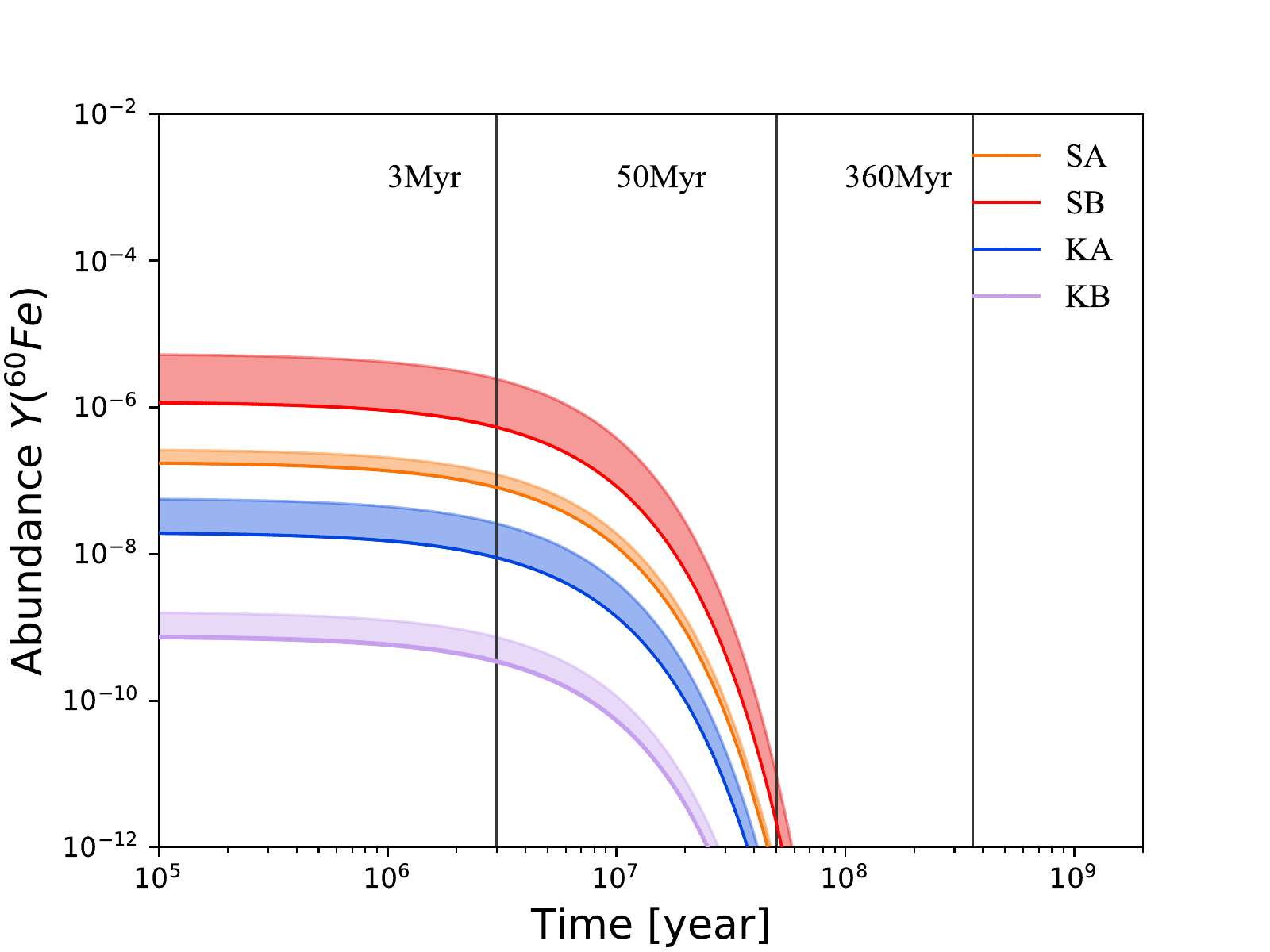}
   \includegraphics[width=7cm]{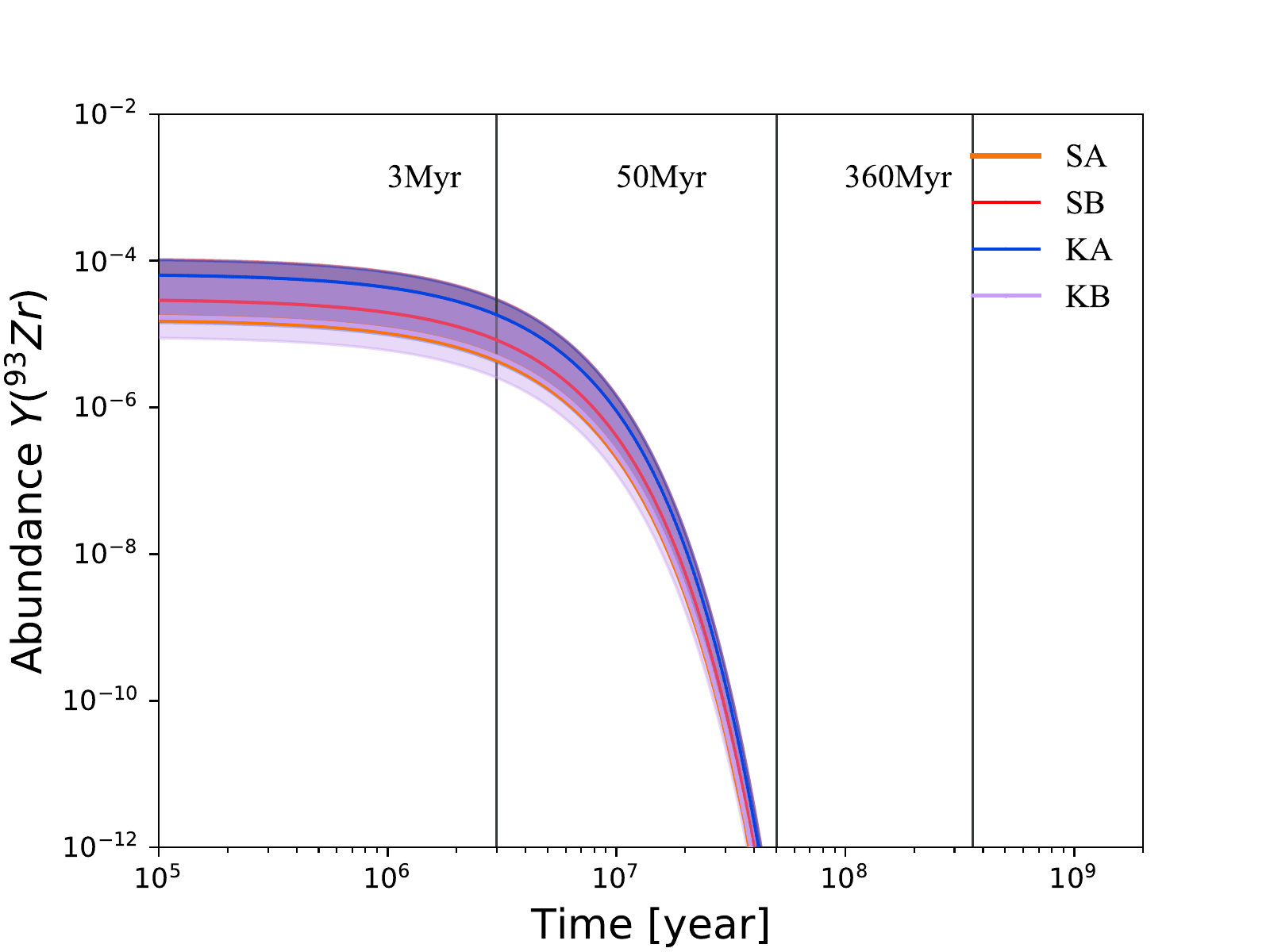}   
   \vspace{-4mm}
   \includegraphics[width=7cm]{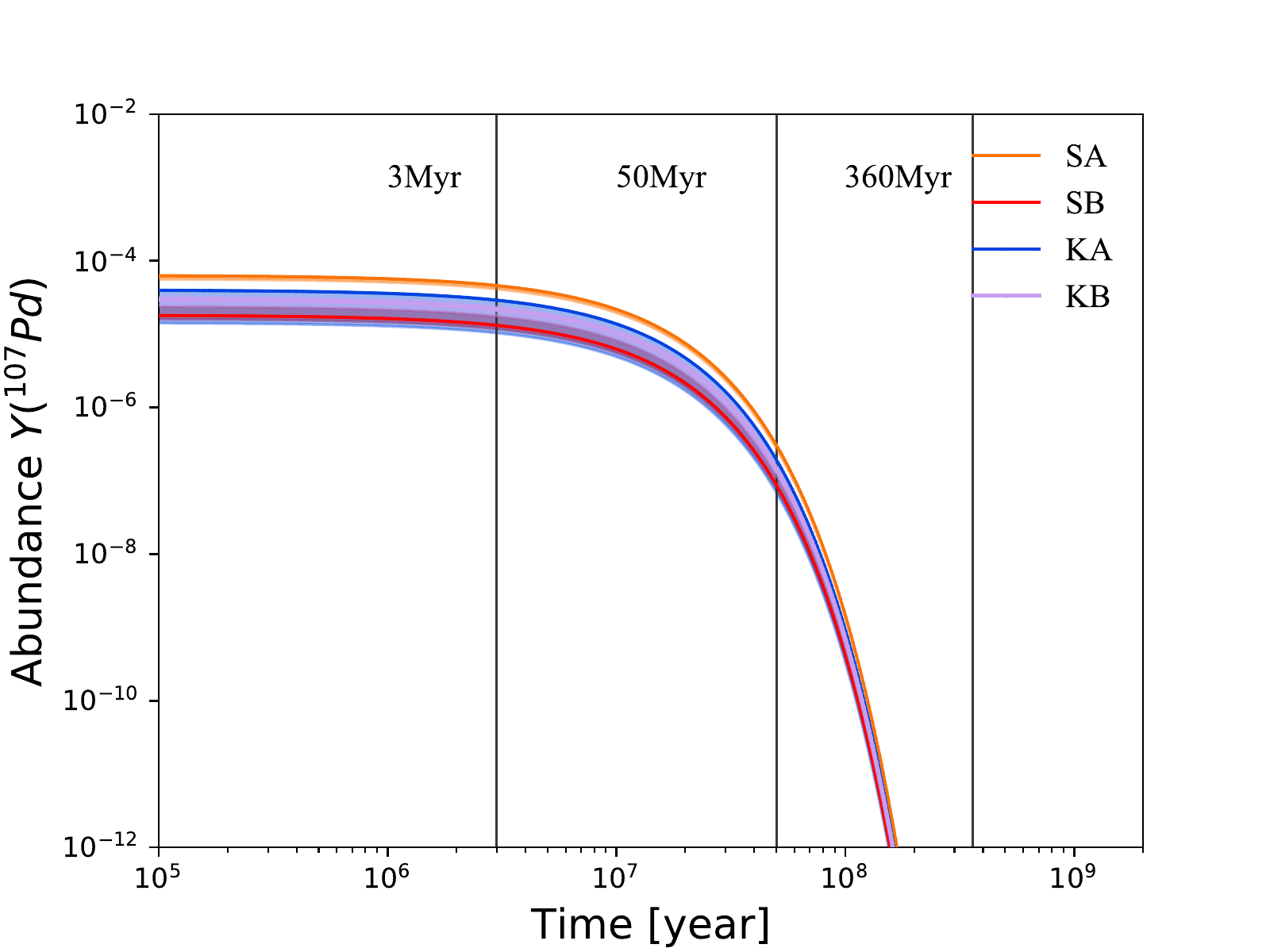}   
   \includegraphics[width=7cm]{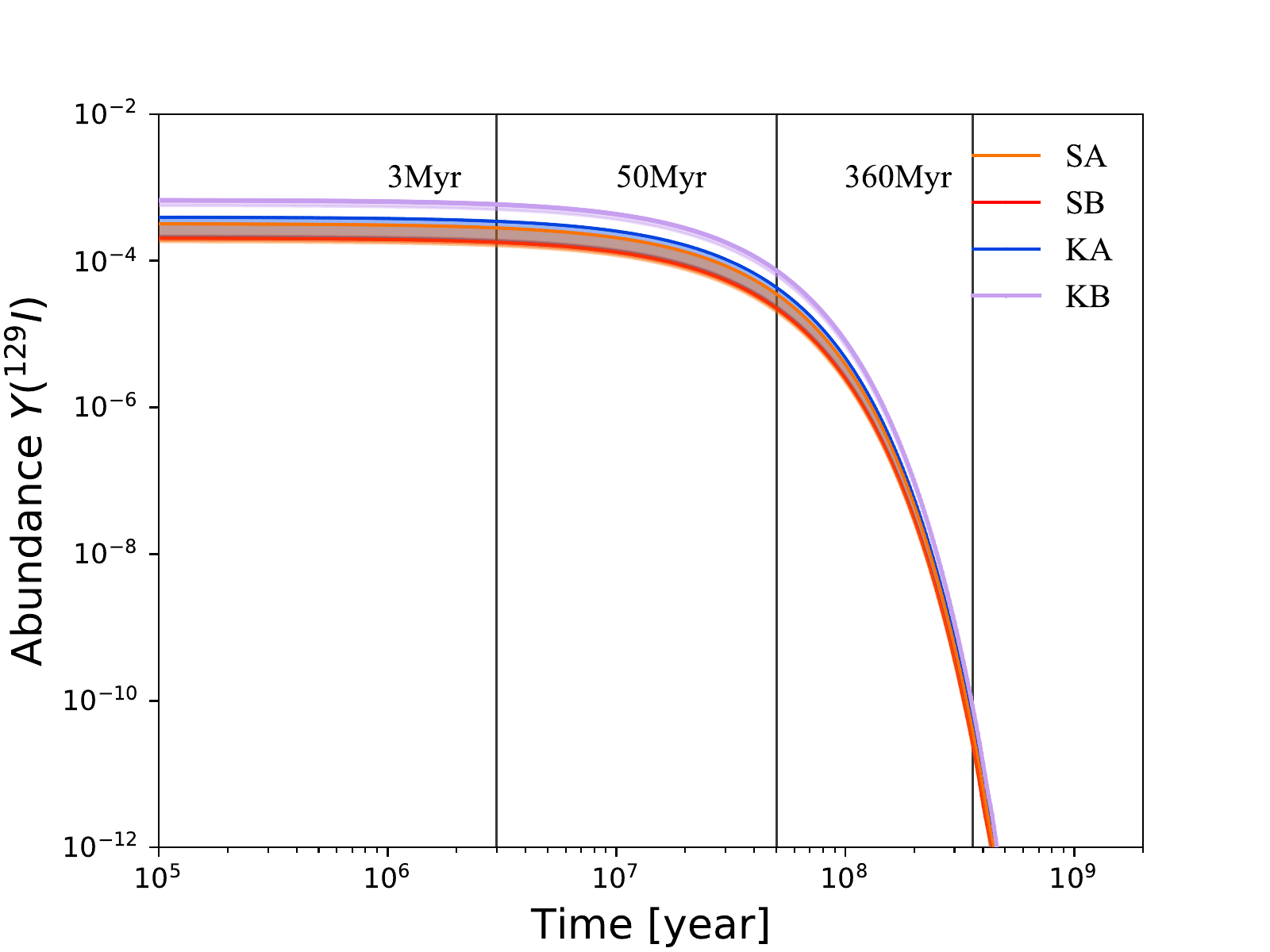}
   \vspace{-4mm}
    \includegraphics[width=7cm]{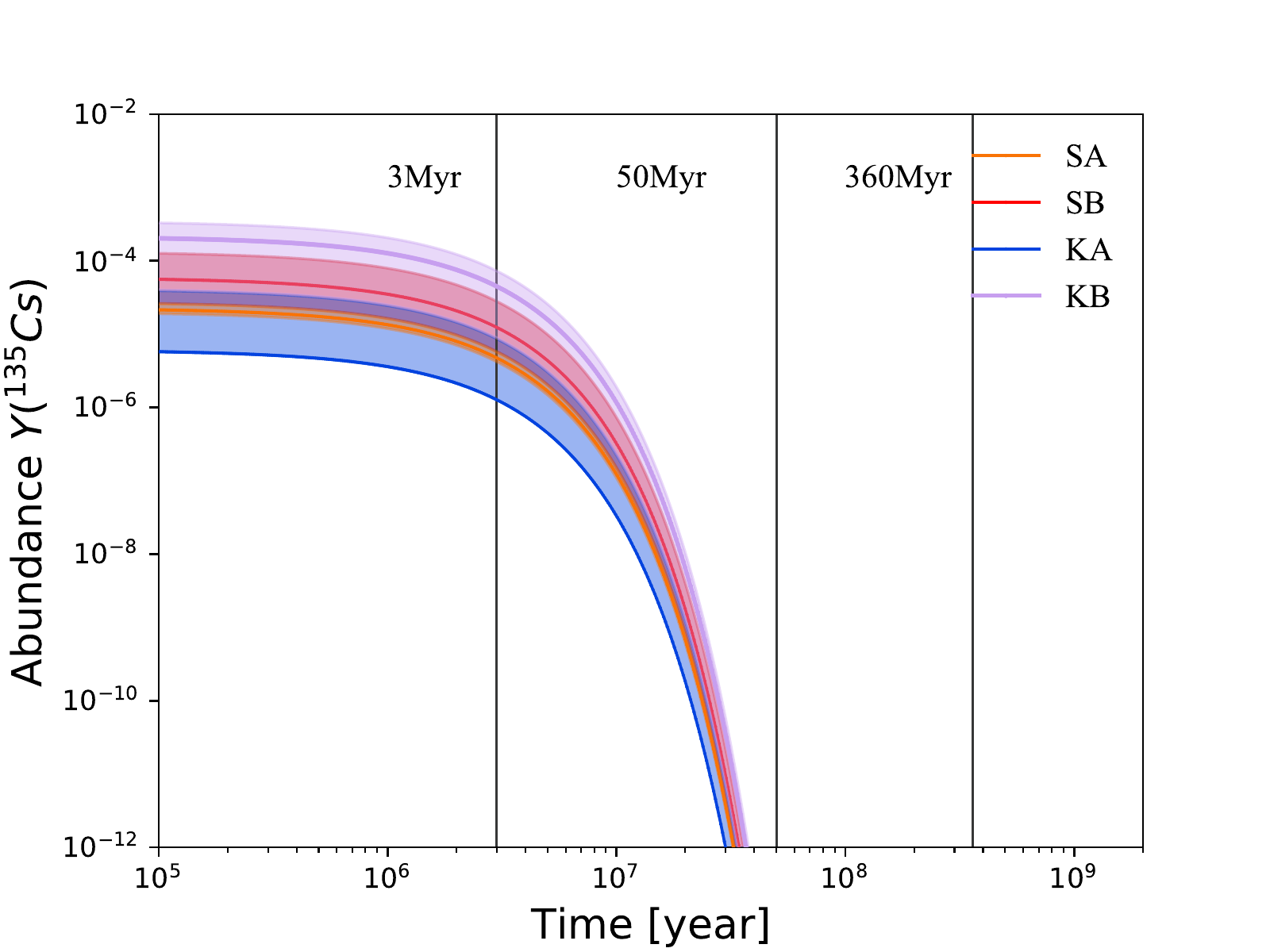}
     \includegraphics[width=7cm]{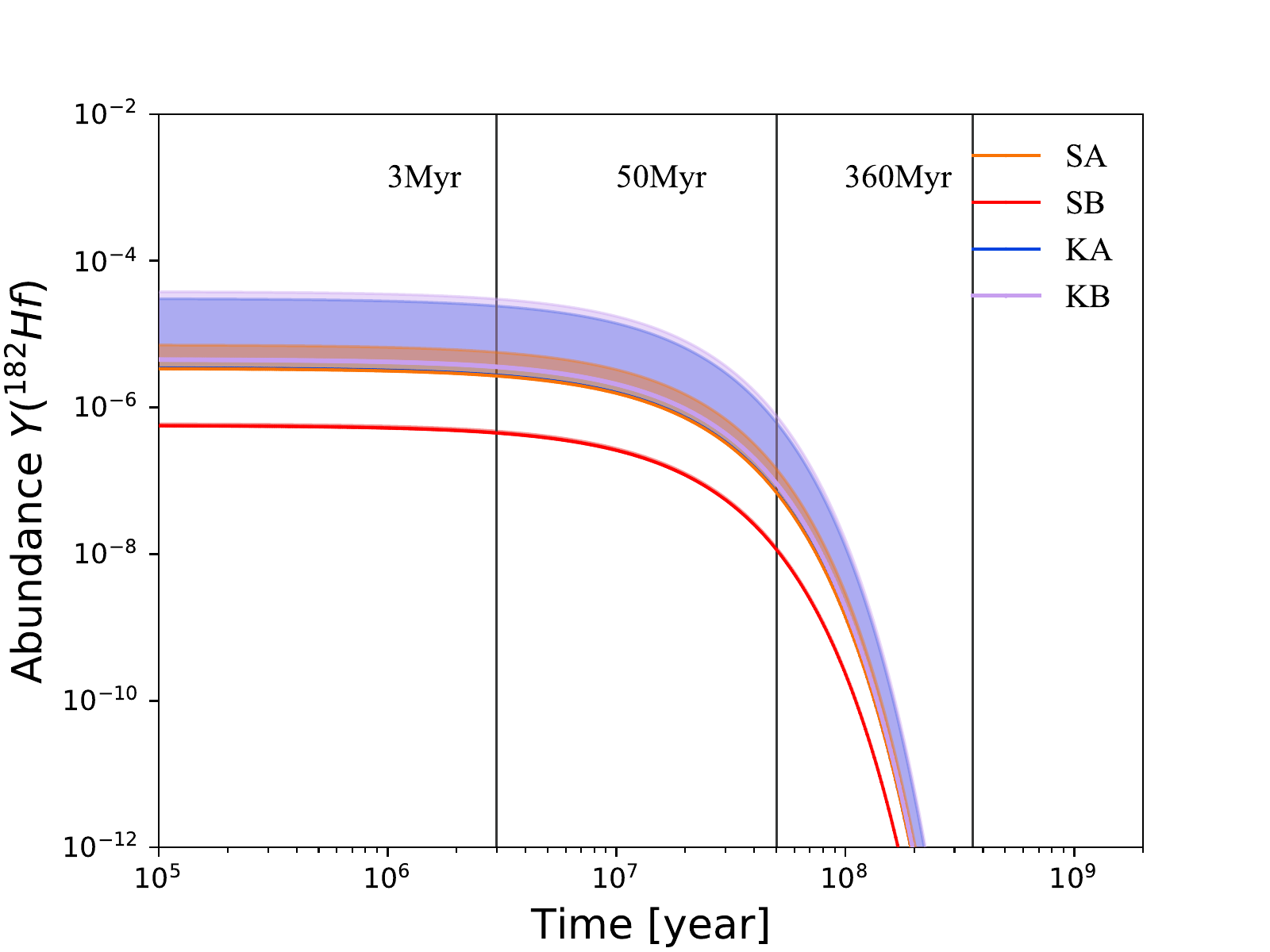}
     \vspace{-4mm}
  \includegraphics[width=7cm]{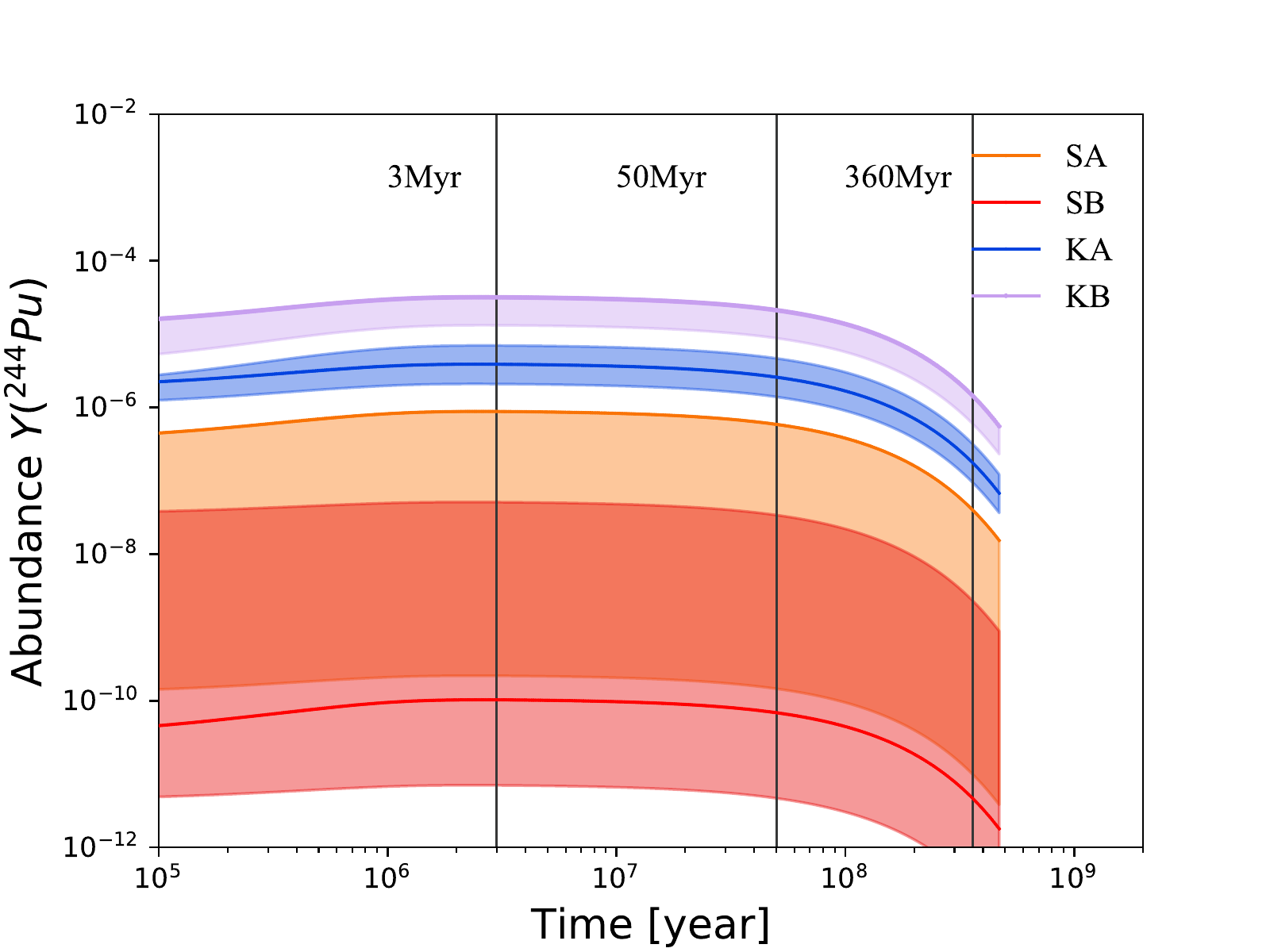}
   \includegraphics[width=7cm]{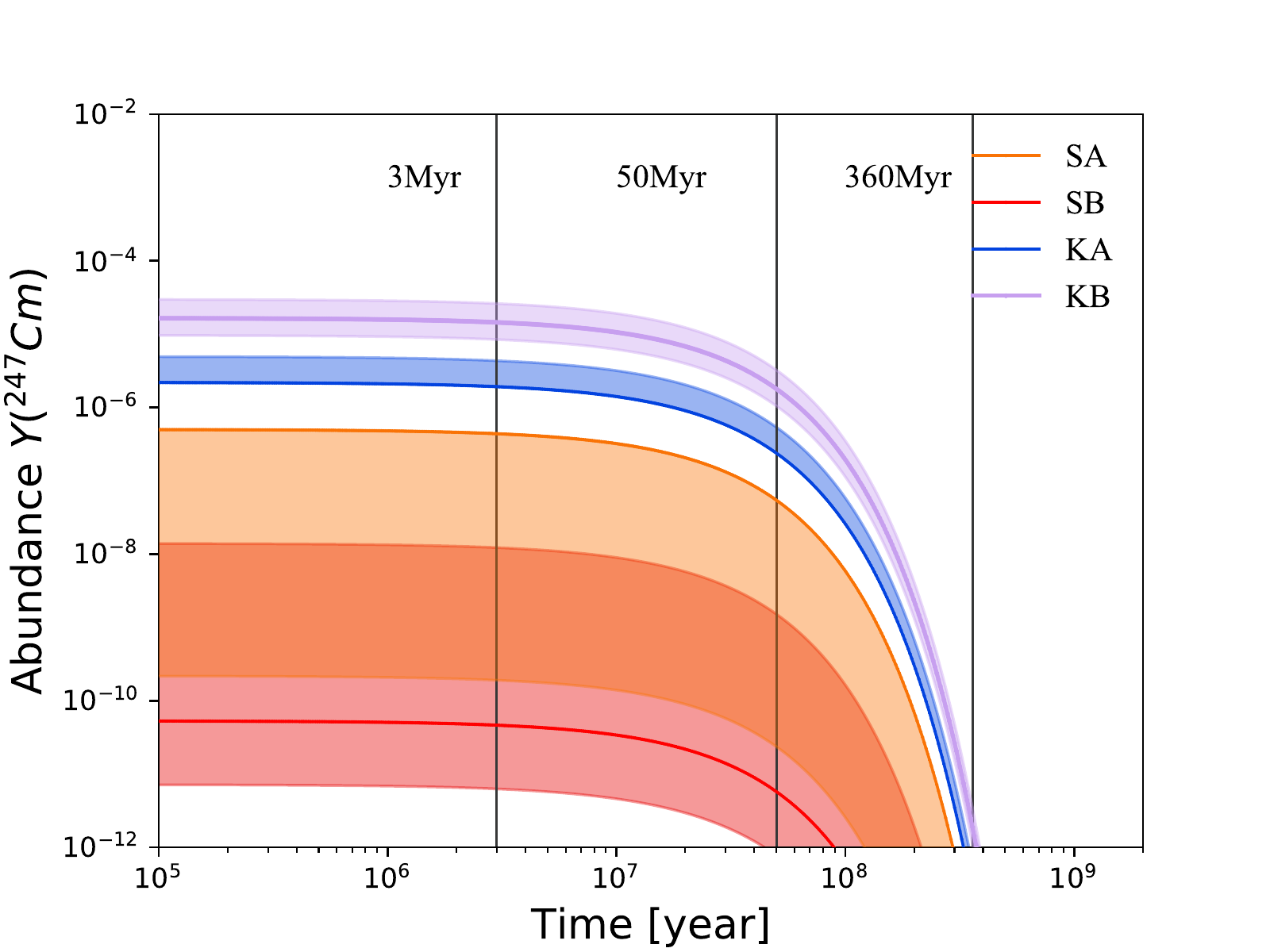}
\vspace{5mm}
	\caption{\it The time evolutions of the relative abundances of \fe60, \zr93, \pd107, \i129, \cs135,\hf182, \pu244, and \cm247 from an $r$-process event as calculated with the
	different astrophysical models described in Table~\ref{tab:models}, with the total mass of $r$-process species normalized to unity for each model. 
	The solid lines were obtained using our baseline $r$-process calculation, 
	and the shaded bands are the ranges due to uncertainties resulting from the adoption of different nuclear data (HFB masses from \citet{HFB17PRL} or $\beta$-decay rates from \citet{MKT} for both SN and KN models, or fission yields from \citet{KT} for KN models, in addition to the baseline nuclear data) in this calculation (orange: SA; red: SB; blue: KA; and light purple: KB).
	The vertical lines indicate the age of the supernova explosion $\sim 3$~Mya attested by discoveries of deposits of \fe60, 
    an age of $50$~Mya comparable with the half-life of \pu244, 	
    and the age of the end-Devonian extinction(s) $\sim 360$~Mya. 
}
	\label{fig:plutonium}
\end{figure}

As we have already discussed, measurements of terrestrial \fe60 deposits indicate
that at least one such explosion took place within $\sim 100$~pc of Earth about
3~Mya, and there has also been deposition of \pu244 on Earth
that may extend back to 25~Mya. The presence of \pu244 indicates that there has
been at least one active $r$-process site close to Earth within the past 80~Myr
or so. In the following, we treat the \pu244 abundance as our reference,
and predict the relative abundances of other $r$-process radioisotopes under 
different hypotheses about the nature of the site(s) and its (their)
timing.

For this purpose, we follow the evolution of interesting
radioisotopes over {$\sim$~Gyr}, using the calculations of the previous section
as starting points and taking account of the possible production of the
radioisotopes via the decays of heavier isotopes as well as their own decays.
The left panels of Fig.~\ref{fig:spaghetti} illustrate the abundances of
the radioisotopes of principal interest, while the right panels show the ratios to \pu244.
The upper panels show the results calculated in SN model SA, and the
lower panels show the results in model SB.
Most of the 
radioisotopes exhibit simple decay curves, but the effects of feedthrough from the 
decays of heavier isotopes are visible in 
\pu244, which is made in $\alpha$-decays of \cm248,
and in \u236, which is a decay product of \pu244.
We have included the $r$-process production of \fe60
in these plots, although the $r$-process is not expected to dominate its
production, at least in core-collapse SNe. Hence the \fe60 curves should 
be regarded as (very) conservative lower bounds on the \fe60 yields and ratios
to \pu244 production.

\begin{figure}[!htb]
	\centering
	\vspace{-5mm}
    \includegraphics[width=8cm]{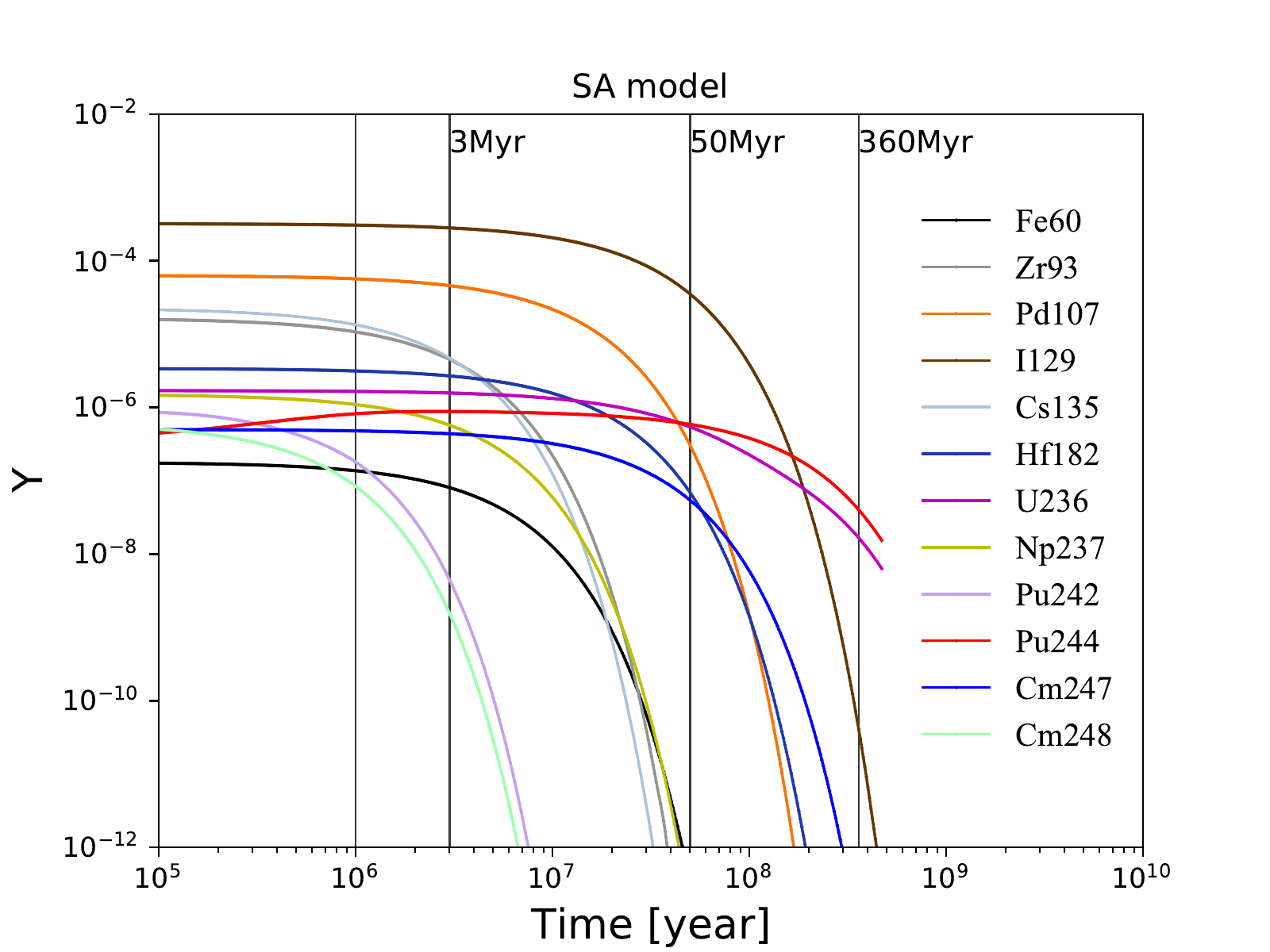}
  \includegraphics[width=8cm]{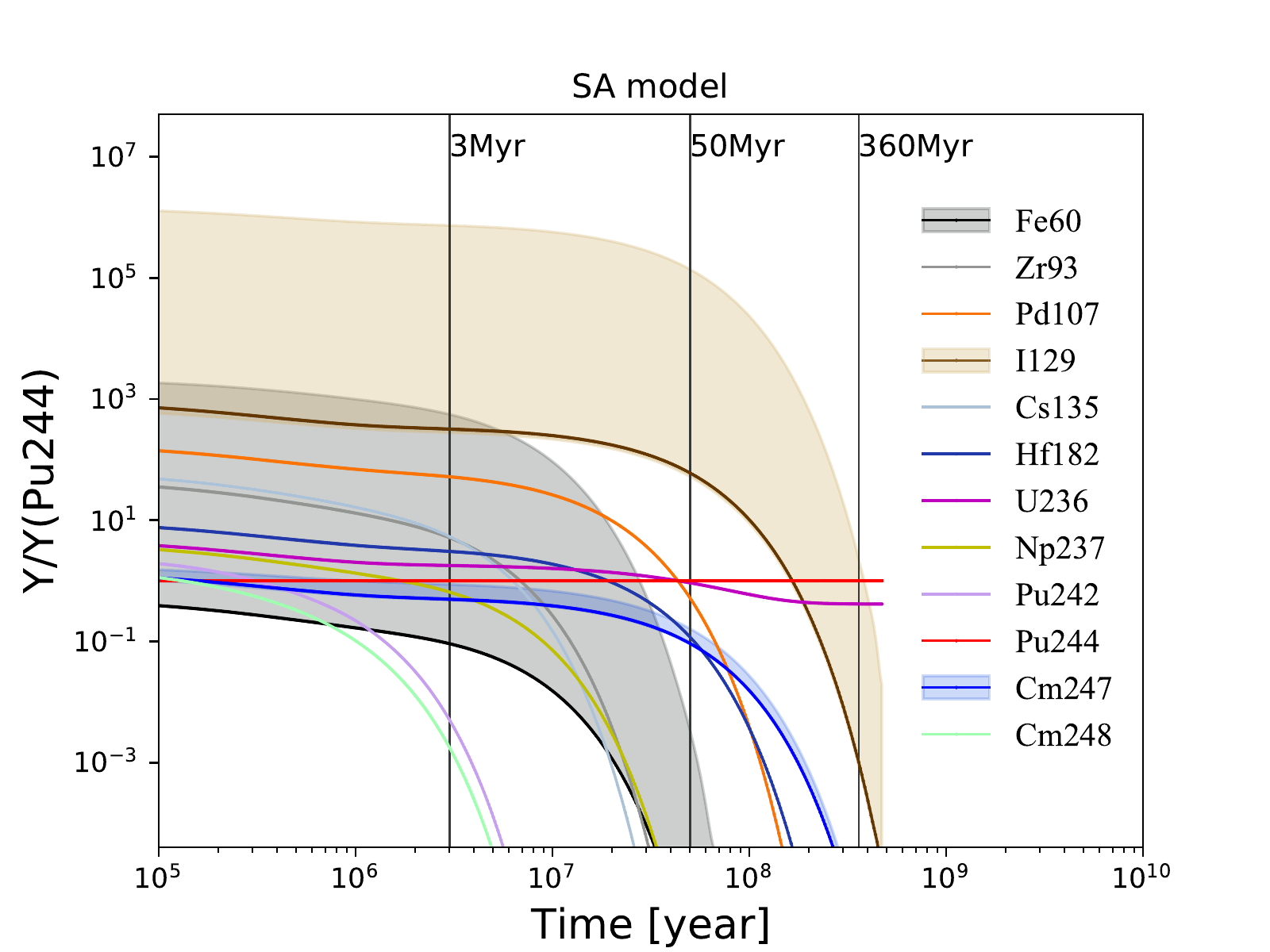}
  \includegraphics[width=8cm]{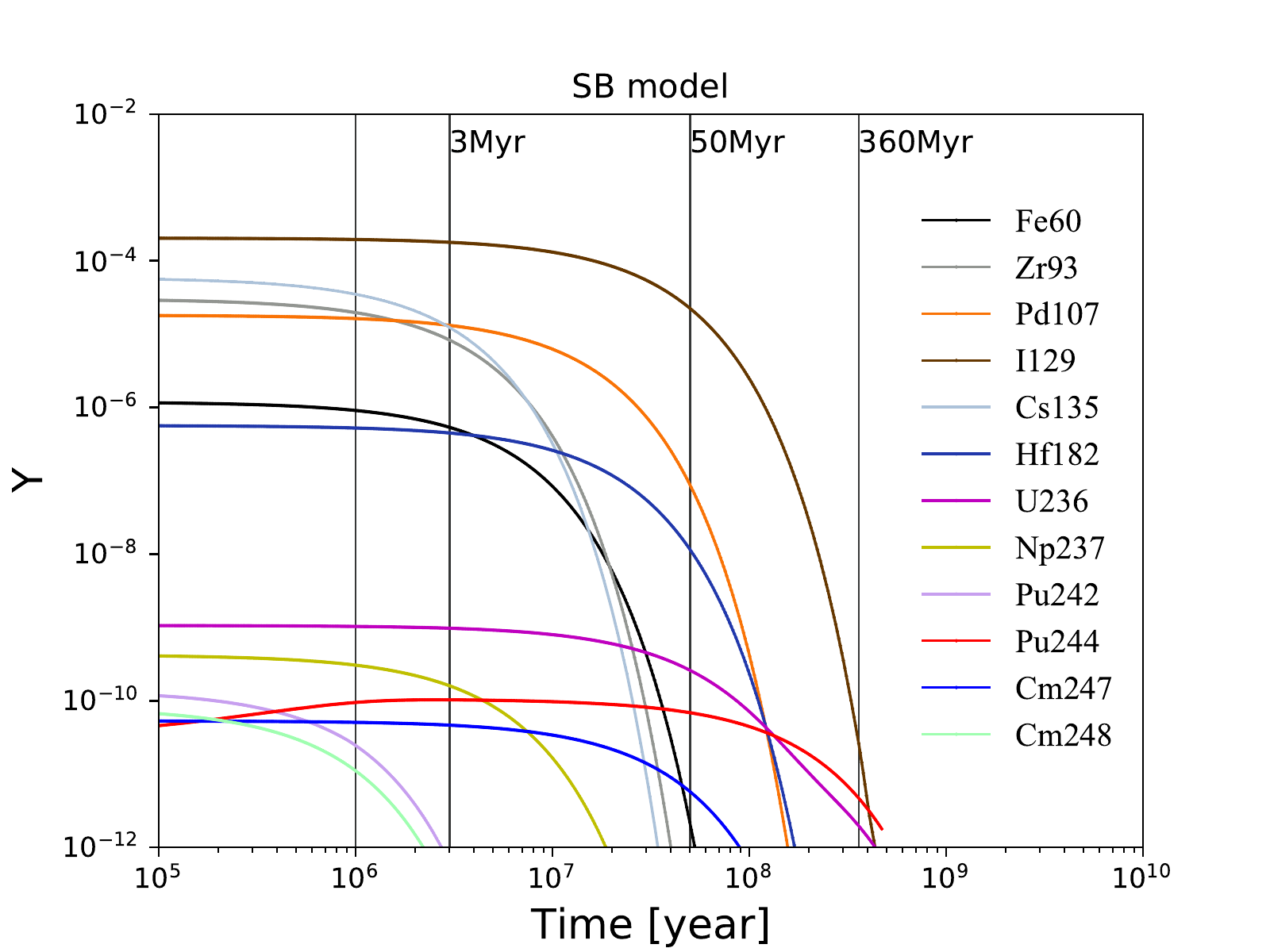}
  \includegraphics[width=8cm]{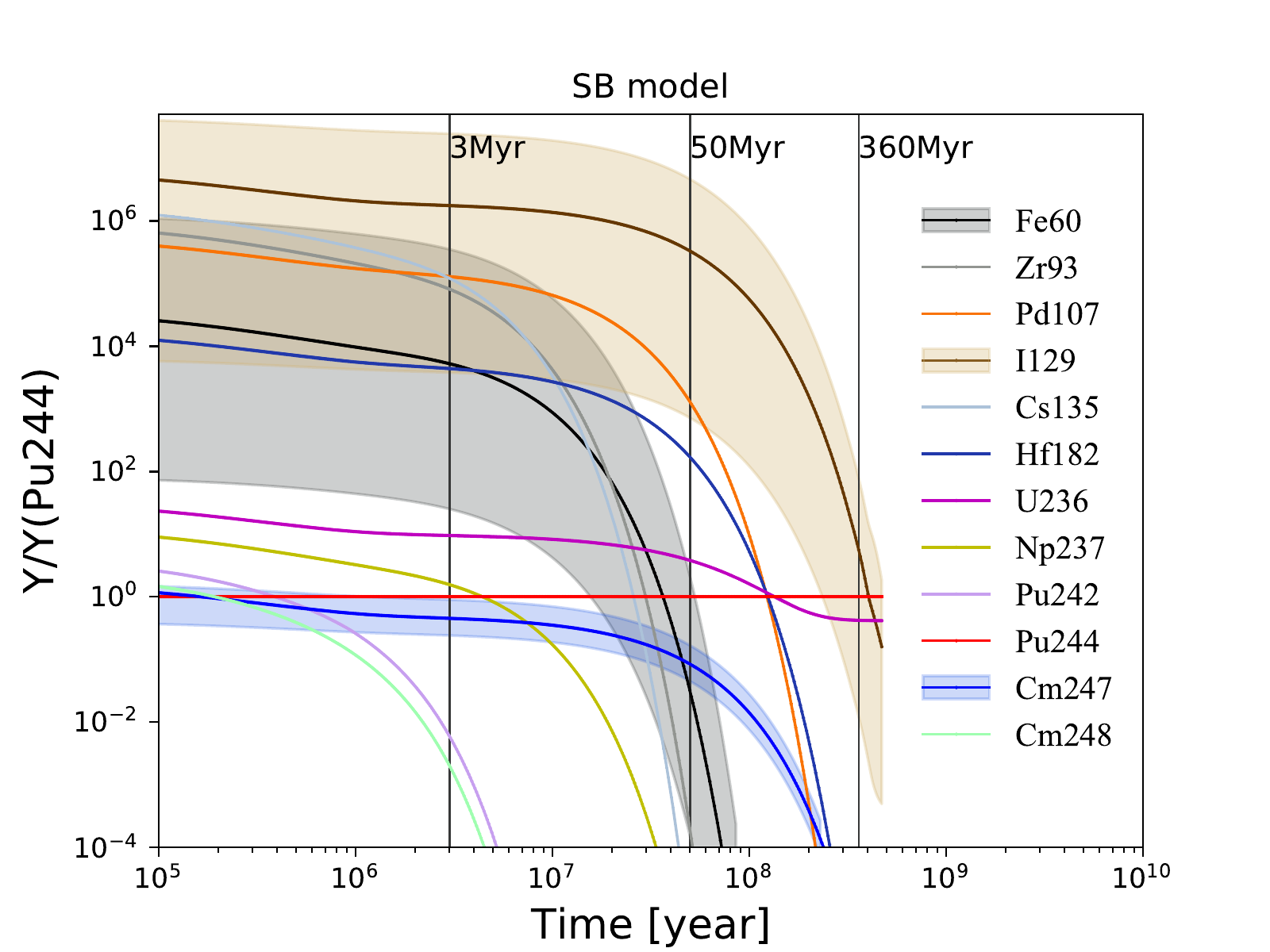}
	\caption{\it The time evolution of the abundances $Y$ (left) and ratios $Y$/$Y(\pu244)$ (right) of $r$-process nuclei of interest from SN model SA (top panels) and MHD SN model SB (bottom panels). The vertical lines indicate the age of the SN explosion $\sim 3$~Mya attested by discoveries of deposits of \fe60, 
	50~Mya comparable with the half-life  of \pu244,
	and the age(s) of the end-Devonian extinction(s) $\sim 360$~Mya. The ratio ranges due to the nuclear variations as described in the text for \fe60, \i129 and \cm247 to \pu244 are shown in shaded bands of black, brown, and blue, respectively.
	}
	\label{fig:spaghetti}
\end{figure}

The vertical lines in Fig.~\ref{fig:spaghetti} are at 3~Myr,
corresponding to the time since the event
that gave rise to the well-attested deep-ocean \fe60 deposition;
50~Myr comparable  with  the  half-life  of \pu244;
and 360~Myr,
corresponding to the time since the end-Devonian extinction(s).
The observation of \pu244 signals overlapping the \fe60 
signals from 3~Mya and 7~Mya suggests that all radioisotopes with yields $Y \gtrsim Y(\pu244)$ would 
be interesting targets for searches in the layers containing the \fe60 signals.
Depending on the model, these may include \zr93, \pd107, 
\i129, \cs135, \hf182, \u236, \np237 and \pu244, and the abundance of \cm247 may
not be much smaller than that of \pu244, as seen in
the right panels of Fig.~\ref{fig:spaghetti} and in Table~\ref{tab:ratios-Plio}.
The ratios \zr93/\pu244, \pd107/\pu244, \i129/\pu244, \cs135/\pu244, 
and \hf182/\pu244 could be particularly useful for discriminating between models, 
followed by \u236/\pu244 and \np237/\pu244.

\begin{table}[htb]
    \centering
    \caption{{\em r}-Process Isotope Ratios after 3 Myr, Corresponding to the Event near the end
    of the Pliocene Era,
    together with Their Ranges $\sigma$ Found in the Different Nuclear Models Adopted (HFB Masses from \citet{HFB17PRL} and $\beta$-Decay Rates from \citet{MKT} for Both SN and KN Models, as well as Fission Yields from \citet{KT} for KN Models), in addition to the Baseline Calculations in the $r$-Process Simulations.}
    \label{tab:ratios-Plio}
\hspace{-2cm}
\begin{tabular}{c|cc|cc}
    \hline \hline
    Radioisotope & \multicolumn{2}{c|}{SN Model} & \multicolumn{2}{c}{KN Model} \\
    Ratio & SA & SB & KA & KB \\
    \hline  
    \fe60/\pu244 &$9.2\times 10^{-2}$  & $5.3\times 10^{3}$  & $2.3\times 10^{-3}$ & $1.1 \times 10^{-5}$ \\
   $\sigma$ (\fe60/\pu244) &$ 9.2\times 10^{-2}$ - $5.6\times 10^{2}$ & 25 - $3.5\times 10^{5}$ & $1.4\times 10^{-3}$ - $1.3\times 10^{-2}$  & $(1.1 -4.0)\times 10^{-5}$ \\
   
     \zr93/\pu244 &5.2  & $8.2\times 10^{4}$ & 4.8 &0.15 \\
   $\sigma$(\zr93/\pu244) &5.2 - $2.4\times 10^{4}$ & 93 - $4.3\times 10^{6}$ & 0.59 - 14  & 0.15 - 0.97 \\
   
    \pd107/\pu244 &52  & $1.3\times 10^{5}$  & 7.5 & 0.69\\
   $\sigma$(\pd107/\pu244) &50 - $2.0\times 10^{5}$ & $3.4\times 10^{2}$ - $1.7\times 10^{6}$ & 2.4 - 7.5 & 0.69 - 1.5 \\
   
    \i129/\pu244 &$3.2\times 10^{2}$ &$1.7\times 10^{6}$ &89 & 19\\
   $\sigma$(\i129/\pu244) & $2.8\times 10^{2}$ - $7.3\times 10^{5}$ & $3.8\times 10^{3}$ - $2.4\times 10^{7}$ & 46 - 89  & 19 - 38 \\
   
    \cs135/\pu244 &5.4 &$1.2\times 10^{5}$ &0.33 & 1.4\\
   $\sigma$(\cs135/\pu244) &5.4 - $1.9\times 10^{4}$ & $1.2\times 10^{2}$ - $4.1\times 10^{6}$ & 0.33 - 1.4 & 0.30 -3.9 \\   
   
    \hf182/\pu244 &3.1 & $4.4\times 10^{3}$ & 0.71  &0.11 \\
   $\sigma$( \hf182/\pu244 ) &3.1 - $2.6\times 10^{4}$ & 8.7 - $6.8\times 10^{4}$ &0.71 - 9.0 & 0.11 - 2.3 \\
   
    \th232/\pu244 &1.5 &11 &1.2  &0.43\\
   $\sigma$(\th232/\pu244) &0.23 - 6.0 & 0.31 - 16 & 0.36 - 1.2 & 0.27 - 0.63\\
   
    \u235/\pu244 &1.6 &7.0 &1.9 & 0.99\\
       $\sigma$(\u235/\pu244 ) &1.2 - 7.2 & 2.0 - 21 & 1.4 - 4.6 & 0.99 - 1.7\\
    
    \u236/\pu244 &1.8 &9.5 & 2.0 & 1.1\\
       $\sigma$(\u236/\pu244) &1.2 - 4.1 & 2.0 - 9.5 & 1.2 - 2.5 & 0.79 - 1.1 \\
       
    \u238/\pu244 &2.1 &5.4 &2.5 &1.8 \\
       $\sigma$(\u238/\pu244) &2.1 - 3.2 & 3.2 - 6.9 & 2.0 - 2.7 & 1.7 - 1.8\\
       
    \np237/\pu244 &0.66 &1.6 & 0.78 & 0.53 \\
    $\sigma$(\np237/\pu244) &0.66 - 1.9 & 1.0 - 2.6 & 0.78 - 1.9 & 0.53 - 2.0\\
    
    \pu242/\pu244 & $5.3\times 10^{-3}$ & $6.2\times 10^{-3}$  & $6.0\times 10^{-3}$  & $5.2\times 10^{-3}$ \\
   $\sigma$(\pu242/\pu244) & $(5.3 - 6.7)\times 10^{-3}$ & $(6.2 - 8.8)\times 10^{-3}$ & $(5.0 - 6.1)\times 10^{-3}$ & $(5.0 - 5.6)\times 10^{-3}$ \\
    
    \cm247/\pu244 &0.50 &0.45 & 0.50  & 0.46\\
    $\sigma$(\cm247/\pu244) &0.48 - 0.87 & 0.24 - 0.90 & 0.62 - 1.2 & 0.46 - 1.4\\
    
    \cm248/\pu244 & $1.9\times 10^{-3}$ & $2.1\times 10^{-3}$  & $1.6\times 10^{-3}$ & $1.9\times 10^{-3}$ \\
   $\sigma$(\cm248/\pu244) & $(1.3 - 1.9 )\times 10^{-3}$ & $(1.0 - 2.1)\times 10^{-3}$ & $(1.5 - 2.3)\times 10^{-3}$ & $(1.9 - 2.2)\times 10^{-3}$ \\
   
    \hline \hline
    \end{tabular}
\end{table}

However, only a fraction of the reported \pu244 may have been generated by the events producing the \fe60 deposition
peaks, with the remainder being due to one or more earlier astrophysical events. In this
case it is natural to compare abundances on a time-scale of
$\sim 50$~Myr, which is comparable  with  the  half-life  of \pu244,
corresponding to the central vertical lines
in the right panels of Fig.~\ref{fig:spaghetti}. On this time scale, as seen in
this figure and in Table~\ref{tab:ratios-pu}, the most interesting remaining
isotopes are \pd107, \i129, \hf182, \u236, and \pu244, with the first three providing
the greatest discriminating power, albeit with similar nuclear model uncertainties 
to those discussed above (not shown).

\begin{table}[htb]
    \centering
    \caption{{\em r}-Process Isotope Ratios after 50 Myr, comparable with the half-life of \pu244.
    \label{tab:ratios-pu}}
\hspace{-2cm}
\begin{tabular}{c|cc|cc}
    \hline \hline
    Radioisotope & \multicolumn{2}{c|}{SN Model} & \multicolumn{2}{c}{KN Model} \\
    Ratio & SA & SB & KA & KB \\
    \hline  
    \fe60/\pu244 & $5.9\times 10^{-7}$ & $ 3.2\times 10^{-2}$ & $1.4\times 10^{-8}$ &$6.6\times 10^{-11}$  \\
    \zr93/\pu244 &$1.5\times 10^{-7}$ &$2.0\times 10^{-4}$ &$1.2\times 10^{-8}$ &$3.8\times 10^{-10}$\\
    \pd107/\pu244 &$0.53$ &$1.3\times 10^{3}$ & $ 7.6\times 10^{-2}$  & $6.9\times 10^{-3}$  \\
    \i129/\pu244 &$60$ & $3.3\times 10^{5}$ & 17 &$3.5$ \\
    \cs135/\pu244 &$2.0\times 10^{-10}$ &$4.5\times 10^{-6}$ &$1.2\times 10^{-11}$ & $5.4\times 10^{-11}$\\
    \hf182/\pu244 & $0.12$ & $1.7\times 10^{2}$ & 0.028 & $ 4.4\times 10^{-3}$ \\
    \th232/\pu244 & $4.6$ & $28$ & $4.2$ & $2.1$\\
    \u235/\pu244 & $2.9$ & $11$ & $3.4$ & $2.0$\\
    \u236/\pu244 & $0.93$ & $3.8$ & $1.0$ & $0.68$ \\
    \u238/\pu244 & $3.2$ & $8.1$ & $3.8$ & $2.6$\\
    \cm247/\pu244 & $9.3\times 10^{-2}$ &  $8.4\times 10^{-2}$ & $9.2\times 10^{-2}$ & $8.5\times 10^{-2}$ \\
    \hline \hline
    \end{tabular}
\end{table}

Finally, after 360~Myr, corresponding to the age of the end-Devonian mass
extinctions, the \pu244 abundance would have decreased by an order of magnitude,
relatively few radioisotopes would have survived, 
{and only uranium isotopes, \th232, and \i129 might have abundances comparable to that of \pu244, as seen in Table~\ref{tab:ratios-Devo}.
Measurements of \i129 should be able to distinguish between SN models,
but not other radioisotopes.}

\begin{table}[htb]
    \centering
    \caption{{\em r}-Process Isotope Ratios after 360 Myr, corresponding
    to the time since the end-Devonian mass extinction(s).
    \label{tab:ratios-Devo}}
    \hspace{-2cm}
    \begin{tabular}{c|cc|cc}
    \hline \hline
    Radioisotope & \multicolumn{2}{c|}{SN Model} & \multicolumn{2}{c}{KN Model} \\
    Ratio & SA & SB  & KA & KB  \\
    \hline
    \i129/\pu244 &$4.1\times 10^{-3}$ &$30$ &$1.3\times 10^{-3}$& $2.3\times 10^{-4}$\\
    \th232/\pu244 &$85$ &$4.4\times 10^{2}$ &$81$&$49$ \\
    \u235/\pu244 &$30$ & $1.1\times 10^{2}$ &$34$&$20$ \\
    \u236/\pu244 &$0.41$ &$0.42$ & $0.41$ & $0.41$ \\
    \u238/\pu244 &$41$ &$1.0\times 10^{2}$ &$48$ &$33$\\
    \hline \hline
    \end{tabular}
\end{table}

The results of analogous calculations for the KN models KA and KB 
are shown in the upper and lower panels of Fig.~\ref{fig:spaghetti_kn_bovard}, respectively,
with yields $Y$ in the left panels and the ratios $Y/Y(\pu244)$ in the
right panels.
As seen in this figure and in Table~\ref{tab:ratios-Plio}, the
radioisotopes \zr93, \pd107, \i129, \cs135, \hf182, \u236, \np237, \pu244, and \cm247 are again
the most promising potential signatures after 3~Myr or 7~Myr, 
with \zr93, \pd107, \i129 and \hf182
offering the greatest discriminating power between KN models, and also
between them and the SNS models. After 50~Myr, as seen in
Table~\ref{tab:ratios-pu}, the most promising radioisotopes for detection
are \i129 and \u236. However, \u236 offers little discriminating power between
the different KN models, nor between the SN and KN models.
After 360~Myr, as also seen in Fig.~\ref{fig:spaghetti_kn_bovard} and in
Table~\ref{tab:ratios-Devo}, the best prospects for detection are again offered by
\u236, which does not discriminate among the KN and SN models.
We do not consider \th232 and \u238 to be promising search targets,
in view of their long half-lives and the consequent large
backgrounds from events before the formation of the solar system.

We recall that there are considerable variations in the isotope ratios calculated using the different nuclear models. In making comparisons, we have used the $\beta$-decay rates from \citet{MKT}, and the HFB model from \citet{HFB17PRL}, which assumes different nuclear masses, in addition to the baseline calculation. We note in brackets in Table~\ref{tab:ratios-Plio} the ranges of isotope ratios {$\sigma$} found in all the models SA, SB, KA and KB.~\footnote{There are similar uncertainties in the abundance ratios {after $\sim$ 1 Myr}, which are omitted for clarity in the corresponding tables.}
The ranges of predictions that come from the nuclear model variations show some overlap between the four models. We also show the ranges due to nuclear data variations in the abundance ratios of the radioisotopes \fe60, \i129, and \cm247 to \pu244 as shaded bands in the right panels of Fig.~\ref{fig:spaghetti} and Fig.~\ref{fig:spaghetti_kn_bovard}, to illustrate the uncertainty evolution for the ratios of light elements and actinides to \pu244. The absolute abundances of \i129 and lighter isotopes are relatively insensitive to the nuclear model used as shown in Fig.~\ref{fig:plutonium}, so the uncertainties in their ratios to \pu244 are largely due to those in the \pu244 yield, and hence correlated. 
While the abundances of isotopes heavier than \u236 are sensitive to the nuclear variations, they have similar uncertainty trends; thus their abundances relative to that of \pu244 vary over smaller uncertainty ranges, as quoted in Table~\ref{tab:ratios-Plio}.

\begin{figure}[!htb]
	\centering
	\vspace{-5mm}
   \includegraphics[width=8cm]{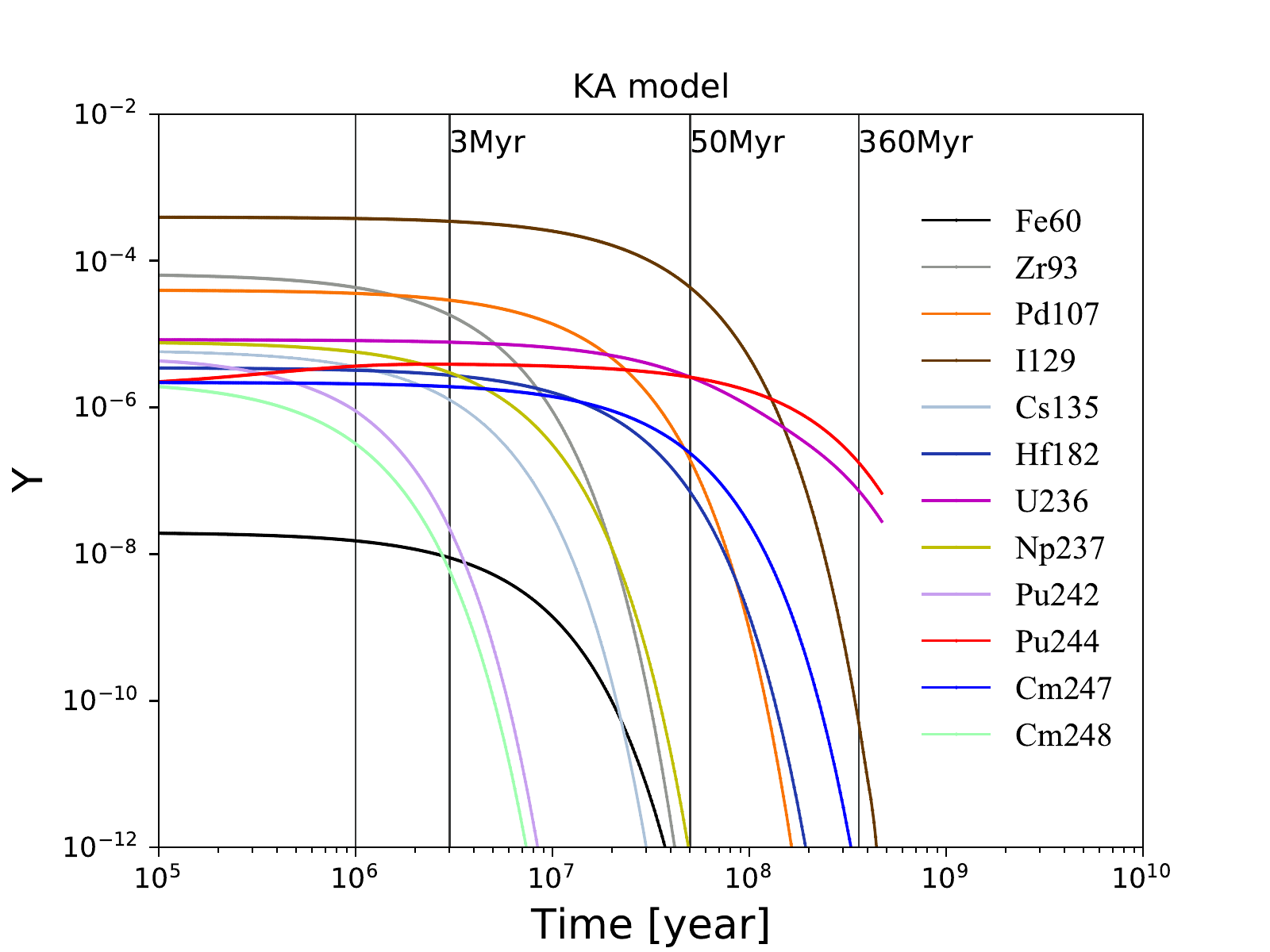}
  \includegraphics[width=8cm]{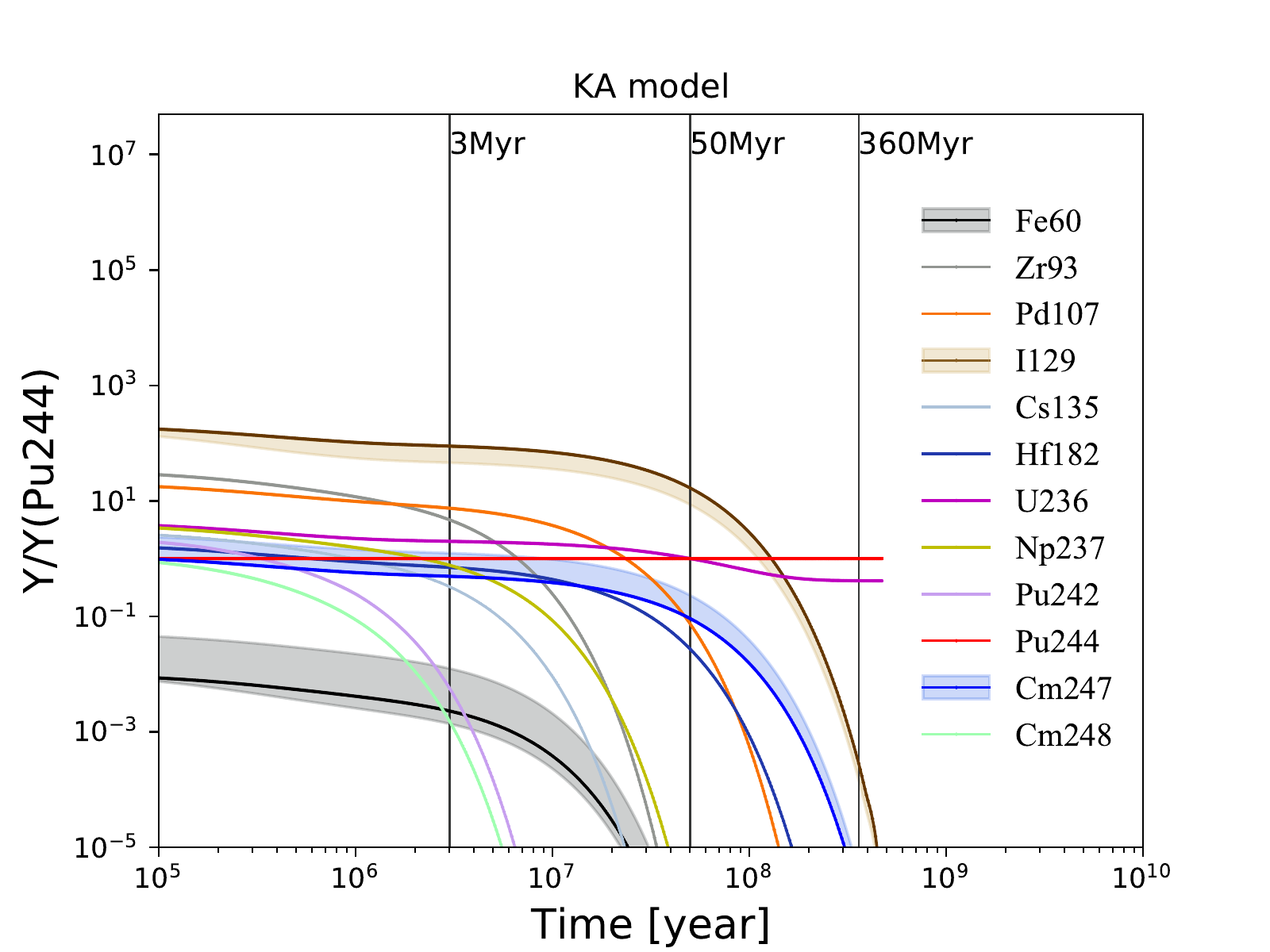}
     \includegraphics[width=8cm]{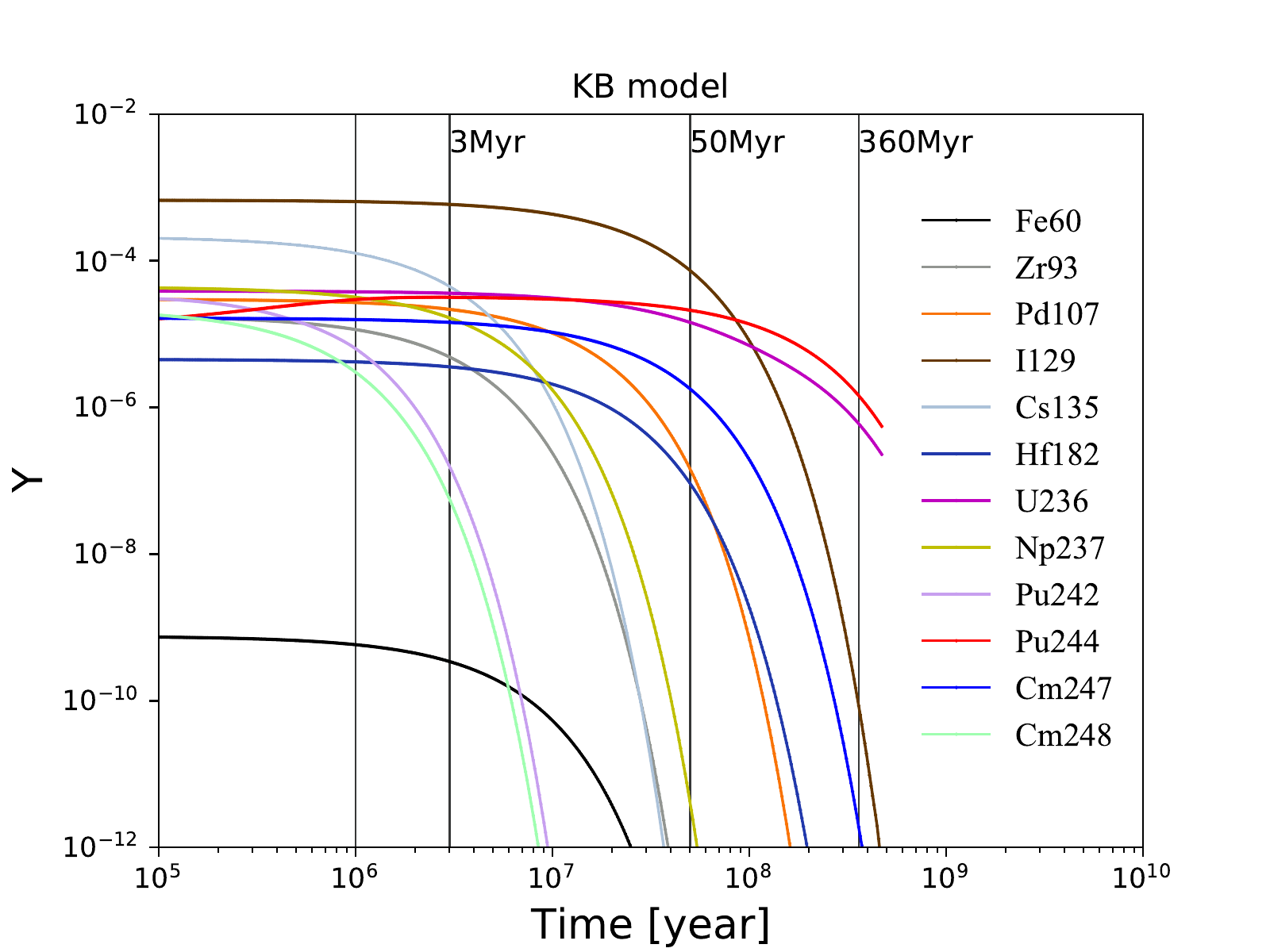}
  \includegraphics[width=8cm]{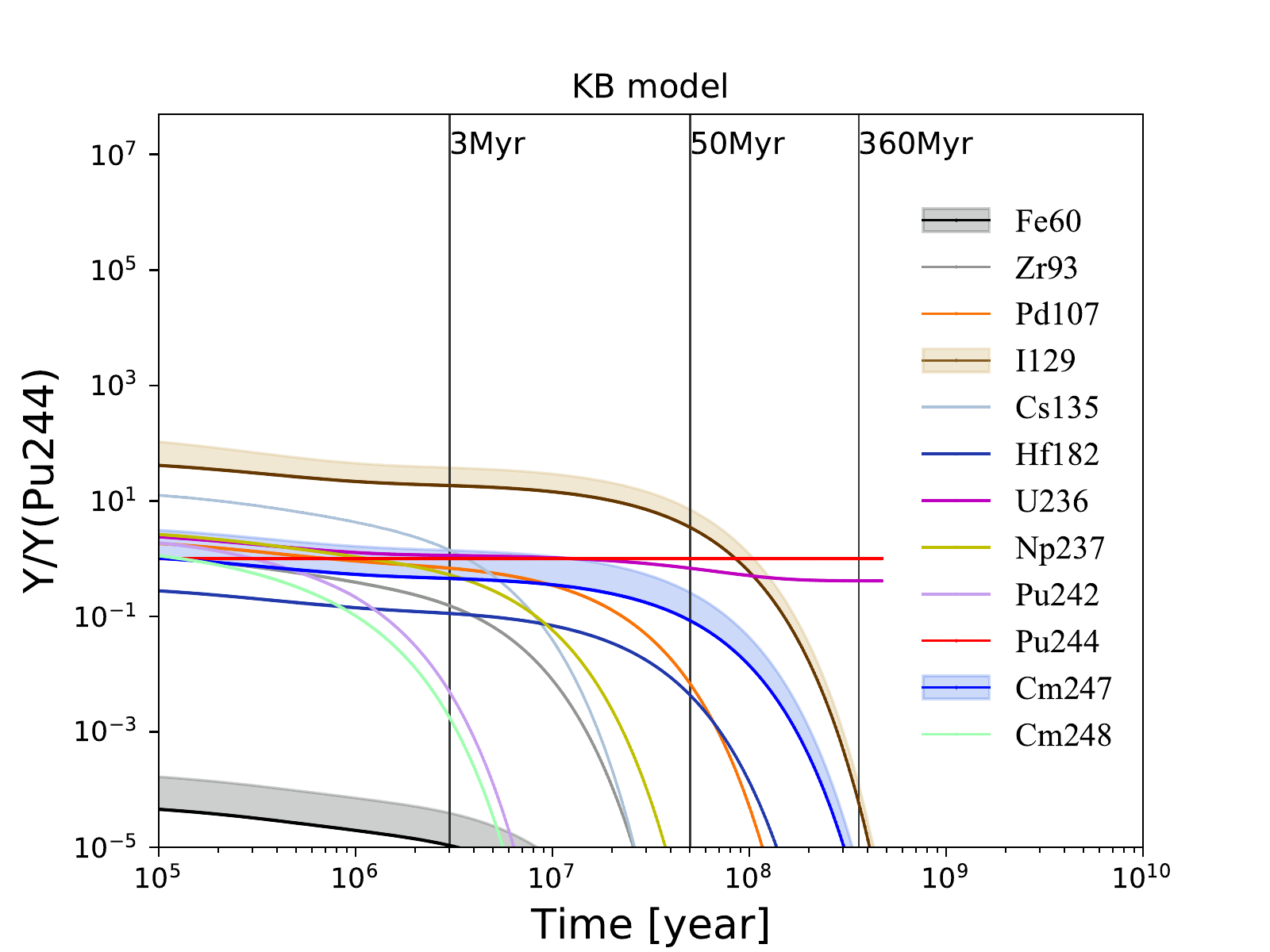}
	\caption{\it The time evolution of the abundances $Y$ (left) and ratios $Y$/$Y(\pu244)$ (right) of $r$-process nuclei of interest from KM models KA (top) and KB (bottom). The vertical lines correspond to the age of the SN explosion $\sim 3$~Mya attested by discoveries of deposits of \fe60,
	$\sim 50$~Mya comparable  with  the  half-life  of \pu244, and the time since the end-Devonian extinction(s) $\sim 360$~Mya. The ratio ranges due to the nuclear variations as described in the text for \fe60, \i129 and \cm247 to \pu244 are shown in shaded bands of black, brown, and blue, respectively.
	}
	\label{fig:spaghetti_kn_bovard}
\end{figure}

The results of our calculations of interesting abundances after 100~kyr, 3~Myr
and 7~Myr (also after 50~Myr for the KN models) for the different astrophysical 
{\em r}-process models described in Table~\ref{tab:models} are summarized in
Fig.~\ref{fig:isotope-ratios-3Myr}, which shows scatterplots of the isotope abundance ratios of
\zr93, \pd107, \i129, \cs135,\hf182, and \cm247 over \pu244 versus \fe60/\pu244.The asymmetric uncertainty bars reflect the ranges of microphysics uncertainties shown in
Table~\ref{tab:ratios-Plio} and discussed above.
Note that these ratios reflect the output purely of the {\em r}-process alone. In the case of a SN there can be additional synthesis of
some of these species, and if there is mixing with multiple 
events this would also change the ratios.

Figure \ref{fig:isotope-ratios-3Myr} shows 
that the ratios of all these isotopes relative to 
\pu244 are highest in the SB SN model.
However, the ordering of the abundance ratios of the other isotopes
in the different models is not universal, with model SA being the lowest
for \u236/\pu244 and \cm247, and KB being the lowest 
for \zr93, \pd107, \i129 and \hf182/\pu244.
These and other differences between the model predictions 
offer prospects for distinguishing between the
different $r$-process models by measurements in deposits up to 50~Myr old.

We emphasize that the predicted ratios in Table \ref{tab:models}
and Fig.~\ref{fig:isotope-ratios-3Myr} are for {\em r}-process
production only.  For the KN models, this should
be indicative of the typical ejected yields
for these explosions.  On the other hand,
for the SN models there will be additional
production of some species due to other processes,
as summarized in Table \ref{tab:radioisotope-inventory}.
For example, \fe60 production in hydrostatic and
explosive burning will far exceed that made in
any SN {\em r}-process, and \zr93, \pd107, \cs135, and \hf182 could be also produced by an $s$-process in an earlier stage of stellar evolution and ejected in the SN event.
Thus for these species
the ratios to \pu244 in the SN models should be
viewed as {\em lower limits}. Even so, they retain their discriminatory power, particularly the \fe60/\pu244 ratio.

The vertical shaded band in Fig.~\ref{fig:isotope-ratios-3Myr} shows the the isotope ratios of \fe60/\pu244 measured at 3 Myr (1.34-4.57 Myr; dark gray) and 7 Myr (4.57-9.0 Myr; gray) time periods from \citet{Wallner2021}.
We see that, as discussed in Section~\ref{subsec:distance}, comparing this measurement to the \fe60/\pu244 ratios from our calculations excludes the KN models as the sole source of both isotopes. This is consistent with \citet{Fry2015}, which 
excluded KNe as the source of the 3 Myr \fe60 pulse.

To lend support to our conclusions, we compare our radioisotope ratios to those from \cite{Goriely2016} and \cite{Cote2021}. Our SA model's \cm247/\pu244 ratios at 1 Myr are consistent with the SN neutrino-driven wind calculations presented in \citet{Goriely2016}. 
We find our SB, KA, and KB model ratios to be largely consistent with the radioisotope abundances from the data sets for the analogous simulations reported in \cite{Cote2021} at 1 Myr, i.e., our combined KN models' radioisotope ratios are well within the uncertainty ranges of those data sets. The one notable difference is in the actinide abundances resulting from the adopted MHD SN models. The MHD SN model from \cite{Cote2021} shows more robust actinide production than our SB model, resulting in somewhat lower ratios to \pu244, while our SB model ratios are more consistent with recent simulation results from \cite{Reichert2021}.
Current MHD SN models exhibit conditions that only marginally reach the actinides, which results in a large astrophysical uncertainty in the \pu244 yield and a distinct contrast to the KN models.

\begin{figure}
    \centering
    \vspace{-5mm}
\includegraphics[width=7cm]{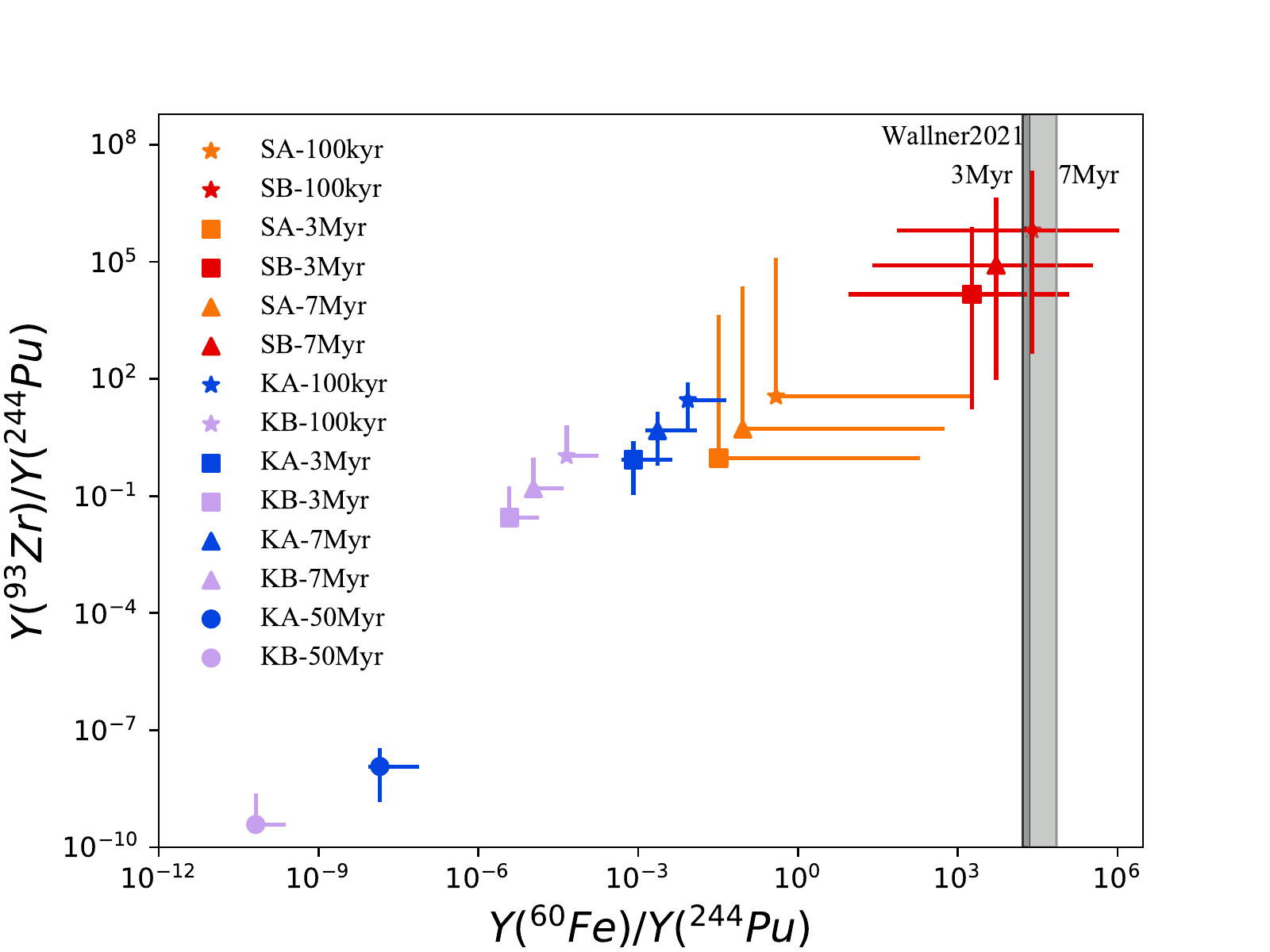}
\includegraphics[width=7cm]{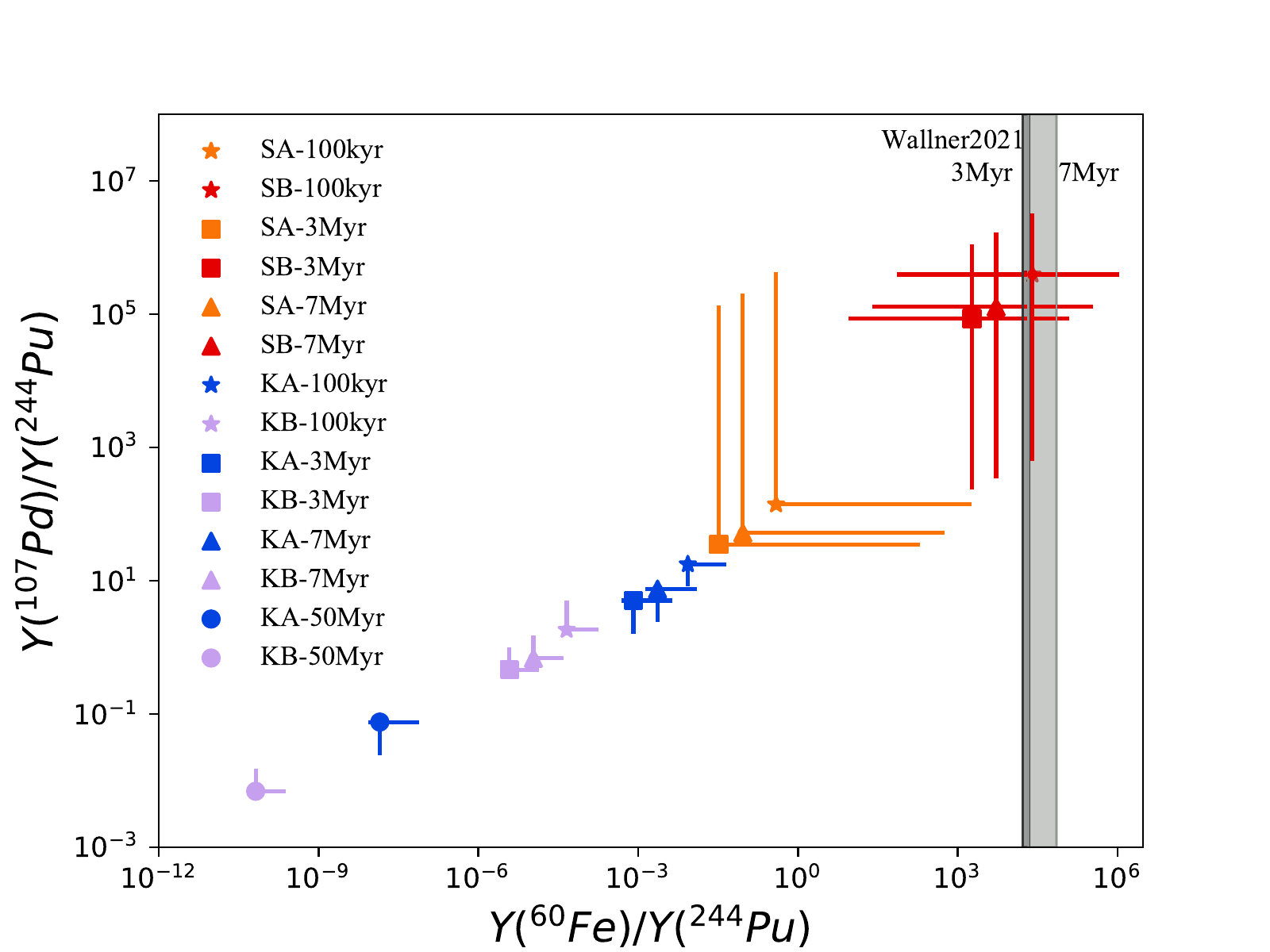}
\includegraphics[width=7cm]{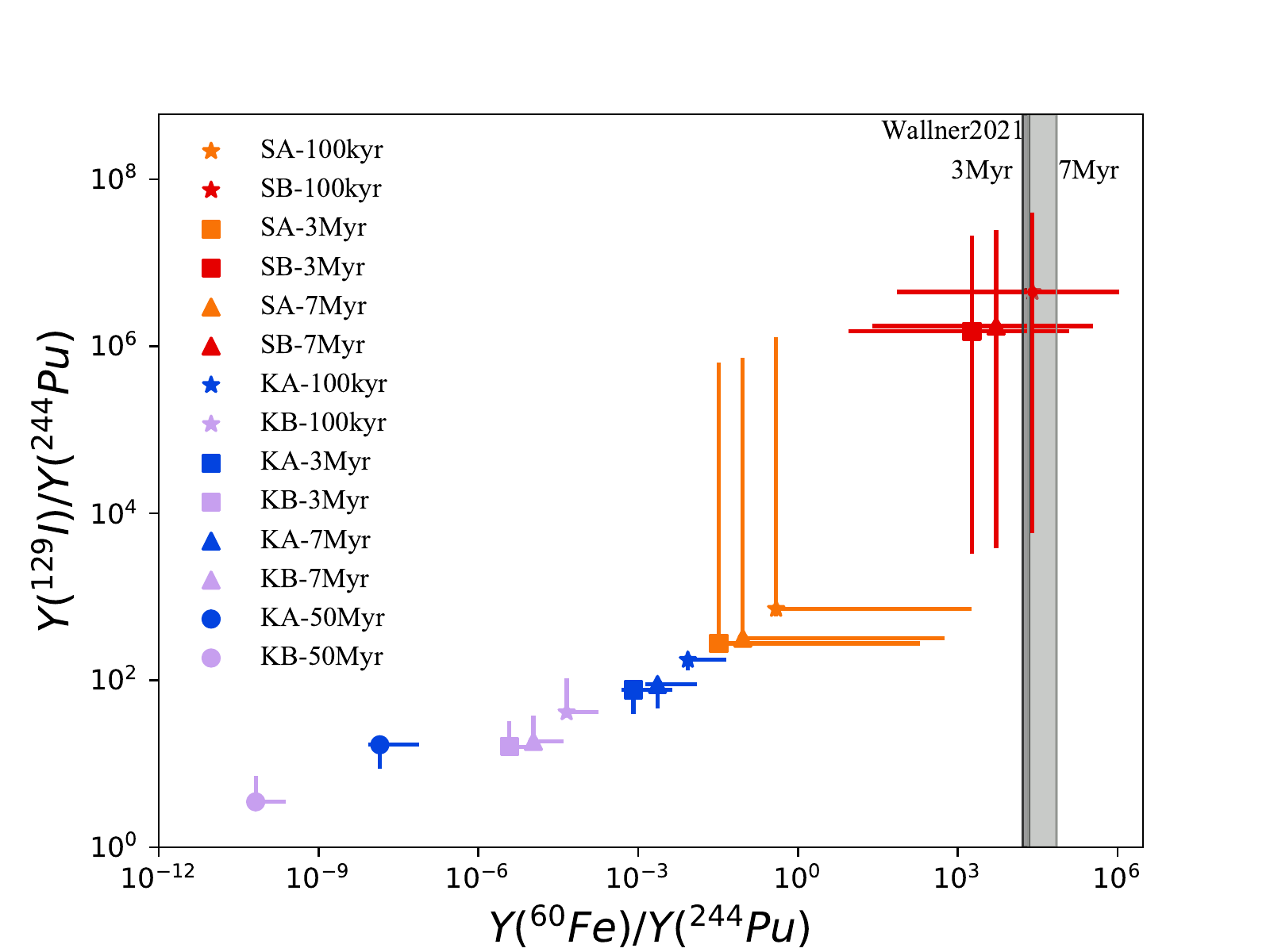}
\includegraphics[width=7cm]{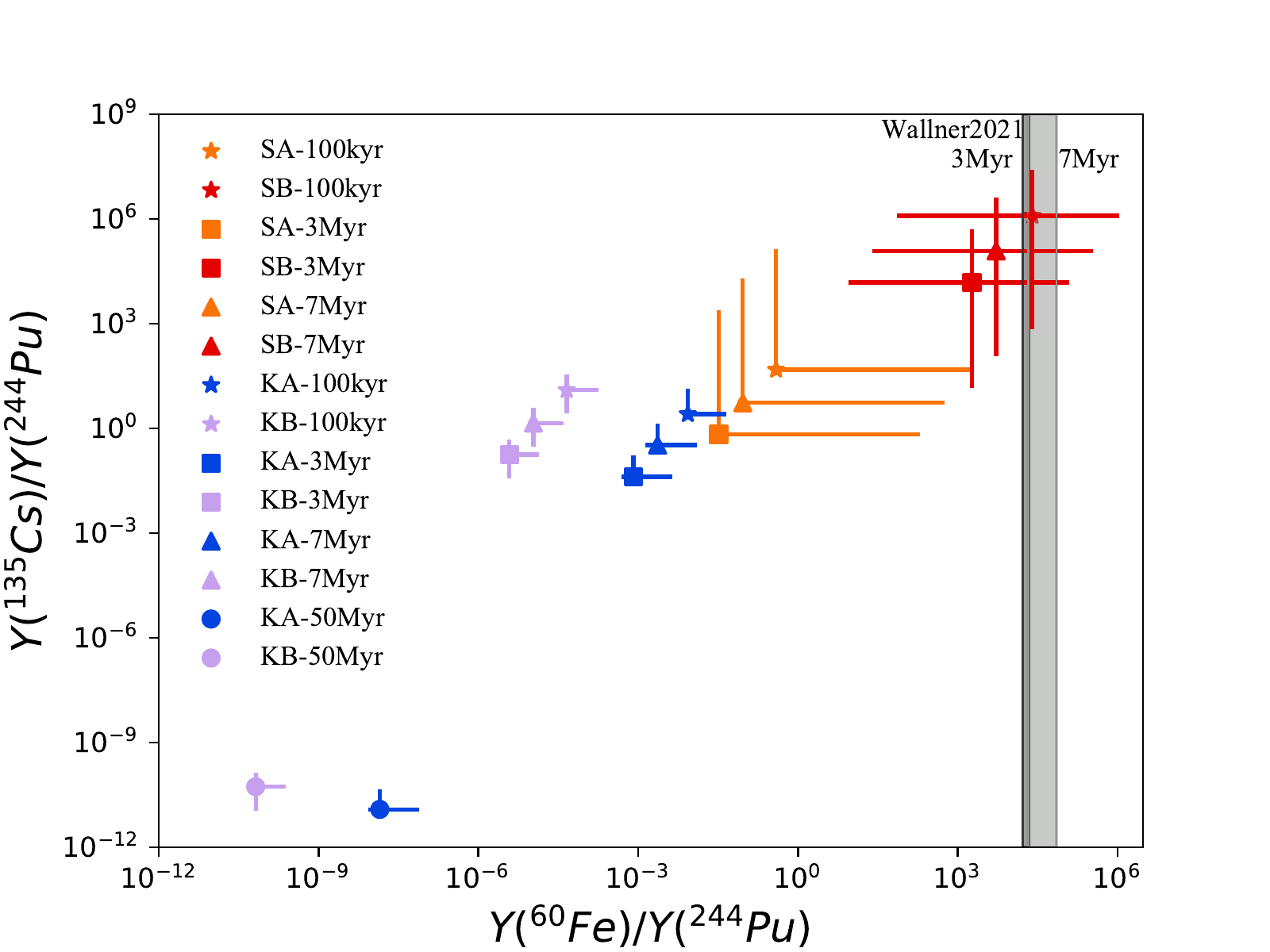}
\includegraphics[width=7cm]{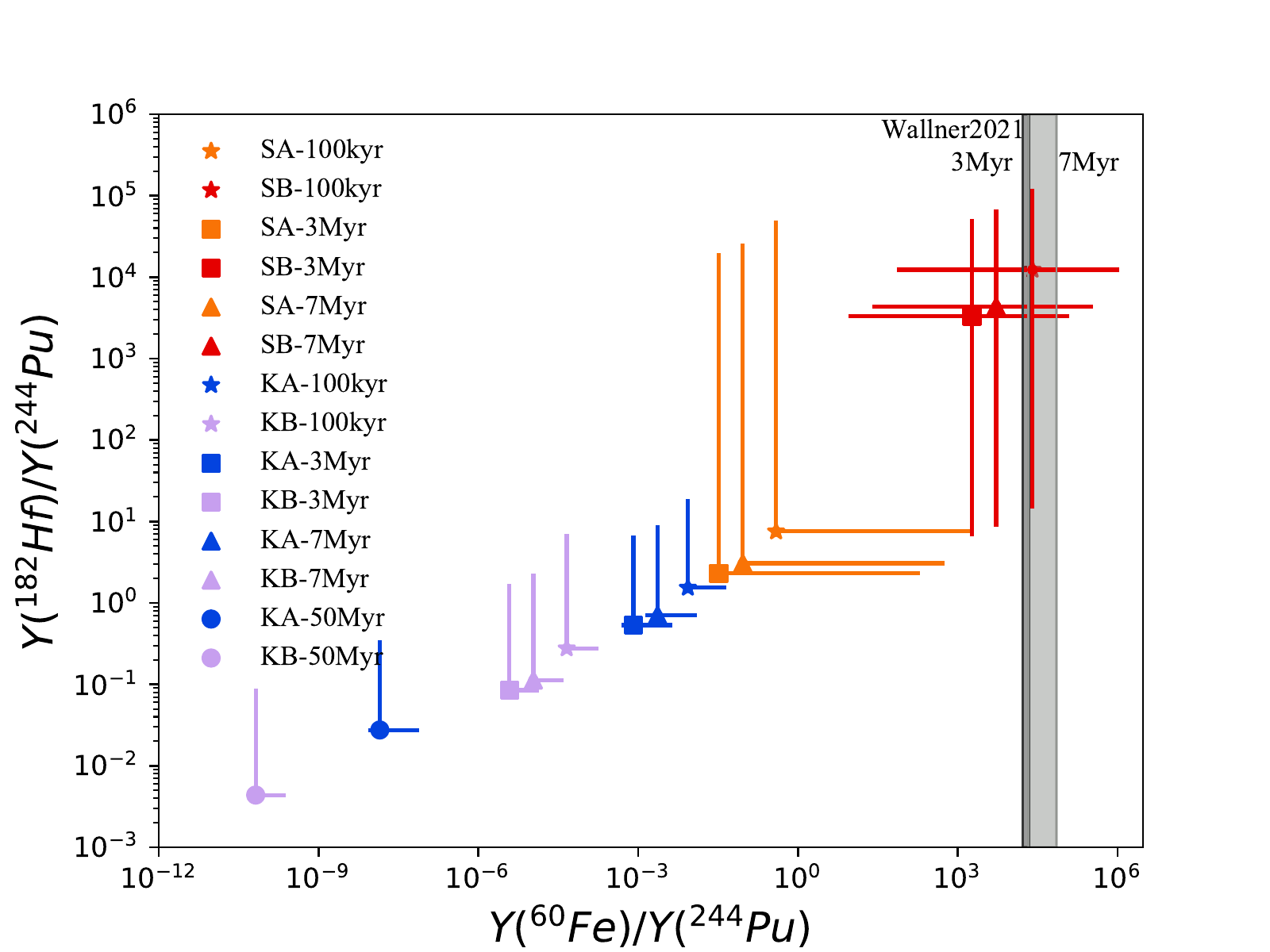}
\includegraphics[width=7cm]{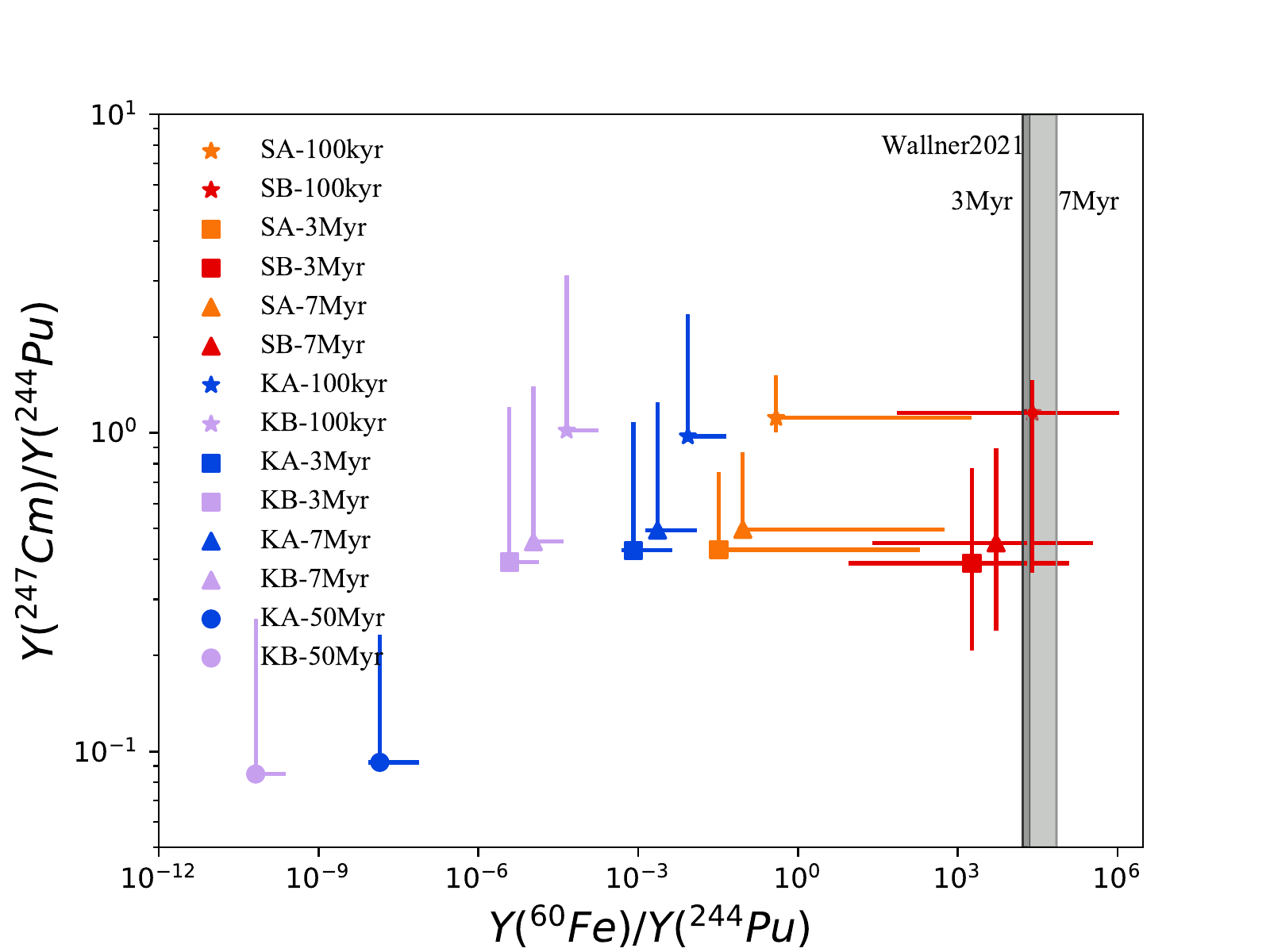}
    \caption{\it Scatterplots of the isotope abundance ratios of \zr93, \pd107, \i129, \cs135,\hf182, and \cm247/\pu244 versus \fe60/\pu244 after times 100~kyr, 3~Myr and 7~Myr (also 50~Myr for the KN models KA and KB), as calculated with 
	the different astrophysical {\em r}-process models described in Table~\ref{tab:models}. The points are the isotope ratios obtained from the baseline $r$-process calculations, and the error bars denote the variations due to the different sets of nuclear models (HFB masses from \citet{HFB17PRL} or $\beta$-decay rates from \citet{MKT} for both SN and KN models, or fission yields from \citet{KT} for KN models, in addition to the baseline nuclear data) adopted in the $r$-process. Note that the isotope ratios presented here are only from the $r$-process nucleosynthesis. For SNe, \fe60 is mainly produced in the pre-SN stage and through explosive nucleosynthesis; thus the isotope ratios of \fe60/\pu244 in the plot are actually the lower limit for SN models SA and SB. The shaded bands are the measured isotope ratios of \fe60/\pu244 at $\sim$3~Myr (1.34-4.57~Myr; dark gray) and $\sim$7~Myr (4.57-9.0~Myr; gray) time periods from \citet{Wallner2021}.
}
\label{fig:isotope-ratios-3Myr}
\end{figure}

In order to make predictions for measurements of radioisotopes like \pu244 we need, in addition to 
their abundances $Y_i(t)$ obtained from the network calculation, estimates of the total $r$-process yields or ejected masses $M_{\rm ej}$ from our SN and KN models, from which we can obtain the absolute yields for the radioisotopes $i$ at time $t$, namely $M_{i}(t)=M_{\rm ej}\times A_i Y_i(t)$.
The nucleosynthetic outcome of our SA model is similar to that of the 2.2 $\msol$ neutrino-driven wind model from \citet{Wanajo2013}, where a total yield of $1.37\times 10^{-5} \msol$ is reported; estimates of the mass ejected in SN neutrino-driven winds vary from $1\times 10^{-5}\msol$ to $5\times 10^{-4} \msol$ \citep[see, e.g.,][]{Wanajo2001, GCE}. The yield from the MHD SN simulation we adopt for our SB model is $\sim0.03 \msol$ \citep{Mosta2018}, while a wide range of MHD SN yields are found in the literature, from $6.72\times 10^{-3}\msol$ in \citet{Winteler2012} to $\sim0.389 \msol$ in \citet{Reichert2021}. For the KN models, the yield from our chosen \citet{Bovard} dynamical ejecta simulation is about $3.53\times 10^{-3} \msol$, which is roughly consistent with the yield in, e.g., \citet{Radice2018} of $\sim 2\times 10^{-3} \msol$. The disk wind yields from \citet{Just} range from $\sim 7\times 10^{-3}$ to $6.68\times 10^{-2} \msol$, similar to the range found in, e.g., \citet{Rodrigo2015}, of $\sim 0.03$ to $0.22 \msol$. For our KB model, the disk and dynamical ejecta masses are similar, giving a total yield of $7\times 10^{-3} \msol$, while for our KA model the disk wind mass is roughly four times that of the dynamical ejecta for a total yield of $1.7\times 10^{-2} \msol$. The theoretical estimates quoted above give ranges for $r$-process yields of $\sim 5\times 10^{-3}\msol - 0.1 \msol$, consistent with the values suggested by the observations of GW170817 \citep{CoteGW170817}.

Following this survey of {\em r}-process radioisotope production 
in SNe and KNe, we now turn to the
prospects for searches in deep-ocean deposits and in the lunar regolith.

\section{Models for Deep-ocean $r$-Process Radioisotopes from Stellar Explosions}
\label{sect:models}

We now have the tools in place to interpret the deep-ocean \pu244 (and \fe60) in light of our SN and KN models.

\subsection{\texorpdfstring{\pu244}{Pu-244} Radioactivity Distance Constraints on an Event 3 Myr Ago}
\label{subsec:distance}

We first consider the possibility that the \pu244 flux coincident with the \fe60 pulse $\sim 3$~Mya is due
to the same event.  Thus, any \pu244 flux outside of this timespan must come from another process,
such as that explored in the following section.  We also assume that the radioisotope delivery is a {\em one-step}
process, i.e., the terrestrial and lunar deposition of these species is a direct consequence of the propagation of
the explosion ejecta.  

Consider an explosion at distance $r$ and time
$t$ in the past.  If the explosion is isotropic, the
time-integrated flux, i.e., the fluence, of radioisotope
$i$ at the Sun's interstellar location is
\begin{equation}
\label{eq:IntFlu}
    {\cal F}_i^{\rm interstellar} = 
    f_i \ \frac{M_{{\rm ej},i}/A_i m_u}{4\pi r^2} 
    \ e^{-t/\tau_i} \, ,
\end{equation}
where the radioactive decay factor includes all decay 
in the interval $t$ between the explosion and the present time $t$,
i.e., including both travel time and the duration since
arrival.
Here the isotope's mass number is $A_i$, its
mean life is $\tau_i$, and $m_u \approx m_p$
is the atomic mass unit.
Also $M_{{\rm ej},i}$ is the yield {at the time of the explosion, i.e., the
total mass of isotope $i$ ejected, 
the total mass of isotopes ejected at explosion time is $M_{{\rm ej}}=\sum_i{M_{{\rm ej,i}}}$}, and
$f_i \le 1$ is
the fraction of atoms of $i$ that are incorporated
into dust particles that arrive at Earth
\citep{Benitez2002,Athanassiadou2011,Fry2016}.

After fallout onto the Earth and accumulation into
natural archives such as deep-ocean sediments and crusts,
the present-day surface density is
\citep{Ellis1996,Fry2015}
\begin{equation}
\label{eq:surfdense}
    N_i = \frac{U_i {\cal F}_i}{4} = U_i f_i 
    \ \frac{M_{{\rm ej},i}/A_i m_u}{16\pi r^2} \,
    \  e^{-t/\tau_i} \, ,
\end{equation}
where the factor of 4 accounts for the ratio of the
Earth's cross section to its surface area and we have included a decay factor.
Here the uptake factor $U_i$ measures the fraction of
incident atoms of $i$ that are incorporated 
into the sample. \citet{Fry2016} note that the fallout
will not be uniform over the Earth, favoring midlatitudes
at the expense of the poles and equator.  Thus the
effective uptake can be different at different sites
for this reason alone, in addition to variations in 
geological conditions.

We can infer the \pu244 yield for a given
explosion distance within this picture.
We focus here on the time interval $\sim 3$~Mya
that contains the better-measured \fe60 pulse, whose duration is at least equal to that of the 
nonzero \fe60 signal seen in sediments, $\Delta t = 1.6 \ \rm Myr$.
Integrating the \pu244 flux from eq.~(\ref{eq:244pu-5Myr}) over this time, we find an interstellar \pu244
fluence ${\cal F}_{244}^{\rm interstellar} = 2670 \pm 560 \ \rm atoms \ cm^{-2}$.
Using this and the \pu244 mass yields $M_{\rm ej,244}$ calculated within the
SN and KN models, we can invert eq.~(\ref{eq:IntFlu})
to infer the explosion distance:
\begin{equation}
\label{eq:rad-dist}
    D_{\rm rad}(\pu244) = \sqrt{ f_{\rm dust} \frac{M_{\rm ej,244}/244 m_{\rm u}}{4\pi {\cal F}_{244}^{\rm interstellar}} } \ \  e^{t_{\rm exp}/2\tau_{\rm 244}} \, .
\end{equation}
This ``radioactivity distance'' is the analog of a standard luminosity
distance, with the yield playing the role of luminosity,
and the fluence playing the role of flux \citep{Ellis1996, Fry2015}.

Figure \ref{fig:rad-dist-3Myr} shows the radioactivity
distance for our four {\em r}-process models,
assuming that most of the \pu244 is incorporated into dust, $f_{\rm dust}^{\rm Pu} \simeq 1$.
We see that the central value of the distance estimate in the MHD model SB
is $\sim 100$~pc, with an uncertainty of about an order of magnitude
in either direction.
The {\em forced} neutrino-driven wind model SA has somewhat larger yields and thus requires larger distances,
but within uncertainties the distance range overlaps with the MHD model.

These \pu244-based distances are quite consistent with the explosion
distance inferred~\citep{Fields2005, Fry2015, Breit2016} for an SN that could have generated the well-established \fe60 pulse $\sim 3$~Mya.  This range, $D_{\rm rad}(\fe60) = 20 - 150 \ \rm  pc$,
is represented by the 
yellow band in Fig.~\ref{fig:rad-dist-3Myr}. Within the
uncertainties, this distance estimate is also compatible with the distances to the Tucana-Horologium and
Scorpius-Centaurus stellar associations, which have been proposed as possible locations of the SN that
generated the \fe60 pulse~\citep{Benitez2002,Breit2012,mamajek2015}. 
Note that these results assume $f_{\rm dust}^{\rm Pu} \simeq 1$. Smaller distances would follow if 
a smaller value for dust efficiency is adopted; within the large yield uncertainties, 
these SN models remain consistent with \fe60 for dust efficiency values down to  $f_{\rm dust} \sim 0.01$.

Thus we see that the \pu244 signal overlapping with the 3 Myr
\fe60 pulse can be explained by an SN explosion, but only one that is {\rm r}-process-enhanced. 
As seen in Fig.~\ref{fig:rad-dist-3Myr}, such an event can give consistent distance estimates or equivalently a consistent \fe60/\pu244 ratio.
We also see that these distances lie between the ``kill'' radius $r_{\rm kill} \sim 10 \ \rm pc$ and the distance at which the remnant would fade away,
$r_{\rm fade} \sim 160 \ \rm pc$,
which provide lower and upper bounds on the distance to a nearby SN event
whose radioisotope signal would be detectable.
We note that the earlier \fe60 pulse at $\sim 7 \ \rm Myr$
has similar but somewhat smaller fluence, and so would 
give a similar but somewhat smaller radioactive distance
range.  This too could also in principle be explained by
an SN, but must again have been an {\rm r}-process-enhanced event.  
However, such events are at best atypical of SN explosions and, as we have noted,
in most modern SN models there is {\em no actinide production at all}.
It is, therefore, particularly unlikely to have two such events 
in close succession.  We will return to this point in the next section.

\begin{figure}
    \centering
    \includegraphics[width=0.6\textwidth]{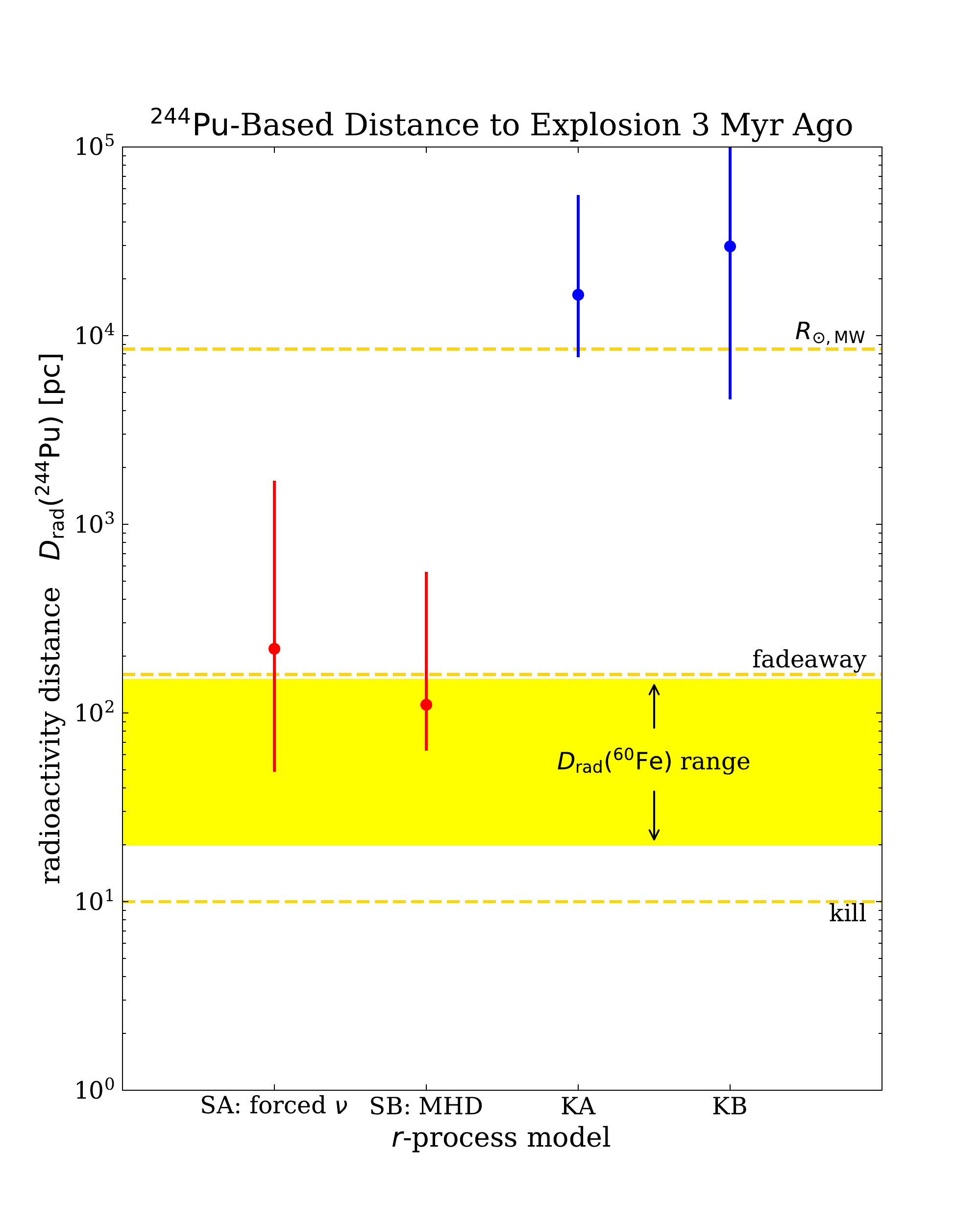}
    \caption{\it The estimated radioactivity distances to possible events that are candidates
    for having produced \pu244 3 Mya.  
    The distance $D_{\rm rad} \sim ( f_{\rm dust} M_{\rm ej,244}/{\cal F}_{244}^{\rm interstellar} )^{1/2}$ 
    depends on the dust fraction, the model yield of \pu244, and the interstellar flux. 
    We see that SN models lead to distances consistent with
    that inferred from \fe60 data, shown in the yellow band, and lie between the ``kill'' and ``fadeaway" distances.  On the other hand, 
    KN models lead to distances far too large to allow for \pu244 transport,
    namely distances similar to or larger than the Sun's distance $R_{\odot,\rm MW}$ from the Galactic center.
    Here we assume $f_{\rm dust}=1$; smaller values lead to smaller inferred distances.}
    \label{fig:rad-dist-3Myr}
\end{figure}

We now turn to the possibility of a KN as the source of
\pu244 3 Mya.  
The yields
for our KA and KB models are $\sim 6$ orders of magnitude larger than those in the SN models.  
As a result, Fig.~\ref{fig:rad-dist-3Myr} shows
that they correspond to much larger radioactivity distances $\gg 1 \ \rm kpc$.  
This would place the explosions implausibly far for any ejecta 
to reach the Earth.  Indeed, the upper range of the distances
extends beyond the size of our Galactic disk.  
Moreover, as seen in Fig.~\ref{fig:isotope-ratios-3Myr}, 
the \fe60/\pu244 ratio for the KN models sharply disagrees with the data.  
This suggests a different scenario is needed for the
KN case, to which we now turn.

\subsection{\pu244 Constraints on a KN Enrichment of the Local Bubble}

As discussed above, the seemingly straightforward association
of the \pu244 signals with the \fe60 pulses would require two consecutive rare, 
{\em r}-process-enhanced SN events. This coincidence seems unlikely, particularly because we know of no reason that {\em r}-process-enhanced SNe would occur in pairs or be clustered.
Lacking such a reason, the two events are independently rare and thus
their occurrence in close succession is exceedingly improbable.
 Moreover, we have seen that the \pu244 data favor a persistent flux that may extend back to as early as 25 Mya, not necessarily an impulsive signal as seen in \fe60.  This leads us to consider an {\em r}-process event that occurred earlier than the \fe60 SNe,
 with a KN explosion being an obvious candidate. We have also seen above that a KN scenario requires that
the \pu244 was not injected directly and impulsively from a single event, but instead suggests
that some form of dilution occurred between the explosion and injection in the solar system.

The requirements of early {\em r}-process creation and subsequent
dilution are both met naturally if a KN exploded prior to or during the early formation of the Local Bubble, enriching the entire star-forming cloud that gave rise to the bubble.  Indeed, \citet{Wallner2015} have proposed such a scenario, 
and here we build on their analysis.

We thus envision a {\em two-step} process in which (1) the KN ejecta propagate to and mix into the proto-Local Bubble, followed by (2) the relative motion of the Earth and {\em r}-process-enriched dust leading to a
flux of $r$-process radioisotopes onto Earth.
Specifically, we envision the following sequence of events.
(1a) More than $\sim 25$~Mya, a KN exploded,
ejecting \pu244-bearing {\em r}-process material.
(1b) Some of the KN ejecta collided with and was mixed into the
molecular cloud giving rise to the Local Bubble.  (1c)
Some of the \pu244 was incorporated into dust grains.
(2) The \pu244-bearing dust subsequently bombarded the Earth.

{\em Step 1: \pu244 injection into the proto-Local Bubble.}
Let a KN explode at distance $r_{\rm KN}$ from the forming Local Bubble, ejecting a yield $M_{\rm ej}(\pu244)$.
If the molecular cloud progenitor of the bubble has radius
$R_{\rm MC}$, then the mass intercepted by the bubble
and becoming dust is
\begin{equation}
\label{eq:M244-LB}
    M_{244}^{\rm LB,dust} = \frac{\pi R_{\rm MC}^2}{4\pi r_{\rm KN}^2} f_{244} M_{\rm ej}(\pu244) 
    = \pi R_{\rm MC}^2 \ m_{244}
    \ {\cal F}_{244}(r_{\rm KN}) \, ,
\end{equation}
with $f_{244}<1$ accounting for the fraction of incident \pu244
stopped by the bubble and ultimately incorporated into grains.
We see that this is just the 
expression in eq.~(\ref{eq:surfdense}),
for the one-step interstellar KN 
fluence ${\cal F}_{244,KN}$ at $r_{\rm KN}$,
multiplied by the cloud cross section and the atomic \pu244 mass $m_{244}$.
Below we use the data, and KN
rate information, to
estimate the distance $r_{\rm KN}$ and 
evaluate its reasonableness.

{\em Step 2:  \pu244 transport to Earth.}
Given the \pu244 mass in eq.~(\ref{eq:M244-LB}), we now
estimate the \pu244 flux onto Earth.
Within the bubble, the
interstellar \pu244 number flux is 
$\Phi_{244} = n_{244} v_{\rm rel}$,
with $n_{244}$ the number density and $v_{\rm rel}$ the relative velocity with respect to Earth.  We expect both factors to vary with time, as \pu244-bearing
dust moves in the turbulent medium constantly
stirred by SNe.
We do not attempt to capture these variations but only estimate average
values.  We then demand that
these match the observed mean flux
$\Phi_{244}^{\rm interstellar}$ in eq.~(\ref{eq:244pu-9Myr}),
and use the results to constrain
this model.

We describe the Local Bubble crudely, as a sphere of radius
$R_{\rm LB}$, so that the \pu244 
average number density is 
$n_{244} = M_{244}^{\rm LB,dust}/(4\pi A_{244} m_u R_{\rm LB}^3/3)$.
Let $v_{\rm rel}$ be the speed of the Local Bubble gas relative to the Sun,~\footnote{Note that this relative speed  encodes not only the Sun's motion with respect to the cloud's center of mass, but also turbulent motions within the cloud,
which have dispersion $\sigma_v \sim 1-10 \ \rm km/s$ for clouds of size $R_{\rm MC} \ga 10 \ \rm pc$.}, so that the interstellar \pu244 number flux is 
$\Phi_{244} = n_{244} v_{\rm rel} \propto M_{244}^{\rm LB,dust}/R_{\rm LB}^3$.  
Since this flux is now measured, we are in a position to evaluate
the \pu244 mass needed to be injected
in the Local Bubble for our model,
given the measured
flux $\Phi_{244}$:
\begin{eqnarray}
\label{eq:M244-LB-infer}
M_{244}^{\rm LB,dust} & = & 
  \frac{4\pi}{3} A_{244} m_{\rm u} R_{\rm LB}^3
  \frac{\Phi_{244}^{\rm obs}}{v_{\rm rel}} \nonumber \\
  & = & 6 \times 10^{-11} M_{\odot} \
  \pfrac{\Phi_{244}^{\rm obs}}{1670 \ \rm cm^{-2} \ Myr^{-1}} \
  \pfrac{R_{\rm LB}}{50 \ \rm pc}^3 \
  \pfrac{30 \ \rm km/s}{v_{\rm rel}} \, .
  \end{eqnarray}
We use for the fiducial velocity in eq.~(\ref{eq:M244-LB-infer})
a value comparable to the Sun's 
present motion with respect to the local standard of rest.  This is conservative in that SN blasts and turbulent motions in the Local Bubble will in general add 
to the relative velocity.

We see from eq.~(\ref{eq:M244-LB-infer}) that the \pu244 inventory is far below the levels of KN yields.
This implies that the explosion was not contained in the Local
Bubble, but occurred outside it with only a fraction of the yield
intercepted by the locally.
This dilution is the main motivation for
the two-step model, allowing it to
avoid the unphysically large 
KN distances seen in the one-step picture
(Fig.~\ref{fig:rad-dist-3Myr}).

We note that it is likely that KN ejecta are unable to form 
significant amounts of dust,
due to their high velocities \citep{Takami2014,Gall2017}.
This is expected as a continuation of
the trend in which Type Ia SNe have smaller
or no dust formation compared to core-collapse
explosions, which have slower ejecta speeds \citep{Nozawa2011, Gomez2012}.
In light of the need for dust grains to deliver ejecta to Earth,
the lack of KN dust production argues against direct, one-step deposition of ejecta from these explosions. 
However, in the two-step model the KN ejecta are
stopped and mixed into the proto-Local Bubble material
and can then be incorporated into dust grains.

We can go further and estimate the distance from the KN
to the Local Bubble.  This is a variant of the radioactivity distance
calculation, similar to the \citet{Looney2006} calculation of
the SN injection of radioisotopes into the pre-solar nebula \citep[see, e.g.,][]{Ouellette2009}. We can then solve for the KN distance, finding
\begin{eqnarray}
r_{\rm KN} & = & \sqrt{ \frac{f_{244} M_{\rm ej}(\pu244)}{4 M_{244}^{\rm LB,dust}} }  \  R_{\rm MC} \\
& = & 2000 \ {\rm pc} \ 
f_{244}^{1/2} \ 
\pfrac{M_{\rm ej}(\pu244)}{10^{-4} M_\odot}^{1/2} \ 
\pfrac{10^{-10} M_\odot}{M_{244}^{\rm LB,dust}}^{1/2} \
\pfrac{R_{\rm MC}}{3 \ \rm pc} \, .
\end{eqnarray}
For our fiducial quantities and $f_{244} < 1$,
we have $r_{\rm KN} \la 2$ kpc.  
This places the KN explosion at a distance that is more consistent with the expectation that KNe
are rarer and further between than SNe.~\footnote{An incorporation efficiency as low as
$f_{244} \sim 10^{-3}$ can be accommodated
and still give a distance comparable to that estimated above for an SN remnant.}
We note that a KN is likely to be so distant that its ejecta blast would be weak when arriving
in the solar neighborhood, and hence unlikely to destroy the local molecular cloud.

Estimates of KN and neutron star merger rates place consistency checks on our distance calculations
\citep{Hartmann2002, Scalo2002}.  The average rate for a KN within distance $r_{\rm KN}$ is 
$\Gamma(r_{\rm KN}) \approx \pi r_{\rm KN}^2 \ Q_{\rm local}$
where $Q_{\rm local} = dN_{\rm KN}/dA \, dt$ is the KN rate per unit area of the Galactic disk at the solar location;
we assume $r_{\rm KN} > h$, the scale height of KN progenitors, so that the
problem is reduced to two dimensions.  Using the usual approximation 
of an exponential disk with scale radius $R_0 = 2.9$~kpc
\citep[see, e.g.,][]{Girardi2005, Murphey2020},
we have $Q(R) = {\cal R}_{\rm KN} e^{-R/R_0}/2 \pi R_0^2$,
where the total Galactic KN rate sets the normalization via
the surface integral 
${\cal R}_{\rm KN} = 2\pi \int Q(R) \ R \ dR$.
We estimate the Galactic KN rate assuming
the ratio to the 
Galactic core-collapse SN rate ${\cal R}_{\rm CC}$
is the same as the ratio of the local cosmic rate densities:
${\cal R}_{\rm KN}/{\cal R}_{\rm CC} = \dot{\rho}_{\rm KN}/\dot{\rho}_{\rm CC}$.
Using ${\cal R}_{\rm CC} \approx 0.032 \ \rm yr^{-1}$
\citep{Adams2013},
$\dot{\rho}_{\rm CC}= 1.1 \times 10^{-4} \ {\rm Mpc}^{-3} \, \rm yr^{-1}$ \citep{Lien2009}, 
and a binary neutron star merger rate as a measure of
the underlying {\em r}-process source rate
$\dot{\rho}_{\rm BNS} \gtrsim \dot{\rho}_{\rm KN} \sim 1500 \ {\rm Mpc}^{-3} \, {\rm yr}^{-1}$ \citep[consistent with estimates from][]{Matteucci2014, Wehmeyer2015, Chruslinska2018, CoteGW170817, DellaValle2018, Jin2018,  Andreoni2020, Andreoni2021}, 
we find $R_{\rm KN} \sim 1.4 \times 10^{-2} {\cal R}_{\rm SN} \sim 440 \ \rm Myr^{-1}$. Finally
\begin{equation}
    \Gamma_{\rm KN} \sim 1.3 \ {\rm events \ Myr^{-1}} \ \pfrac{r_{\rm KN}}{1 \ \rm kpc}^2
\end{equation}
at the solar distance from the center of the galaxy, i.e., $r_{\rm KN}=R_\odot = 8.7$~kpc.
We see that the typical KN recurrence time is
$\Gamma_{\rm KN}^{-1} \sim 0.8 \ \rm Myr \ (1 \, \rm kpc/r_{\rm KN})^2$.  Thus KN explosion
times of $t_{\rm KN} = (10, 50) \ \rm Myr$
correspond to mean distances of $r_{\rm KN} = (280, 120)$~pc. Our two-step \pu244-based radioactivity distances lie comfortably within this range, 
demonstrating overall consistency.

Two other issues constrain the timing of the KN explosion and
subsequent {\em r}-process rain.  One is the
timescale for the Local Bubble to assemble and form stars;
this timescale plus massive-star lifetimes
sets an upper limit to the KN injection time.
The \citet{Fuchs2006}
fit to massive-star lifetimes gives
23 (18) Myr for masses $\ge 8 (10) M_\odot$ at the threshold of core collapse,
and the lifespans are nearly linearly dependent on the inverse of the mass.} 
Estimates of the lifetimes of molecular clouds span a significant range from local solar neighborhood values of a few Myr \citep{Hartmann2001}
to $\sim 30 \ \rm Myr$ \citep{Murray2011}.
For the Local Bubble itself, based on lifetimes 
of extant stars and pulsars, as well as on expansion dynamics,
\citet{Breitschwerdt2009,Breit2016} estimate an age $\sim 15 \ \rm Myr$ and $\sim 17$ SNe,
which is consistent with 
the \citet{MA2001}  argument
for $\sim 20$ SNe during the last 10-12 Myr. On the other hand,
\citet{Abt2011} argues that some portions of the bubble could
be as old as $\sim 50 \ \rm Myr$, whereas  \cite{Smith2001} consider models with ages $\la 10  \ \rm Myr$ and three SNe.
We therefore consider a range of KN injection timescales of $(10, 20, 50)$ Myr, as noted in eq.~(\ref{eq:t-LB}).

Finally, the {\em r}-process bombardment on Earth can begin only when the Sun enters the Local Bubble.  This time is uncertain and depends on the evolving bubble morphology, but also sets an upper limit on the injection time.

We observe that this two-step process incurs larger uncertainties overall than the one-step SN mechanism.
One of them is the timing of the KN: the further in the past, the greater the losses of shorter-lived species.
If the \pu244 event in the 12.5--25 Mya crust layer is real, it sets a lower limit on this time.
The longest-lived $r$-process radioisotopes of interest, apart from \pu244, are \i129, \hf182, and  \cm247.
These live long enough for their production in a KN to be potentially observable in this layer; also the secular equilibrium abundance of \u236 potentially provides a check
on anthropogenic sources of actinides.

\subsection{Predicting {\em r}-Process Radioisotope Signatures in Terrestrial Archives}

The previous two subsections show that it is possible to construct  
both SN and KN scenarios for some or all of the \pu244 signal.
We conclude that the \pu244 flux in eqs.~(\ref{eq:244pu-5Myr}) and (\ref{eq:244pu-9Myr})  could have an astrophysical origin, and
use these to normalize our subsequent predictions for possible live isotope searches using AMS techniques.
In this section we compute terrestrial signals for additional {\em r}-process radioisotopes, with a particular focus on Fe-Mn crusts. 
Our strategy is to pursue the consequences of the \pu244 detection, using
it to infer the abundances of other {\em r}-process radioisotopes predicted by
our SN and KN models.

The available natural terrestrial archives that accumulate gradually over the longest periods of time, and thereby
give the most complete dating information, are deep-ocean crusts and sediments.
In the natural archives of interest, radioisotope abundances
are usually presented as an isotope fraction $(N_i/N_j)_{\rm obs}$,
the ratio of a radioisotope $i$ to a stable isotope or element
$j$.  
Here we derive predictions for the ratio
$N_i^{\rm astro}/N_j^{\rm bg}$, i.e.,
the astrophysical signal relative to a stable ``background" isotope.

The astrophysical signal can be expressed as the
incident interstellar number flux $\Phi_i^{\rm interstellar} = dN_i/dA dt$,
usually given without decay losses included.
The measurable flux onto a terrestrial sample is
\begin{equation}
    \Phi_i^{\rm sample} = \frac{f_i U_i}{4}\Phi_i^{\rm interstellar} e^{-t/\tau_i} \, ,
\end{equation}
where the factor of 4 accounts for the ratio of the Earth's cross section to 
its surface area, $f_i$ is the fraction of $i$ in dust arriving at Earth,
and $U_i$ accounts for uptake into the sample.
We have also accounted for radioactive decay.
If the flux is measured for a time interval $\Delta t$, the 
corresponding surface density in the sample is $N_i^{\rm astro} = \Phi_i^{\rm sample} \, \Delta t$.

During this time interval, the sample accumulates a surface mass density
$\rho \ \dot{h} \ \Delta t$, where $\rho$ is the total density, and
$\dot{h} = dh/dt$ is the rate of growth of the thickness.
Let the background species $j$ have mass fraction $X_j = \rho_j/\rho$
in the sample,
and mass number $A_j$.  Then the background atoms $j$
have surface density $N_j = X_j \rho \, \dot{h} \ \Delta t/A_j m_u$,
with $m_u = 1/N_{\rm Avo}$ the atomic mass unit.  
Thus 
\begin{equation}
  \frac{N_i^{\rm astro}}{N_j^{\rm bg}}
  = U_i f_i \frac{\Phi_i^{\rm interstellar}}{4 X_j \rho \, \dot{h}/A_j m_u}
  e^{-t/\tau_i}
\end{equation}
is the desired isotopic ratio.

We anchor our predictions to the detections of \pu244 discussed above.
In this case we have
\begin{equation}
 \frac{N_i^{\rm astro}}{N_j^{\rm bg}}  
  = U_i \ \frac{f_i}{f_{\rm Pu}} \ \pfrac{{\cal N}_i}{{\cal N}_{244}}_{\rm astro}  
    \ \frac{\Phi_{244,\rm astro}^{\rm interstellar}}{4 X_j \rho \, \dot{h}/A_j m_u}  e^{-t/\tau_i} \, ,
\end{equation}
where we use the fact that the interstellar flux of a species is proportional
to its (number) yield,
$\Phi_i^{\rm interstellar} \propto {\cal N}_i$.
Using values typical for crusts, we have
\begin{eqnarray}
\label{eq:obsratio}
     \frac{N_i^{\rm astro}}{N_j^{\rm bg}}  
 & = & 2 \times 10^{-13}  \ U_i \ \frac{f_i}{f_{\rm Pu}} \ \pfrac{{\cal N}_i}{{\cal N}_{244}}_{\rm astro}  e^{-t/\tau_i} \\
 & &  \times \pfrac{\Phi_{244}^{\rm interstellar}}{1670 \ \rm cm^{-2} \, Myr^{-1}} 
 \ \pfrac{1 \ \rm ppm}{X_j}
 \ \pfrac{A_j}{150}
 \ \pfrac{2 \, \rm g/cm^3}{\rho}
 \ \pfrac{3 \, \rm mm/Myr}{\dot{h}} \, .
\end{eqnarray}
Note that the signal is inversely proportional to both the growth rate and the background abundance in the sample:  $N_i^{\rm astro}/N_j^{\rm bg} \propto (X_j \dot{h})^{-1}$.  This favors samples with low growth rates and hence Fe-Mn crusts.  It also favors elements that are rare in crusts (so long as the uptake is not too small).

\begin{table}[!htb]
    \caption{Predicted Interstellar Fluences for {\em r}-Process Radioisotopes based on \pu244}
    \centering
    \hspace{-3cm}
    \begin{tabular}{c|cccc|cc|cc|cc}
    \hline \hline
        & \multicolumn{4}{c|}{$\fluence({\rm 3 \, Myr)}$} & \multicolumn{2}{c|}{$\fluence({\rm  10 \, \rm Myr})$} & \multicolumn{2}{c|}{$\fluence({\rm  20 \, \rm Myr})$} & \multicolumn{2}{c}{$\fluence({\rm  50 \, \rm Myr})$} \\
        Isotope &  SA & SB & KA & KB & KA & KB & KA & KB & KA & KB\\
        \hline
        \zr93 & 5.1(3) & 8.0(7) & 4.7(3) & 1.5(2) & 2.3(3) & 7.5(1) & 7.1(1) & 2.3(0) & 5.8(-4) & 1.9(-5) \\
        \pd107 & 5.1(4) &  1.3(8) & 7.4(3) & 6.7(2) & 3.7(4) & 3.4(3) & 2.8(4) & 2.5(3) & 3.7(3) & 3.4(2) \\
        \i129 & 3.1(5) & 1.7(9) & 8.7(4) & 1.8(4) & 6.8(5) & 1.4(5) & 9.6(5) & 2.0(5) & 8.3(5)  & 1.7(5) \\
        \cs135 & 5.3(3) & 1.2(8) & 3.2(2)& 1.4(3) & 8.5(1) & 3.7(2) & 1.1(0) & 4.7(0) & 6.0(-7) & 2.7(-6) \\
        \hf182 & 3.0(3) & 4.3(6) & 7.0(2) & 1.1(2) & 4.3(3) & 6.7(2) & 4.3(3) & 6.8(2) & 1.4(3) & 2.1(2) \\
        \u236 & 1.8(3) & 9.3(3) & 2.0(3) & 1.1(3) & 1.7(4) & 1.0(4) & 3.0(4) & 1.8(4) & 4.9(4) & 3.4(5) \\
        \np237 & 6.4(2) & 1.5(3) & 7.6(2) & 5.2(3) & 8.1(2) & 5.5(2) & 7.2(1) & 4.9(1) & 1.5(-2) & 1.0(-2) \\
        \cm247 & 4.9(2) & 4.4(2) & 4.9(2) & 4.5(2) & 3.8(3) & 3.5(3) & 5.3(3) & 4.9(3) & 4.5(3) & 4.2(3) \\
         \hline \hline  
    \end{tabular}
    \label{tab:fluences} \\
        \vspace {3mm}
    \parbox{\textwidth}{\it
    Notes. The interstellar fluences $\fluence^{\rm interstellar}$ are normalized relative to the fluence of 
    \pu244, and are expressed in units of $[\rm atoms \ cm^{-2}]$, using the notation
        $A(B) \equiv A \times 10^B$.
    All results are for the present day,
    and include the effects of decay.
    The signal at 3 Mya assumes a flux duration of 1 Myr, and the
    others assume a constant flux after the event.
    
    }
\end{table}

\begin{table}[!htb]
    \caption{Predictions for {\rm r}-Process Radioisotopes in Fe-Mn Crusts based on \pu244 after 3 Myr}
        \vspace {3mm}
    \centering
    \hspace{-2.5cm}
\resizebox{1.13\columnwidth}{!}{%
\begin{tabular}{ccccc|ccccccc}
    \hline \hline
    & \multicolumn{4}{c|}{Concentration [atoms $g^{-1}$]} & Ratio & Measurement & Crust Elemental &  \multicolumn{4}{c}{Predicted Signal Ratio } \\
    Isotope & SA & SB & KA & KB  & $N_i^{\rm astro}/N_j^{\rm bg}$ & Limit & Abundance $X_j$ & SA & SB & KA & KB \\
    \hline 
    \zr93 & 6.5(2) & 1.0(7) & 5.9(2) & 1.9(1) & \zr93/\zr92 & $2\times 10^{-11}$ & $6.5 \times 10^{-4}$ &  8.9(-16) & 1.4(-11) & 8.2(-16) & 2.7(-17) \\
    \pd107 & 6.5(3) & 1.6(7) & 9.4(2) & 8.6(1) & \pd107/\pd106 & $10^{-8}$ & $1.7 \times 10^{-9}$ & 2.5(-9) & 6.1(-6) & 3.6(-10) & 3.3(-11) \\
    \i129 & 4.0(4) & 2.2(8) & 1.1(4) & 2.3(3) & \i129/\i127 & $10^{-14}$ & $5 \times 10^{-6}$ & 1.7(-11) & 9.2(-8) & 4.7(-12) & 9.8(-13) \\
    \cs135 & 6.8(2) & 1.5(7) & 4.2(1) & 1.8(2) & \cs135/\cs133 & $9 \times 10^{-12}$ & $1.5\times 10^{-6}$ & 1.0(-13) & 2.2(-9) & 6.1(-15) & 2.6(-14) \\
    \hf182 & 3.8(2) & 5.5(5) & 8.9(1) & 1.4(1) & \hf182/\hf180  & $10^{-11}$ & $8 \times 10^{-6}$ & 4.5(-14) & 6.4(-11) & 1.0(-14) & 1.7(-15) \\
    \hline
    \u236 & 2.2(2) & 1.2(3) & 2.5(2) & 1.4(2) & \u236/\u238 & $10^{-13}$ & $10^{-5}$ & 8.9(-15) & 4.7(-14) & 1.0(-14) &  5.7(-15) \\
    \np237 & 8.2(1) & 2.0(2) & 9.7(1) & 6.6(1) & \np237/\u238 & $5 \times 10^{-11}$ & $10^{-5}$ & 3.3(-15) & 7.8(-15) & 3.9(-15) & 2.6(-15) \\
    \cm247 & 6.2(1) & 5.6(1) & 6.2(1) & 5.7(1) & \cm247/\u238 & - & $10^{-5}$ & 2.5(-15) & 2.3(-15) & 2.5(-15) & 2.3(-15) \\
     \hline \hline
    \end{tabular}} \\
        \vspace {3mm}
    \parbox{\textwidth}{\it 
    Notes. The values are expressed as $A(B) \equiv A \times 10^B$.
    The crust abundances are taken from \citet{Hein2000}, and 
    we assume $\Phi^{\rm interstellar}(\pu244)$
    fluxes in eqs.~(\ref{eq:244pu-5Myr})
    and (\ref{eq:244pu-9Myr}).
    The uptake is assumed to be $U=0.10$ in all cases.
    The isotopic concentration $n_i^{\rm astro}/\rho$ is the number of atoms of isotope $i$ per gram of crust.  The crust elemental abundances $X_j$ are mass fractions; for \np237 and \cm247, the uranium abundance is given. 
    The measurement limits are the
    AMS sensitivities for \zr93 \citep{Hain2018, Martschini2019, Pavetich2019}, \pd107 \citep{Korschinek1994}, 
    \i129 \citep{Vockenhuber2015},
    \cs135 \citep{Yin2015} and \hf182 \citep{Vockenhuber2004}; see discussion in \S\ref{sect:terrestrial}.
    }
    \label{tab:crust-predict-3Myr}
\end{table}

\begin{figure}
    \centering
    \includegraphics[height=0.7\textheight]{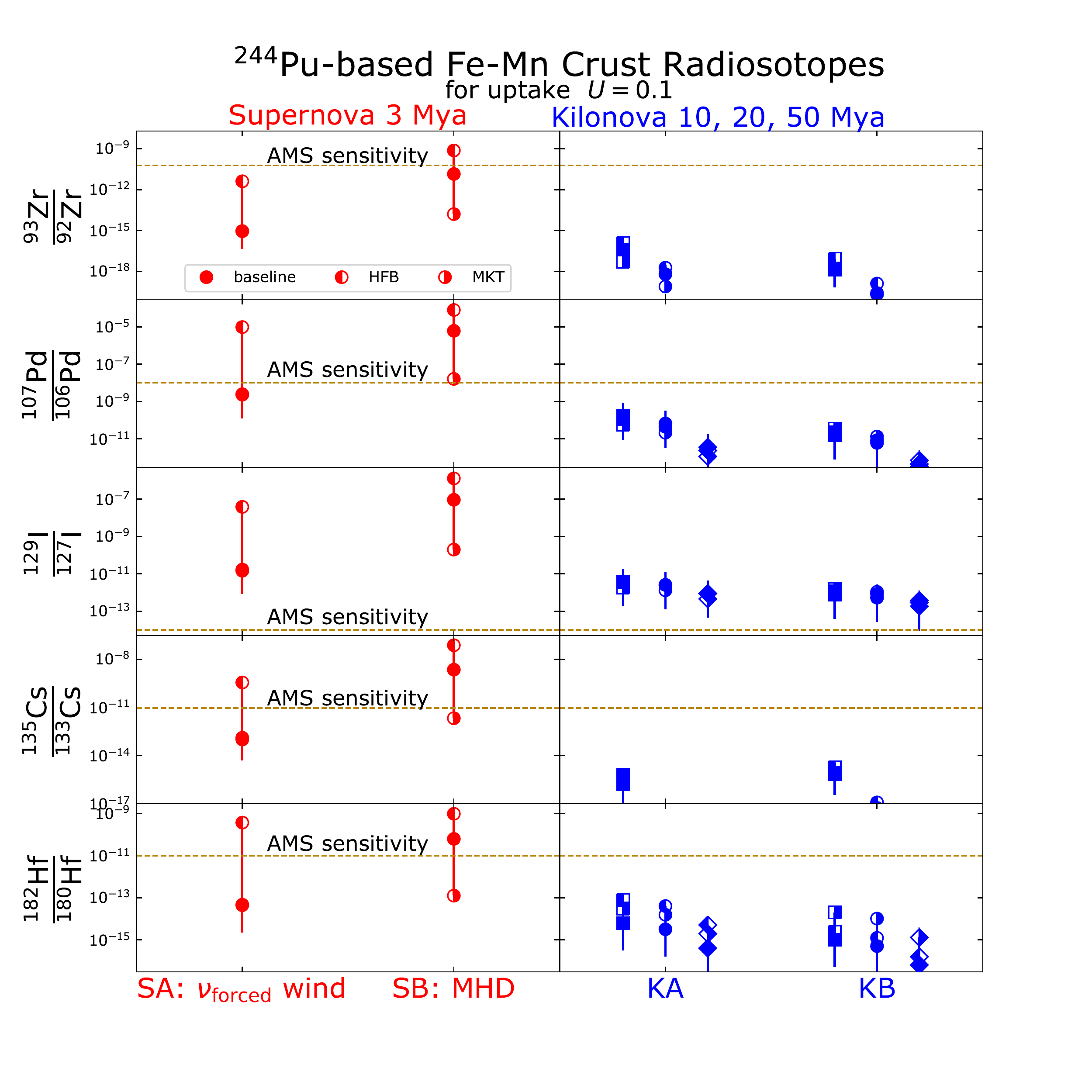}
    \vspace{-1cm}
    \caption{\it Radioisotope predictions 
    normalized to the \citet{Wallner2015} \pu244 flux from
    Eqs.~(\ref{eq:244pu-5Myr}) and (\ref{eq:244pu-9Myr}).
    The predictions are normalized to a 10\% uptake,
    and assume dust efficiency equal to that of Pu; for other values multiply by $U_i f_i/f_{\rm Pu}$.  The uncertainties 
    are only those due to the \pu244 flux, and the measurement limits are those listed in Table \ref{tab:crust-predict-3Myr}. The
    predictions are from (left panels) the SN {\em r}-process models after 3 Myr and (right panels)  the KN {\em r}-process models after 10, 20, and 50 Myr. The results
    from our baseline calculations are shown as full symbols, and the half-filled symbols are
    results from HFB and MKT calculations. A minor change to the figure should be made for \zr93/\zr92 from $6 \times 10^{-11}$ to a new value of $2 \times 10^{-11}$ in light of a new citation that has been added to the text.
    }
    \label{fig:geo-ratios-3Myr}
\end{figure}

Tables \ref{tab:ratios-Plio}-\ref{tab:ratios-Devo}
give the number ratios for various isotopes of interest
$({\cal N}_i/{\cal N}_{244})_{\rm astro} e^{-t/\tau_i}$,
where we incorporate the decay factors corresponding to the different time scales of interest.
For the lighter species with stable isotopes, 
the candidates of particular interest are those with high ratios to \pu244.

Table \ref{tab:fluences} shows the time-integrated interstellar flux or {\em fluence},
as defined in eq.~(\ref{eq:IntFlu}),
for selected {\em r}-process radioisotopes in specific scenarios. The values are for the present day, and include the effects of decay.  
For the 3 Mya signal associated 
with an SN, the duration is 1 Myr.
For the KN signal at 10 Myr, the duration
is assumed to be 10 Myr.
The fluences are the most direct results from the
astrophysical calculations,
but connecting them to geological and lunar
measurements requires one to specify the abundances
in particular samples.

Table~\ref{tab:crust-predict-3Myr} shows our predictions for selected radioisotope ratios following deposition in Fe-Mn crusts 3 Mya,
normalized to a 10\% uptake and $f_i/f_{\rm Pu}=1$.
These results also appear in Fig.~\ref{fig:geo-ratios-3Myr}, where use the \pu244 flux 
inferred from sediments $0.5-2.2 \, \rm Myr$ old~\citep{Wallner2015} for normalization.
In general we see that the SN predictions are as large as, and sometimes much larger than, those of the KN models.  This reflects the difficulty of making actinides in SN models, which means that lower-mass species are more abundant relative to the \pu244 isotope that anchors our results. 
Variations between the KN model predictions are generally within a factor of 10. In contrast, the SN predictions can span several 
orders of magnitude, with the SB model generally leading to larger low-mass abundances due to its particularly low \pu244 output.
Also shown in Table~\ref{tab:crust-predict-3Myr} and Fig.~\ref{fig:geo-ratios-3Myr} are the prospective
AMS sensitivities for \zr93 \citep{Hain2018, Martschini2019, Pavetich2019}, \pd107 \citep{Korschinek1994}, 
\i129 \citep{Vockenhuber2015},
\cs135 \citep{Yin2015}, and \hf182 \citep{Vockenhuber2004}. 
We see that, whereas predictions for the \zr93 signal generally lie below
the AMS sensitivity, those for \pd107, \cs135, and possibly \hf182 in model SB lie above the AMS sensitivity, and the predictions for the \i129 signal
lie above the AMS sensitivity in both SN scenarios and possibly also in the KN scenarios.

\begin{figure}
    \centering
    \includegraphics[height=0.8\textheight]{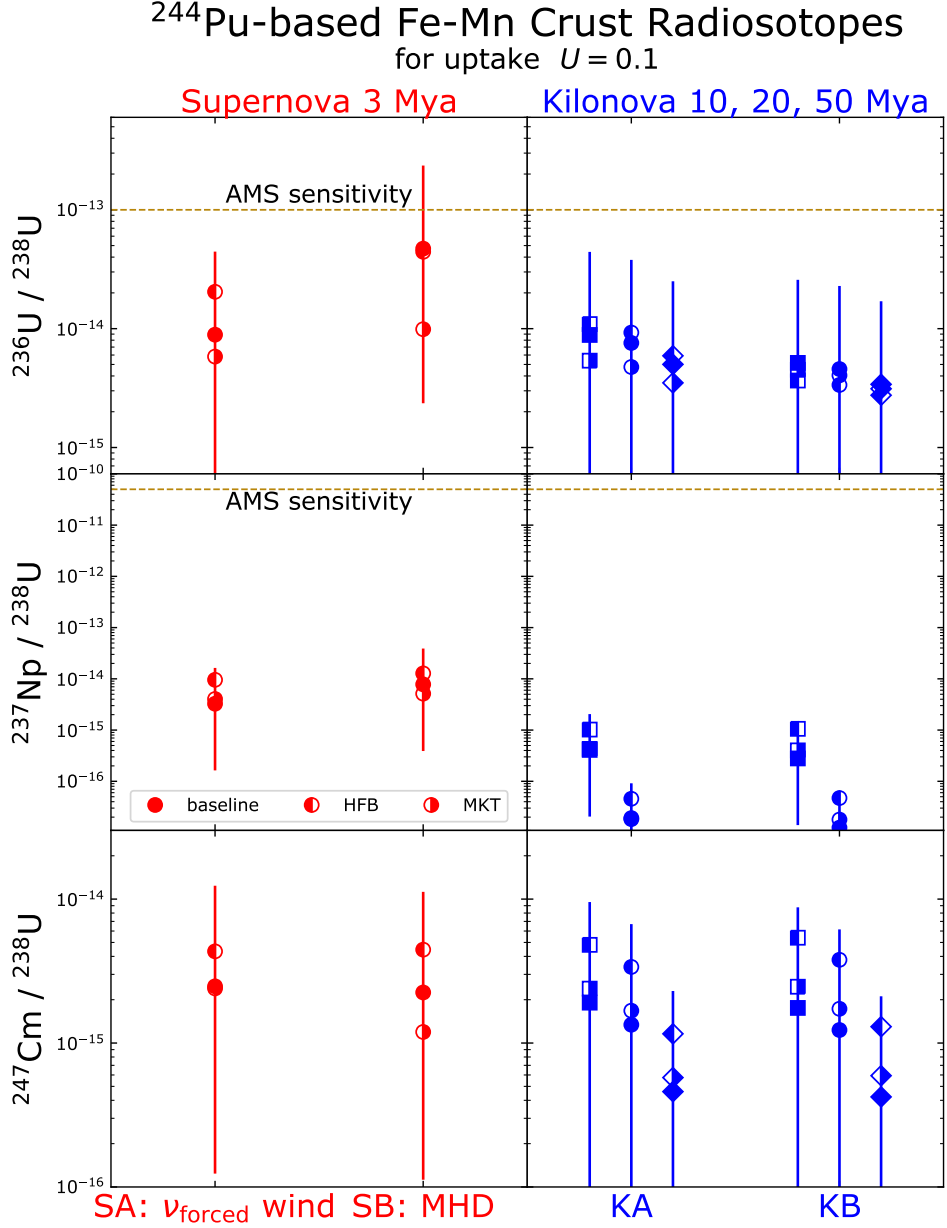}
    \vspace{-1cm}
    \caption{\it Predicted abundances of the actinides \u236, \np237, and \cm247 in Fe-Mn crusts, shown for
    the same SN and KN models as in Fig.~\ref{fig:geo-ratios-3Myr}.
    We use \u238 as the reference standard for all three species.
    The results
    from our baseline calculations are shown as full symbols, and the half-filled symbols are
    results from HFB and MKT calculations.}
    \label{fig:crust-transU}
\end{figure}

Figure \ref{fig:crust-transU} shows the
Fe-Mn crust abundances for the actinides \u236, 
\np237, and \cm247, for the same models
as in Fig.~\ref{fig:geo-ratios-3Myr}.
These are all
elements with no stable isotopes, 
so we use \u238 as a reference standard
for the ratios presented; we note that
\u238 is also the source of \u236 background. We recall that natural uranium will far overwhelm any 
astrophysical contribution. For this reason, the \u236 signal may be masked by other production mechanisms. 
For \np237, suppression of the \u238 signal may be necessary before a signal can be seen. 
We thus conclude that \u236 and \np237 detections likely await improvements in measurement techniques.
For \cm247, uranium is not anticipated to be a limitation, but detection will require establishing a pure curium reference spike.
We urge further study of the experimental possibilities for all of these species.

To summarize, we have found that
the \pu244 evidence implies
that the geological record should contain other
{\em r}-process radioisotope signals.
The abundances are in some cases within or close to the reach of existing AMS techniques,
whereas improvements in the AMS sensitivity to key species are necessary in other cases.

We emphasize that the experimental program for these new species, as well as additional \fe60 and \pu244 searches, should have two related but distinct goals:
sensitivity and time resolution.  
\begin{itemize}

\item 
{\em Sensitivity.}  High sensitivity, i.e., the ability to detect small extraterrestrial
radioisotope abundances, obviously is critical.  It allows for detection of additional species
and more accurate abundances, both of which sharpen the probes of the potential nucleosynthesis
sites.

\item 
{\em Time resolution.} Time resolution offers another powerful means of distinguishing between models.  A better determination of the \pu244 time history can clarify whether it is exclusively coincident with the two \fe60 pulses, suggesting a common SN origin, or it appears at other epochs, suggesting a distinct origin likely in a KN.  The time behavior of additional radioisotopes can then be compared with that of \fe60 and \pu244, which can further distinguish between these scenarios and potentially distinguish between SN and KN models.

\end{itemize}
These goals are in some tension, as finer time resolution implies smaller samples per time bin, reducing the available signal at a fixed sensitivity.  As with the development of the \fe60 measurements, a sensible strategy may be to first strive for the detection of new species, then
follow up with better time resolution.

We now turn to considerations of sample collection
for each of the radioisotopes of interest.

\subsection{Terrestrial Searches}

\label{sect:terrestrial}

In order to detect and preserve these signals, we emphasize that samples must be collected from
well-chosen sites and suitable media, in the sense that
the samples should have maintained a detectable abundance of the SN debris from the initial deposit to the present and
that the samples should be accessible for collection and delivery to the laboratory.
We consider these issues for the Earth in this
section, and for the Moon in the next section.

Marine samples are the primary focus of interest
as natural terrestrial archives, since
continental material is subject to erosion, and radioisotope studies in ice cores
do not extend so far into the past.~\footnote{However, we recall that 
Antarctic ice dating back to 2.7~Mya has been recovered~\citep{Yan2019}, potentially opening up
prospects for the future.}
In general, deep-ocean deposits are essentially undisturbed, with very little disturbance from creatures at the ocean floor (bioturbation) to blur the time structure, and are largely free of
anthropogenic contamination from nuclear debris, as well as backgrounds from cosmic-ray spallation in the atmosphere. 
Debris at the sea floor accumulates in sediments that stretch back beyond the $\sim 10$~Myr that has been studied so far,
rendering it suitable for exploring possible astrophysical triggers of mass extinctions that occurred $\gtrsim 10^8$~ya ago
in addition to the well-attested \fe60 signal, evidence for \mn53 from $\sim 3$~Mya, and reports of \pu244 deposition
over the past 25~Myr.
Ultimately, deep-ocean sediments may harden to make sedimentary rock in geological strata reaching back to much longer timescales.
In this section and Table~\ref{tab:oceanic} we highlight some of the key
considerations from chemical oceanography
that determine which astrophysical radioisotopes 
may have geologically favorable conditions for detection.
We make heavy use in the following discussion of \citet{Brocker1982}, \citet{Nozaki2001},
and the Geotraces database~\footnote{\href{https://www.geotraces.org/}{https://www.geotraces.org/}} and private communications
with Craig Lundstrom and Tom Johnson.

In considering the prospects for detecting marine deposits of radionuclides, one must consider the sedimentation rates
of the radionuclides of interest, focusing on those with a short residence time and discarding those that 
remain dissolved in seawater. Table~\ref{tab:oceanic} lists the elements discussed earlier in this paper, together with
their mean oceanic concentrations and the characteristics of their distributions, 
which provide the basis for our
discussion of the possibilities of detecting their radioisotopes in view of the SN and KN model calculations above.\\

\begin{table}[]
    \centering{
    \caption{Estimated mean oceanic concentrations of selected elements, adapted from
    \citet{Nozaki2001}
    \label{tab:oceanic}
    }
    \hspace{-2cm}
    \begin{tabular}{c|c|c|c}
    \hline \hline
    Element &  Mean oceanic  & Type of  & Leaves Solution \\
     &  mass fraction ($\times 10^{-12}$) & Distribution & Rapidly \\
    \hline \hline 
    Mn & 20 & s & yes\\
    Fe & 30 & s + n & yes\\
    Zr & 15 & s + n & yes\\
    Pd & 0.06 & n & yes\\
    I (IO$^-_3$) & $5.8 \times 10^4$ & c & no\\
    I (I$^-$) & 4.4 & r + s & yes\\
    Cs & 310 & c & no\\
    Hf & 0.07 & s + n & yes\\
    U & $3.2 \times 10^3$ & c & no\\
    Np & $\sim 10^{-5}$ & c & no \\
    Pu & - & r + s & yes \\
    Cm & - & {s?} & {?} \\
    \hline \hline
    \end{tabular}} \\
        \vspace {3mm}
\parbox{\textwidth}{\it Notes.  In this Table c denotes an element whose distribution is conservative, in the sense that it follows the salinity of the ocean water,
    n denotes a distribution that increases with depth, like that of typical nutrients in the water,
    s denotes an element that attaches to particulate matter and may be concentrated (``scavenged") at the water/sediment interface,
    and r denotes an element whose distribution is controlled by reduction and oxidation
    reactions (``redox-controlled") and may be formed in oceanic basins that are depleted in oxygen (``anoxic")~\citep{Nozaki2001}. 
    \np237 results are from \citet{Lindahl2005}.  
    We note also that there may be other geochemical and biological processes
that affect the rates of uptake of different radionuclides in various materials. For example,
``magnetotactic" bacteria that orientate themselves along magnetic fields} secrete iron and therefore concentrate \fe60, as discussed in~\citet{Ludwig2016},
and certain sorts of organic matter may absorb plutonium effectively (J.~Marshall, private communication, 2021), specifically
in anoxic environments. \citet{Schneider1984} 
measured anthropogenic 
\cm242 and \cm244 in Scottish coastal sediments, and summarized data indicating
that \cm244 has also been detected in
fish and seaweed but not in seawater; we also note that curium
adheres very tightly to soil particles.
    \label{oceanic}
\end{table}    

$\bullet$
Among the elements that may be produced by the $r$-process and are identified in the previous section as being of interest, 
we note that \underline{zirconium} sinks in the ocean, 
attaches to solid particles, and may be scavenged at the water/sediment interface,
similarly to iron. The half-life of \zr93 ($1.53 \times 10^6$~yr) is somewhat less than those of the isotopes 
\fe60 and \mn53 that have been reported in layers $\sim 2.5 \times 10^6$~yr old.
Figs.~\ref{fig:spaghetti} and \ref{fig:spaghetti_kn_bovard} 
and Table~\ref{tab:ratios-Plio} indicate that its $r$-process production
rate, which is quite model-dependent, 
may be orders of magnitude higher than that of \fe60, or considerably lower. However,
Table~\ref{tab:crust-predict-3Myr} and Fig.~\ref{fig:geo-ratios-3Myr} indicate that the \zr93
signal in 3 Myr old samples may lie below the estimated AMS sensitivity.\\

$\bullet$
\underline{Palladium} has a distribution that increases with depth, but is not expected to bind to particles or be
concentrated at the water/sediment interface. 
It has been detected in Fe-Mn crusts, but with low abundances (see Table \ref{tab:crust-predict-3Myr}),
which is advantageous for our purposes. These low abundances imply that,
as seen in Table~\ref{tab:crust-predict-3Myr} and Fig.~\ref{fig:geo-ratios-3Myr}, the relative abundances of \pd107 predicted in some $r$-process models may be within reach of the AMS technique if a suitable 3 Myr old sample can be found.\\

$\bullet$
Most \underline{iodine} is in the form of the IO$^-_3$ ion, which is distributed conservatively, 
but a small fraction is in the form of the I$^-$ ion, which may be scavenged, 
preferentially in anoxic basins such as the Black Sea. 
\citet{Fitoussi2007} attempted to measure \i129 in a relatively recent but pre-anthropogenic sediment, and found that care is needed to avoid  anthropogenic contamination.
\citet{Ji2015ams} have presented the first AMS measurements
of \i129 in Fe-Mn crusts.  These data already reveal the power
of \i129 to probe near-Earth explosions, particularly the 
possible KN enrichment of the Local Bubble,
and the corresponding constraints are discussed in detail in the following
Section~\ref{sect:129I}. \\

$\bullet$
\underline{Cesium} dissolves in water and has a conservative oceanic distribution, 
so \cs135 offers poor prospects for deep-ocean detection. Moreover, as seen in Table~\ref{tab:crust-predict-3Myr} and Fig.~\ref{fig:geo-ratios-3Myr}, most of the models studied indicate that
the fraction of \cs135 may lie below the AMS sensitivity.\\

$\bullet$
\underline{Hafnium} is an interesting target, particularly if production occurred more than 3~Mya,
thanks to the \hf182 half-life of 8.9 Myr.
Most of our $r$-process models produce less \hf182 than \zr93, 
while the oceanic distribution of hafnium is similar to that of zirconium.  In a sediment,
\citet{Vockenhuber2004} have established the upper limit $\hf182/\hf180 < 10^{-6}$ using a 45~g sample, and
quoted a total hafnium elemental abundance by dry weight of 8 ppm. Combining these numbers, we find an upper limit on the 
fraction by weight of \hf182 of $8 \times 10^{-12}$. 
\citet{Martschini2020} showed that these limits could in
principle be improved dramatically using a
laser technique to remove stable isobars.  They found
$\hf182/\hf180 = (3.4 \pm 2.1) \times 10^{-14}$,
on the basis of which they estimated a sensitivity of $(\hf182/\hf180)_{\rm min} \approx 6 \times 10^{-14}$.
This would correspond to a mass fraction of $\sim 4 \times 10^{-18}$,
which could be achieved if the stable tungsten isobar \iso{W}{182}
is absent or can be removed.

For comparison, \citet{Wallner2021} estimated a range of $(0.1-100) \times 10^{-22}$ \pu244 atoms per atom in the crust samples. It is important to note that one major improvement for these measurements was the use of the 1 MV accelerator Vega at Australia's Nuclear Science and Technology Organisation, which provided an improved \pu244 detection efficiency from $1 \times 10^{-4}$ \citep{Wallner2015} to $1.5\%$. These measurements correspond to a sensitivity to the fraction by weight of $\sim 10^{-22}$. 
In Table~\ref{tab:ratios-Plio} we see ratios $\hf182/\pu244 < 8.7 \times 10^4$
(reached in model SB), suggesting that the model sensitivity of the \pu244 detection by
\citet{Wallner2021} is better than that of the \hf182 search by \citet{Vockenhuber2004},
though a more sensitive \hf182 search may have interesting prospects. \citet{Vockenhuber2004} used their
data to quote an upper limit on the \hf182 flux into the sediment of $2 \times 10^5$ cm$^{-2}$yr$^{-1}$, to be compared with
the estimated ISM flux of 0.05 cm$^{-2}$yr$^{-1}$ assuming a global distribution of the infalling ISM and that
all of the material is deposited in the sediment. Thus there is considerable scope for a more sensitive
measurement to observe a signal above the expected background.
Table~\ref{tab:crust-predict-3Myr} and Fig.~\ref{fig:geo-ratios-3Myr} indicate that the fraction of \hf182
may be within the AMS sensitivity range in model SB.\\

$\bullet$
\underline{Uranium} has a natural background 
of \u235 and \u238 that far overwhelms any
astrophysical perturbations we might hope to detect.
The shorter-lived \u236 has no remaining proto-solar component,
and so potentially could serve as an astrophysical signature.
However, significant obstacles exist.  In uranium ores,
neutrons from fission can capture on \u235 to create \u236.
The resulting \u236/\u235 ratio will be strongly sensitive to local conditions, making this background challenging to estimate.  Anthropogenic \u236 contamination
is also a concern.  
Finally, uranium has a conservative oceanic distribution, so \u236 is likely difficult to use in sediments and
crusts would be preferred.\\

$\bullet$
\underline{Neptunium}
is poorly studied as compared to other transuranic elements. Indeed, it has
been called the ``neglected actinide'' \citep{Thompson1982}.
Table~\ref{tab:ratios-Plio} shows it is produced in abundances comparable to those of
\pu244 and thus is of great interest to 
search for as a cross-check on \pu244 and
potentially as a probe of the details of actinide synthesis.
As for other transuranic elements,
there are no stable isotopes, so elemental
searches are viable. However, anthropogenic contamination is an issue.
\citet{Lindahl2005} found that
\np237 
has a conservative oceanic distribution,
and so does not readily precipitate.
Thus, Fe-Mn crusts would seem to be a preferred
terrestrial target.  Future lunar measurements are also
of interest; unfortunately, the initial report of \np237 evidence in {\em Apollo} 
return samples \citep{Fields1972} was subsequently found to be compromised by airborne anthropogenic contamination \citep{Fields1976}. \\

$\bullet$
Like iodine, \underline{plutonium} may be scavenged, and it is encouraging that \pu244 detection has been reported
by multiple experiments, as seen in Table~\ref{tab:244Pu-obs} and Fig.~\ref{fig:plutonium}.~\footnote{We note, however, 
that most of the shorter-lived isotope \pu242 would
have decayed on the time-scales discussed here.} 
Measurements with higher sensitivity 
using deep-ocean deposits stretching back to ages comparable to the half-life of \pu244
($t_{1/2} = 81$~Myr) or further would be very interesting, as would sufficient time resolution to
distinguish a possible \pu244 pulse coincident with that observed for \fe60, and/or any earlier similar pulses.
We encourage searches in anoxic basins where the concentration of \pu244
may be higher. \\

$\bullet$
The model calculations in Table~\ref{tab:ratios-Plio} indicate
that an SN $\sim 3$~Mya could have deposited \underline{curium} on Earth, 
with {\cm247}  
at levels comparable to those for \pu244, whereas most of the shorter-lived {\cm248} would
have decayed.  This isotope thus makes an important target for geological and lunar 
searches.

\subsection{\i129 in Fe-Mn Crusts and Near-Earth KNe: Present Constraints and Future Opportunities}
\label{sect:129I}

\begin{table}[]
    \caption{Ferromanganese Crusts with Measured \i129 Profiles
    \citep{Ji2015ams,ji2015crusts}}
    \hspace{-0.7cm}
    \begin{tabular}{c|c|c|c|c|c}
    \hline \hline
    Crust &  Crust & \multicolumn{2}{c|}{Elemental Mass Fractions} & \i129/\i127 & \i129/Fe \\
    Name  & Location & $X({\rm Fe})$ & $X({\rm I})$ & Ratio/$10^{-12}$ \\
    \hline \hline 
    CDX80-1 & $(19^\circ57.9^\prime \, \rm N, 172^\circ55.1^\prime \, E)$
    & $0.05-0.20$ & $(2.7-7.7) \times 10^{-5}$ & $\le 1.27$ & $\le 10^{-15}$ \\
    MDP5D44 & $(10^\circ 20.24^\prime \, \rm N, 167^\circ 26.5^\prime \, W)$ 
    & $0.10-0.20$ & $(3.8-10.9) \times 10^{-5}$ & $\le 0.48$ & $\le 10^{-15}$\\
             \hline \hline  
   \end{tabular}      \vspace {3mm} \\
    {\it The ranges of the iron and iodine fractions 
    in the crust sample CDX80-1 (MDP5D44) shown are for 
    depths $\in (0.5, 4)$~cm ($\in (1.6, 5)$~cm)
    and the iodine isotope ratios shown are for 
    depths $> 0.5$~cm ($> 1$~cm).}
    \label{tab:i129crusts}
\end{table}

\begin{figure}[!htb]
	\centering
    \includegraphics[height=0.5\textheight]{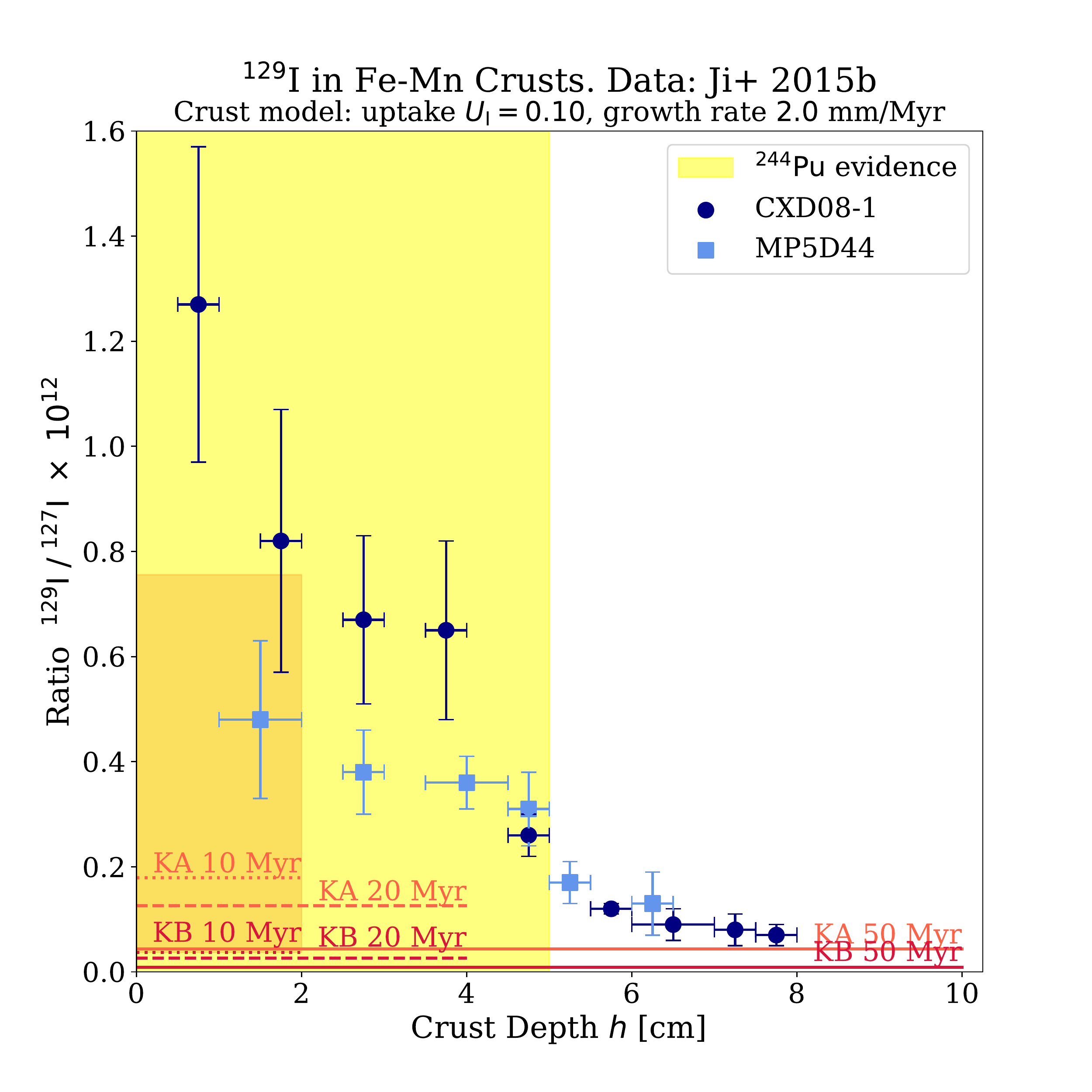}
	\caption{\it An example of the potential power of \i129 measurements.  We show an estimate of the evolution with depth of the \i129/\i127 isotope ratio, based on the AMS data of~\citet{ji2015crusts}. The horizontal lines represent the ratio \i129/\i127 that would be generated within the KA and KB models occurring at the the indicated times $t_{\rm KN}$ in the past; these represent a floor below which the natural \i129 background cannot fall. The KA curves for $t_{\rm KN}=(10,20) \ \rm  Myr$ are off-scale.  These model lines assume a crust iodine uptake $U_{\rm I} = 0.1$ and a constant growth rate $\dot{h}_0 = 2 \, \rm mm/Myr$. Model uncertainties for $t({\rm KB})=10 \ \rm Myr$ are shown as a red band; similar ranges apply to the other cases.  Model line heights scale as $U_{\rm I} \, \exp(-t_{\rm KN}/\tau_{129})$, and maximum depth scales as $h_{\rm max} = \dot{h}_0 t_{\rm KN}$.  The yellow band shows the extent of \pu244 evidence using the assumed $\dot{h}_0$.}
	\label{fig:Idecay}
\end{figure}

The AMS measurements of \i129 by \citet{Ji2015ams} and \citet{ji2015crusts} are from three ferromanganese crusts.  
For two of these, \citet{Ji2015ams} tabulated \i129 profiles to depths of 8 cm. Table \ref{tab:i129crusts}
gives the locations and elemental compositions of these crusts, which
are both in the mid-Pacific, at locations separated by about 2300 km. 
The natural production of \i129 includes uranium fission products, 
as well as the daughters of spallation events between cosmic rays
and atmospheric xenon. These give rise to a persistent
flux of \i129 into the crusts, which represents an irreducible background.

\citet{Ji2015ams} and \citet{ji2015crusts} found that the crusts
show an \i129/\i127 isotopic profile that drops
with depth in a manner consistent with
a background component undergoing
radioactive decay, as seen in Fig.~\ref{fig:Idecay}. 
Moreover, the most recent layer has an abundance consistent 
with the pre-bomb level found in sediments~\citep{Moran1998}.
\citet{Ji2015ams} argue that this drop-off with depth suggests
that the results are free of anthropogenic contamination.
We note that the crust data show no strong evidence for
a ``floor'' of \i129 that persists as the background diminishes
at increasing depth.  

The crusts in which \citet{Ji2015ams} measured \i129
lack independent measures of
their growth rates, without which the depth profile
cannot be transformed into a precise time history.
Nonetheless we can use typical Fe-Mn growth rates
to estimate the epochs probed:
for a low growth rate $\dot{h}=1 \ \rm mm/Myr$,
the crusts span 5--80 Myr, while 
a high growth rate $\dot{h} = 6 \ \rm mm/Myr$
would correspond to a range of 1-13 Myr.
It is therefore not clear that these data
include the 2-3 Mya range of the \fe60 pulse,
so that the SN (one-step) scenario may not be
probed directly by these data. 
However, as seen in Fig.~\ref{fig:geo-ratios-3Myr}
and Table \ref{tab:crust-predict-3Myr},
the SB model predicts \i129 many orders of magnitude
above the natural background, so that such a signal
could be easily ruled out (or detected!) in
a search coincident with the \fe60 pulses in a crust.
This is the case despite the significant model uncertainties, which are
indicated in Fig.~\ref{fig:geo-ratios-3Myr} but omitted in Fig.~\ref{fig:Idecay} for clarity.
On the other hand, the SA model uncertainties are large enough that their lower extreme overlaps with the KN models.

On the other hand, it is clear that these data
cover the timespan probed by the extended \pu244 flux,
and thus probe the two-step scenario of
KN enrichment of the proto-Local Bubble.
We illustrate the power of \i129 data in Fe-Mn crusts using
the \citet{Ji2015ams} data while emphasizing that,
due to the lack of information on growth rate and uptake, 
our quantitative results are only crude estimates.
Nevertheless, we hope that such order-of-magnitude estimates
may stimulate further experimental work.
In this spirit, we again assume a constant growth rate
$\dot{h}_0$, so that a time $t$ in the past corresponds
to a depth $h(t) = \dot{h}_0 \, t$.  
If the KN event occurred at a time $t_{\rm KN}$ ago,
the signal can extend to a maximum depth
$h_{\rm max} = \dot{h}_0 \, t_{\rm KN}$.

All KN-produced radioisotopes would have
decayed for the (unknown) $t_{\rm KN}$ duration since the 
explosion, regardless of the 
time history of their flux on Earth. 
Thus any KN-created flux will {\em not} show a radioactive decay pattern versus depth.
This is in contrast to the 
natural \i129 background, which is due to ongoing
production and so should show the effects of
decay versus depth. Just such a pattern is evident
in Fig.~\ref{fig:Idecay}; any KN-produced signal would
represent a ``floor'' underneath this natural background.  
Thus the deepest measurements have the most constraining power.

The KN signal in the crust
is straightforward to calculate in the case where the
growth rate and {\em r}-process flux are both constant. 
In this case the \i129 profile
versus depth $h$
for KN model $\alpha = (\rm KA,KB)$ has a steplike structure: 
\begin{equation}
\label{eq:I129-vs-depth}
\frac{\i129}{\i127} 
\left( h \right)_{\alpha}          
= \left\{
    \begin{array}{cl}
    (\i129/\i127)_{\alpha,0} \ \ e^{-t_{\rm KN}/\tau_{129}} \ \ , \ \  & h \le h_{\rm max} = t_{\rm KN}/\dot{h}_0 \, , \\
    0 \ \ , \ \ & h > h_{\rm max} \, ,
    \end{array}
\right.
\end{equation}
where  $(\i129/\i127)_{\alpha,0}=
N_{129}^{\alpha}/N_{127}^{\rm bg}$
from Eq.~(\ref{eq:obsratio}), which
includes a factor $f_{\rm I}$ for
dust formation and transport, and the uptake factor $U_{\rm I}$.
We see that the isotope ratio is constant with depth,
and for increasing $t_{\rm KN}$ the level 
is progressively smaller but extends to greater depths.

In Fig.~\ref{fig:Idecay} we confront the data of~\citet{Ji2015ams,ji2015crusts}
with some results from calculations based on the KN models KA and KB.
For each of these, we show predictions for KN
explosion time $t_{\rm KN}=(10,20,50) \ \rm Myr$, corresponding to a scenario in which
an explosion at this time enriches and stirs
the proto-Local Bubble,
initiating a flux of {\em r}-process dust onto the Earth.
The \pu244 data in Fig.~\ref{fig:plutonium} shows that
this flux extended at least to the earliest measured bin $12-25$ Mya, but neither excludes nor requires an earlier flux.
Thus our constraints should apply only within this time window,
though the explosion would have occurred earlier.

To predict the \i129/\i127 ratio in the crust, we
assume a constant growth rate $\dot{h}_0=2 \ \rm mm/Myr$ and
and a value $U_{\rm I} = 0.1$ of the uptake factor for iodine.~\footnote{According
to~\citet{ji2015crusts}, iodine is enriched in crusts relative to seawater, though
much less so than iron. This would allow a range that includes the value of $U_{\rm I}$ that we assume, but with
considerable uncertainty.  For comparison, in Fe-Mn crusts the estimates of the iron uptake in different crusts have varied from $U_{\rm Fe}=0.6\%$ \citep{Knie2004} to 7\% and 17\% \citep{Wallner2016}, while \citet{Wallner2015} estimated the \pu244  incorporation efficiency at $U_{\rm Pu}=21\% \pm 5\%$. \citet{Wallner2021} found for their crust
a uptake efficiency of $17\% \pm 3\%$ for \fe60, and of $12\% \pm 4 \%$  for \pu244; they adopted $U=17\%$ for both.}
With these $(\dot{h}_0,U_{\rm I})$ values, model KA would be excluded if the KN occurred $\lesssim 50$~Mya,
whereas model KB would be allowed for explosions $\gtrsim 20$~Mya.
There are clearly many uncertainties in this analysis: the line heights scale as $(f_i/f_{\rm Pu}) U_{\rm I} \, \exp(-t_{\rm KN}/\tau_{129})$
and they reach to maximum depths that scale as $h_{\rm max} = \dot{h}_0 \; t_{\rm KN}$.  Thus KA models can be accommodated with,
for example, smaller uptake values.

The larger lesson is that it is of great interest to make 
additional \i129
measurements in Fe-Mn crusts, which could offer an important new probe of possible sources of $r$-process isotopes.
Also, it would be of considerable interest to search at even greater depths and thus earlier times than
those shown in Fig.~\ref{eq:I129-vs-depth},
to see if the \i129 signal continues to drop as expected 
from a natural background, or whether
a ``floor'' reveals itself and thereby indicates
a scenario with a early injection time
and thus a distant production epoch $t_{\rm KN}$.
It would be of particular interest to measure \i129 in crusts with \fe60 data, searching both inside and beyond the \fe60 pulses;
similarly,
it would be useful to search for \i129 in the same crusts that
show \pu244 signals.  
Finally, we stress that independent measures of the growth rate,
and of the iodine uptake, are as important as the additional \i129 data themselves.  The ability to limit astrophysical perturbations to \i129 hinges 
on the precision and reliability of the \i129 profiles with time.

\subsection{Lunar Searches}

A great advantage of lunar material is that it avoids geological and oceanographic
effects that transport, mix, and dilute the signal. 
Moreover, the uptake is likely to be high, so that all of the favored radioisotopes among those that can be produced via
the $r$-process, namely \zr93, \pd107, \i129, \cs135, \hf182, \u236, \np237, \pu244, and \cm247, are {\it a priori}
interesting targets for searches in the lunar regolith (i.e., unconsolidated surface material). 
\citet{Fry2015} cautioned that some fast dust particles could lead to the vaporization and escape of some of the ejecta mass. However, the lunar detections of \fe60~\citep{Fimiani2016}, with a fluence that is
relatively high compared to terrestrial results, suggest that these effects are small.  This could imply that much of the 
astrophysical dust slows to lower speeds by the time the particles arrive.

Thus the surface fluences of SN species
should be close to the interstellar fluences
estimated in Table \ref{tab:fluences}, modulo a geometric factor of the cosine of the vertical angle.
However, converting these values into predictions
in, say, atoms per gram, requires one understands
lunar surface processes over Myr timescales.  Such a detailed analysis is beyond the scope of this work;
we intend to visit this issue in a separate paper.  Here we summarize some important considerations.

As impactors strike the lunar surface, the regolith gets continually reworked, a process known as gardening. 
Gardening mixes material more deeply over time, so that deeper material is less likely to have been disturbed recently,
though it would have been disturbed in the more distant past. Shallow soil is continually turned over by the large flux 
of very small impactors. \citet{Costello2018} showed that there has been significant reworking down to $\sim$10 cm over the past 3 Myr,
and significant reworking down to $\sim$40 cm over the past 400 Myr. 
Tables~\ref{tab:ratios-pu} and \ref{tab:ratios-Devo} suggest that \i129, \u236, and \pu244 may be the most interesting search targets
at lower depths. However, the gardening process implies that lunar samples cannot
be time-stamped accurately, and we recall that the {\it Apollo} lunar samples~\citep{Fimiani2016} have no direct timing information.

On the other hand, lunar samples may provide valuable information on the possible direction of any astrophysical source of
live radioisotopes. If these arrive in dust grains that travel ballistically, their distributions could depend on the lunar
latitude. The {\it Apollo} landing sites were all relatively close to the lunar equator. Therefore, in view also of their limited statistics,
they provide limited information in this regard. However, the recent {\it Chang'e-5} sample return mission landed at a higher latitude, $43.1^o$~N~\citep{Change5},
the {\it Artemis} program envisions landing at the lunar south pole~\citep{Artemis}, and various commercial lunar sample return missions
are also planned. More information on the latitude distribution may therefore be available in the coming years, and could discriminate
between origins in the Scorpius-Centaurus~\citep{Benitez2002,Breit2012} and Tucana-Horologium~\citep{mamajek2015} associations.  

Even if dust propagation is not ballistic \citep[as suggested by][]{Fry2020}, the lunar surface distribution of
radioisotopes in general offers a unique measure of the directionality of the dust velocity distribution.  This
directly probes SN dust propagation that is otherwise inaccessible observationally.

We note, however, that there is an important source of radioisotope background on the Moon, namely cosmic-ray spallation on the lunar surface.  
This is an irreducible background, so the signal must be found above it. Spallation is less important for the heaviest isotopes,
and is most effective when only one or a few nucleons are removed from a target nucleus.
Thus a key issue is whether there are relatively abundant stable isotopes that have one or a few more nucleons than the radioisotope of interest.
We note also that, although there is no anthropogenic background on the Moon, care must be taken after bringing samples to Earth, so as to
avoid anthropogenic contamination, which was an issue for the evidence for \u236 and \np237 on the Moon reported by \citet{Fields1972}.

Cosmic-ray irradiation of the lunar regolith 
also produces a substantial neutron flux at depth.  Neutron exposure would lead to
a \u236 background that makes any extrasolar \u236 difficult to find, but \u236 searches could nonetheless be useful to establish 
which if either of these components is present.
We note also that neutron captures on \u236 will lead to \np237, creating a background
for that species.

It is therefore encouraging that the {\it Chang'e-5} mission has recently returned to Earth a new sample of lunar material, and that
the future {\it Artemis} and other lunar missions have similar objectives. We advocate efforts to replicate the {\it Apollo}
results of \fe60 and urge searches for the other radioisotopes discussed in this paper.
The {\it Apollo} samples were gathered relatively
close to the lunar equator, whereas {\it Chang'e-5} landed in the northern hemisphere and {\it Artemis} is planned to land near the south pole.
Measurements in their samples may therefore provide 
a measure of the dust arrival direction(s), and perhaps
some indication of the latitude of their astrophysical sources.
This would herald the dawn of radioisotope astronomy.

\section{Discussion}
\label{sect:disc}

We have found that the \pu244 detection implies that there should also be measurable traces
of other {\em r}-process radioisotopes, whose abundances and time history can shed
important new light.
Of these, \zr93, \pd107, \cs135, and \hf182 are not only {\em r}-process species but can also be made in the {\em s}-process.  Were these the only isotopes detected, this would lead to ambiguity about the nucleosynthesis site that led to their injection on Earth.  But by comparing the time signature with those of \fe60 and \pu244, one can see if they are associated with the production of these isotopes.   It is conceivable as well that some species such as these could arise from a recent event distinct from the origins of both the \fe60 and \pu244.   For example, the {\em s}-process output of an AGB star could be delivered to Earth, perhaps via a two-step process within the proto-Local Bubble similar to what we have proposed for the KN injection of \pu244.  Here again,
the abundance patterns and time history of these species can test for such a model, and it would 
be of interest to investigate this scenario in more detail.

Our studies of live radioisotopes ejected from recent nearby explosions are closely linked with the long-studied question of now-extinct radioisotopes injected into the proto-solar nebula.
As described in the excellent reviews by \citet{Meyer2000,Adams2010, Lugaro2018}, 
meteorites show evidence that more than a dozen of what we call ``medium-lived'' 
radioisotopes (``short-lived'' in the cosmochemistry literature) were present in the nascent solar system.  These include some {\em r}-process species
\citep{Meyer1993,Cote2021}. Our calculations could be used to provide a new evaluation of pre-decay abundances
for the early Solar System.

The discovery of \fe60 in deep-ocean deposits heralded a new era of laboratory astrophysics using relic live
radioisotopes to explore events in the solar neighborhood, using them to understand quantitatively their
contributions to nucleosynthesis. Previous studies have focused on nearby SNe and their roles in
forming the Local Bubble. The hints of discovery of \pu244 carry these studies to a new level, including
the possibility of gathering experimental information about sites of the $r$-process, extending the
catalog of events of interest to include KNe, possibly out to a larger range surrounding the
Local Bubble. This development also extends and expands the investigation of the potential repercussions 
on Earth of astrophysical events, including possible impacts on the biosphere. The longer half-life of
\pu244 (81~Myr) compared to \fe60 (2.5~Myr) offers the possibility of studying the implications of such
events in the more distant past, and exploring directly possible links to mass terrestrial extinctions,
opening up a new frontier in astrobiology.

\section{Conclusions and Future Directions}
\label{sect:Conx}

The \citet{Wallner2021} detection of a substantial \pu244 signal in the deep ocean dramatically broadens the study of near-Earth explosions,
because it demands that an {\em r}-process event occurred relatively recently and close by.
Indeed, \pu244 not only originates exclusively in the {\em r} process, but also is one of the heaviest nuclei
it can produce.  Thus the firm detection of \pu244 not only opens a new window into the elusive {\em r}-process astrophysical site(s),
but also points to an engine capable of synthesizing the fullest possible complement of species, extending to actinides.
The \pu244 coincidence in time with \fe60--now seen in two distinct events--suggests a common origin in SNe,
yet modern SN models often struggle to host an {\em r}-process site, and typically yield no \pu244 at all.
Observationally, {\em r}-process abundances in halo stars and dwarf galaxies also demand that most
SNe do not make the {\em r}-process.  Thus one is driven also to consider a separate event--a neutron star merger,
where theoretical calculations show robust synthesis of actinides including \pu244, now supported by evidence
of substantial {\em r}-process production in the GW170817 KN.

We have therefore presented pairs of SN and KN models for {\em r}-process actinide production and delivery to Earth.
The former illustrate scenarios for SN actinide production in a forced neutrino-driven wind model
and in an MHD model; these are adjusted to fit the data on abundances measured in the metal-poor star HD160617.
We have also presented two representative KN models, 
with different combinations of dynamical ejecta and disk wind contributions, chosen to fit the data on HD160617 and an actinide-boost star
J0954+5246. These models indicate that radioisotopes of interest with half-lives between about 1 and 100~Myr include \zr93, \pd107, \i129, \cs135, and \hf182.
As seen in Fig.~\ref{fig:geo-ratios-3Myr}, at least some of
these may be detectable using AMS techniques, and their presence or non-detection could constrain
significantly models of $r$-process nucleosynthesis. 

The reported
observations of \pu244 are nicely compatible with the the direct (one-step) deposition of
explosion debris of an {\em r}-process-enhanced SN within
${\cal O}(100)$~pc of Earth, the same distance range postulated to explain the observed \fe60 signal. 
But most SNe do not make {\em r}-process 
radioisotopes and certainly not {\em r}-process actinides, 
so if the \pu244 had been produced by two SNe, 
both must be rare events 
(and it must be possible for some SNe to produce actinides).  
If instead we try to account  for the \pu244 by direct (one-step) deposition by a KN,
the required distances are unfeasibly large. As an alternative, we have proposed a two-step scenario, in which some of the \pu244 produced
by a more distant KN was absorbed into the 
ISM within the Local Bubble, before subsequently
reaching Earth among the debris from a nearby SN.\\

We now summarize our conclusions.
First, we reviewed the state of present and future {\em r}-process radioisotope observations.
\begin{itemize}

    \item We have compiled a comprehensive list of
     possible signatures of near-Earth explosions
     in the form of medium-lived radioisotopes, summarizing their possible astrophysical origins and geological prospects, and 
     assessing their detectability via AMS. Thereafter
     we have focused on {\em r}-process species, and presented new calculations of {\em r}-process radioisotope yields and uncertainties for selected SN and KN scenarios constrained to reproduce solar and actinide-boosted halo star abundance patterns.
     
     \item  Turning to geological data, we reviewed data on deep-ocean \pu244 in Fe-Mn crusts and sediments, combining searches by four groups.  The published indications are
     intriguing: The published evidence for \pu244 deposits spans an extended time period from 1 to between 12 and 25 Mya.  The largest fluxes overlap the times of SN \fe60 deposition, but the sustained flux over a much longer interval (if real) points toward a separate mechanism for {\em r}-process deposition on Earth.
\end{itemize}

We then performed {\em r}-process nucleosynthesis calculations, and linked them to astrophysical models for radioisotope delivery to Earth.
\begin{itemize}     
          \item
     We find that, for both SNe and KNe, \i129 is the most abundant product, with {\em r}-process mass fractions that are
     comparable in the different models we address here.  The production of \pd107 is similarly robust, but there are progressively larger variations in the yields of \zr93, \cs135, and \hf182. The rates of actinide production show the largest variations between the models, but their production ratios are relatively stable. Motivated by the discovery
     of live \pu244 in the deep ocean, we have used these observations to anchor predictions for other actinide radioisotopes.

     \item For SNe, we have studied {\em r}-process radioisotope production in both a forced neutrino-driven wind scenario and a magnetohydrodynamic model. These SN scenarios struggle to make actinides at all unless neutrino processes are able to create low $Y_e$ conditions.
     Thus the SN models, and particularly the unmodified MHD SN model, give high ratios of the lighter {\em r}-process radioisotopes relative to \pu244. On the other hand,
     KNe resulting from neutron star mergers show robust {\em r}-process actinide 
     synthesis, and therefore have lower ratios of the lighter
     {\em r}-process radioisotopes relative to \pu244.

     \item  We have studied the delivery to Earth by a nearby SN 3~Mya of \pu244 and other {\em r}-process radioisotopes along with \fe60.  In this one-step model, the inferred \pu244 yields can be accommodated in both the forced neutrino-driven wind and MHD models.
     By contrast, such a one-step direct deposition of KN products is strongly inconsistent with both \fe60 and \pu244 data.
     We therefore constructed a two-step model for {\em r}-process delivery to Earth in which a KN explosion $\sim 10$ to $\sim 50$ Mya enriched the molecular cloud that gave rise to the Local Bubble.  Dust grains seeded thereby with {\em r}-process radioiostopes would have bombarded the Earth thereafter.  This model can account for the observed \pu244 with plausible KN distances and rates.

\end{itemize}

Finally we presented a series of predictions for specific isotope ratios and discussed their observability.
\begin{itemize}
     \item We have presented predictions for AMS searches
     for {\em r}-process radioisotopes in deep-ocean Fe-Mn crusts, in both the one-step SN and two-step KN scenarios.  With the current AMS sensitivities, \i129 is the most promising. Potential AMS advances could make \hf182 another very interesting target.  Additional AMS improvements would be needed to detect \zr93, \pd107, and \cs135. Whilst alternative production mechanisms may elevate the \u236 background, measurements of \u236, \np237, and \cm247 may serve as important complementary measurements to the \pu244 data. 
     
     \item  We have also reviewed the Moon's fossil record of radioisotopes,
     where the systematics are very different from,
     and complementary to, those of geological searches. We recall that \fe60 has already been discovered in the {\it Apollo}
     samples of the lunar regolith, and have discussed opportunities for the new generation of lunar sample return missions to
     yield evidence for $r$-process radioisotopes.

\end{itemize}

Our work and its larger context suggest
many directions for future work, which we summarize here.\\

\textit{Astrophysical models}:
\begin{itemize}     
     
    \item
    Explosion models remain essential, and 
    we urge that radioisotope yields be reported for calculations within
    models of both SNe and KNe and in studies of their {\em r}-process outputs.
    An improved treatment of neutrino interactions is critical, particularly for evaluating
    whether realistic neutrino physics allows for SN actinide production at all in the absence of jets.
    Similarly, more realistic SN jet models are needed to assess their ability and frequency of
    actinide production.
    
 \item
     We urge continued work on
     models of SN dust formation,
     propagation, and injection into the heliosphere.
    Astrophysical models of the Local Bubble are critical,
    and it would be of particular interest to study the effects of injection of debris from a nearby KN.
    To establish the broader context of the Local Bubble and recent radioisotope production, we urge continued
    studies of chemical evolution models for the larger solar neighborhood.  Of particular importance is the distribution of radioisotopes including
    {\em r}-process species, as well as their variation.

\end{itemize}

\textit{Geological and lunar studies with AMS}:
\begin{itemize}

    \item 
    We urge geological searches for {\em r}-process
    radioisotopes in deep-ocean Fe-Mn crusts, most pressingly \i129.  There is also a clear need for more sensitive searches in deep-ocean sediments for other live radioisotopes, coincident
    with the \fe60 pulses and the putative \pu244 signals in sediments.
    Improved time resolution for \pu244, including additional searches earlier than 10 Mya, is critical to determine if its flux history is coincident with or distinct from that of \fe60.  Time resolution for additional radioisotopes will be similarly illuminating.

    \item  We urge searches for {\em r}-process isotopes in the regolith samples brought to Earth recently by the {\em Chang'e-5}
    lunar mission and upcoming missions including {\em Artemis}.

     \item
     We urge efforts to improve the  AMS sensitivities for {\em r}-process radioisotopes, especially for \zr93, \pd107, \hf182, and \pu244.
     
     \item 
    Probes of possible live radioisotope deposits in the more distant past would also be interesting. In particular, 
    the end of the Devonian epoch $\sim 360$~Mya 
    experienced an extinction event coincident
    with radiation UV-B damage to plant spores~\citep{Marshall2020}. 
    This could have been due to     
    destruction of the ozone layer during cosmic-ray bombardment following an SN explosion $\sim 20$~pc away~\citep{Fields:2020nmv}. 
    \pu244 and possibly also \u236 
    can provide evidence of an SN connection to the 
    extinction if the event produced the {\em r}-process.

\end{itemize}

\textit{Nuclear experiments and theory}:
\begin{itemize}
     
     \item
     We look forward to new data on unstable, neutron-rich nuclides whose properties set the abundance ratios described here, from experiments at current and upcoming radioactive beam facilities such as CARIBU~\citep{Savard+2008} and the $N=126$ factory~\citep{Savard2020} at ATLAS, RIBF at RIKEN~\citep{Motobayashi+2012}, ISAC/ARIEL at TRIUMF~\citep{Dilling+2014a,Dilling+2014b}, FAIR at GSI~\citep{Kester+2016}, and FRIB~\citep{FRIB,Horowitz+2019}.~\footnote{See also \url{https://groups.nscl.msu.edu/frib/rates/fribrates.html}.}
     
     \item
    Some key nuclear properties---particularly those of very neutron-rich species close to the neutron drip line---will remain inaccessible to experiment for the foreseeable future. Thus, advances in nuclear structure and reaction theory will also be required to improve estimates of masses, reaction rates, and fission properties and reduce yield prediction uncertainties.
     
\end{itemize}

We have entered an era of multi-messenger astrophysics, with the advent of gravitational wave, cosmic ray, and neutrino measurements that
complement the many decades of the electromagnetic spectrum that were already being explored. 
In particular, a new window into {\em r}-process nucleosynthesis opened with the simultaneous detection of a neutron star merger 
gravitational-wave event GW170817 and the electromagnetic discovery of the associated KN outburst~\citep{GBM:2017lvd,Monitor:2017mdv}.
To these multi-messenger observations may be added the detection of live radioisotopes, which cast light not only on nucleosynthesis, but also on the history of the solar neighborhood and the potential impacts of nearby astrophysical
events on the terrestrial environment and life. The first observations of deep-ocean deposits of live \fe60 have been confirmed
by many other experiments, extending to measurements of \fe60 in samples of the lunar regolith, Antarctic ice, and cosmic rays.
This wealth of data firmly establishes that nearby SNe
occurred about 3 and 7~Myr ago, and \pu244 deposition on similar time scales has now been established.  

The detection of deep-ocean \pu244 represents a novel probe of the {\em r}-process, one that is complementary to observations of neutron star mergers and stellar abundance patterns.  Geophysical samples of live \pu244, and potentially other {\em r}-process radioisotopes, offer new laboratory-based probes of fresh nucleosynthesis products.
These measurements are highly sensitive and isotopically specific, tied to a single nucleosynthesis event,  and capable of helping us to infer the nature of the source--though only indirectly.
In contrast, in neutron star merger observations the basic nature of the event is known, but the nucleosynthesis output may only be inferred indirectly and without isotopic or even elemental specificity.
The abundances of {\em r}-process species in low-metallicity halo and dwarf galaxy stars
offer a wealth of elemental information but generally no isotopic
information. In general, stellar abundances sum over multiple synthesis events
and, like geo-radioisotopes, do not directly identify the source(s).
Clearly the most fruitful strategy is to use all of these {\em r}-process probes together,
and now deep-ocean data contributes to this holistic approach.

\acknowledgments
We are grateful for illuminating discussions on geological 
issues with Tom Johnson and Craig Lundstrom,
on AMS with Philippe Collon,
and on star formation with Tony Wong and Leslie Looney.
The work of X.W. was supported by the U.S. National Science Foundation (NSF) under grants No. PHY-1630782 and PHY-2020275 for the Network for Neutrinos, Nuclear Astrophysics, and Symmetries (N3AS) and by the Heising-Simons Foundation under award 00F1C7.
The work of A.M.C. was supported by the U.S. Nuclear Regulatory Committee Award 31310019M0037 and the National Science Foundation under grant number  PHY-2011890.
The work of J.E. was supported partly by the United Kingdom STFC Grant ST/T000759/1 
and partly by the Estonian Research Council via a Mobilitas Pluss grant. 
The work of A.E. and B.D.F. was supported in part by the NSF under grant number AST-2108589.
The work of J.A.M. was supported by the Future Investigators in NASA Earth and Space Science and Technology (FINESST) program under grant number NNH19ZDA001N-FINESST.
The work of R.S. was supported by N3AS as well as the U.S. Department of Energy under Nuclear Theory Contract No. DE-FG02-95-ER40934.

\noindent
\software{\\
 Matplotlib \citep[][http://dx.doi.org/10.1109/MCSE.2007.55]{matplotlib}, \\
 Numpy \citep[][https://doi.org/10.1109/MCSE.2011.37]{numpy1, numpy2}, \\
 Portable Routines for Integrated nucleoSynthesis Modeling (PRISM)~\citep{Mumpower2018,Sprouse2020}.}

\bibliography{SNr}
\bibliographystyle{aasjournal}

\end{document}